\documentclass[preprint,12pt,letterpaper]{elsarticle}

\pdfoutput=1 

\usepackage{amsbsy}
\usepackage{amssymb}
\usepackage{amsthm}
\usepackage{amsmath}
\usepackage{stackrel}
\usepackage{float}
\usepackage{bm}
\usepackage{graphicx}
\usepackage{multirow}
\usepackage{rotating}
\usepackage{subfigure}
\usepackage{captcont}
\usepackage{url}
\usepackage{color}
\usepackage{xspace}
\usepackage{natbib}
\usepackage{hypernat}
\usepackage{ifpdf}
\ifpdf
  \usepackage{epstopdf} 
\fi
\usepackage[T1]{fontenc}

\usepackage[letterpaper]{geometry}
\usepackage[colorlinks=true, citecolor=blue, filecolor=black,linkcolor=red, urlcolor=blue, pdftex]{hyperref}
\usepackage[ruled]{algorithm2e}

\graphicspath{{figures/}}

\renewcommand{\[}{\left[}
\renewcommand{\]}{\right]}
\renewcommand{\(}{\left(}
\renewcommand{\)}{\right)}

\newcommand{\pp}[2]{\frac{\partial #1}{\partial #2}}

\newcommand{\be}{\begin{equation}}
\newcommand{\ee}{\end{equation}}
\newcommand{\bea}{\begin{eqnarray}}
\newcommand{\eea}{\end{eqnarray}}
\newcommand{\ben}{\begin{equation*}}
\newcommand{\een}{\end{equation*}}
\newcommand{\bean}{\begin{eqnarray*}}
\newcommand{\eean}{\end{eqnarray*}}

\newcommand{\EE}{\mathbb{E}}

\newcommand{\PP}{\mathbb{P}}
\newcommand{\RR}{\mathbb{R}}

\newcommand{\CF}{\mathcal{F}}

\newcommand{\pt}{\tilde{p}}

\newcommand{\pd}{\partial}

\DeclareMathOperator{\Var}{Var}
\DeclareMathOperator{\V}{Var}
\DeclareMathOperator{\Cov}{Cov}
\DeclareMathOperator{\card}{card}
\DeclareMathOperator{\trace}{trace}

\DeclareMathOperator{\erf}{erf}

\newtheorem{theorem}{Theorem}[section]
\newtheorem{problem}[theorem]{Problem}

\SetKwHangingKw{KwSolve}{Solve}

\journal{Journal of Computational Physics}

\begin{document}


\begin{frontmatter}
  
  \title{Bayesian Inference with Optimal Maps}

  \author{Tarek A.\ El Moselhy}
  \ead{tmoselhy@mit.edu}

  \author{Youssef M.\ Marzouk}
  \ead{ymarz@mit.edu}
  
  \address{Massachusetts Institute of Technology, Cambridge, MA 02139, USA}
  
  
  \begin{abstract}

  We present a new approach to Bayesian inference that entirely avoids
  Markov chain simulation, by constructing a map that \textit{pushes
    forward} the prior measure to the posterior measure. Existence and
  uniqueness of a suitable measure-preserving map is established by
  formulating the problem in the context of optimal transport theory.
  We discuss various means of explicitly parameterizing the map and
  computing it efficiently through solution of an optimization
  problem, exploiting gradient information from the forward model when
  possible. The resulting algorithm overcomes many of the
  computational bottlenecks associated with Markov chain Monte Carlo.
  Advantages of a map-based representation of the posterior include
  analytical expressions for posterior moments and the ability to
  generate arbitrary numbers of independent posterior samples without
  additional likelihood evaluations or forward solves. The
  optimization approach also provides clear convergence criteria for
  posterior approximation and facilitates model selection through
  automatic evaluation of the marginal likelihood.
  We demonstrate the accuracy and efficiency of the approach on
  nonlinear inverse problems of varying dimension, involving the
  inference of parameters appearing in ordinary and partial
  differential equations.

\end{abstract}










\end{frontmatter}


\section{Introduction}







The estimation of model parameters from observations is a ubiquitous
problem in science and engineering. Inference or ``inversion'' are
central challenges in geophysics and atmospheric science, chemical
kinetics, quantitative biology, and a host of additional domains. In
all these settings, observational data may be indirect, noisy, and
limited in number or resolution.
Quantifying the resulting \textit{uncertainty} in parameters is then
an essential part of the inference process. Uncertainty in model
parameters in turn drives uncertainty in predictions; characterizing
the latter is vital to the use of computational models in design and
decision-making, and even to subsequent rounds of data collection.
%

The Bayesian statistical approach provides a foundation for learning
from noisy and incomplete data, a natural mechanism for incorporating
heterogeneous sources of information, and a complete assessment of
uncertainty in model predictions
\cite{stuart:2010:ipa, gelman:2003:bda, kaipio:2004:bip,
  sivia:2006:daa, bernardo:1994:bt}. Uncertain quantities are
treated as random variables, and the \textit{posterior} distribution,
i.e., the probability distribution of any quantity of interest (be it
a parameter or a model prediction) conditioned on data, represents
one's state of knowledge about that quantity. Characterizing the
posterior---simulating the distribution with samples; marginalizing;
evaluating moments, quantiles, or credibility intervals---thus becomes
one of the central computational challenges in Bayesian inference.

In specialized cases (e.g., simple statistical models with conjugate
priors) expectations with respect to the posterior may be evaluated in
closed form. But in the general setting---and particularly when
complex physical models enter the likelihood function---computational
approaches are needed. By far the most widespread and versatile method
for posterior simulation in this context is Markov chain Monte Carlo
(MCMC) \cite{metropolis:1953:eos, hastings:1970:mcs, gelman:2003:bda, gilks:1996:mcm,robert:1999:mcs,junliu:2008:mcs}. MCMC requires
only pointwise evaluations of an unnormalized density, and generates a
stream of samples that can be used to evaluate posterior expectations.

Despite its power and practical utility, MCMC suffers from many
limitations. Samples generated by the algorithm are necessarily
correlated; strong correlation among successive samples leads
to smaller effective sample sizes and larger errors in posterior
estimates \cite{robert:1999:mcs}. An efficient MCMC algorithm endeavors to make
the effective sample size after $N$ steps as close to $N$ as possible,
as the posterior evaluation(s) required at each step may be
costly. Efficient sampling in Metropolis-Hastings MCMC \cite{gilks:1996:mcm}
rests on the design of effective \textit{proposal distributions}, but
this task is difficult for target distributions that contain strong
correlations or localized structure, particularly in high dimensions.
Improvements to MCMC proposal mechanisms are therefore the subject of
intense current interest. A non-exhaustive list of recent
methodological advances includes adaptivity based on past samples
\cite{haario:2006:dra,roberts:2009:eof}, Langevin methods \cite{stramer:1999:ltm,apte:2007:stp}, the
use of Hessian information \cite{girolami:2011:rml} and Hessians with low-rank
structure \cite{martin:2011:asn}, Hamiltonian dynamics \cite{neal:2011:mcm}, the
exchange of information among multiple chains \cite{higdon:2002:aba, vrugt:2009:amc,craiu:2009:lft},
multi-stage proposals incorporating approximate models \cite{christen:2005:mcm,
  efendiev:2006:pmc}, and the use of surrogate or reduced models to accelerate
likelihood evaluations \cite{marzouk:2007:ssm, marzouk:2009:dra,
  lieberman:2010:psm, frangos:2011:sar, wang:2005:ubs}.

Even with these advances, MCMC remains a computationally intensive
process; typical problems require many thousands or even millions of
posterior evaluations. Moreover, the sampling process does not come
with a clear convergence criterion, indicating when the chain has
adequately explored the distribution or how many initial samples to
discard as ``burn-in.'' Convergence diagnostics for MCMC are largely
heuristic and remain something of an art \cite{gelman:2011:ifs}.
Finally, the nature of MCMC is to represent the posterior distribution
with samples; this may seem an obvious remark, but a sample
representation immediately favors the use of Monte Carlo methods for
subsequent uncertainty propagation steps. For computationally
intensive models, however, sampling may be impractical. More efficient
uncertainty propagation methods already
exist~\cite{ghanem:2003:sfe,maître:2010:smf,xiu:2010:nmf} and it would be desirable to
use them. Even sequential Monte Carlo methods \cite{doucet:2001:smc}, designed
with recursive inference and forecasting in mind, still rely on a
weighted-sample representation of the posterior.

\vspace{1em}

In this paper, we present a novel approach to Bayesian inference that
relies on the construction of \textit{maps}. We entirely avoid Markov
chain simulation, and instead explicitly construct a map that
\textit{pushes forward} the prior measure to the posterior measure. In
other words, the map transforms a random variable $X$, distributed
according to the prior, into a random variable $Z$, distributed
according to the posterior. Existence and uniqueness of a monotone map
is assured under rather weak conditions \cite{mccann:1995:eau}. 
The map is actually found through solution of an optimization
problem, which requires only the ability to evaluate the posterior density up to a
normalizing constant (exactly as in Metropolis-Hastings MCMC). If
gradients of the likelihood function or forward model are available,
however, then we immediately take advantage of them. To make the
optimization problem finite dimensional, the map is described by
multivariate orthogonal polynomials; it therefore becomes a polynomial
chaos expansion \cite{ghanem:2003:sfe,xiu:2002:twa} of the posterior.

As a byproduct of the optimization procedure, the scheme automatically
calculates the posterior normalizing constant, i.e., the marginal
likelihood or evidence, which is an essential quantity in Bayesian
model selection \cite{kass:1995:bf}. We show that the optimization
procedure also provides an unambiguous convergence criterion, with a
quantity that can be monitored in order to decide whether to terminate
iterations or to enrich the polynomial space used to describe the map.
With a map in hand, one can cheaply generate independent posterior samples by
simulating from the prior. Also, as the map is represented by
polynomials orthogonal with respect to standard measures, 
posterior moments may be evaluated algebraically.

To place the map-based scheme in context, we note that variational
Bayesian methods \cite{jaakkola:2000:bpe} also convert inference into
an optimization problem, by approximating the posterior with a simpler
distribution from a parameterized family, perhaps also chosen to have
a particular conditional independence structure. These are
approximations of the posterior (or posterior predictive)
\textit{distribution} function, however. Distributions must be chosen
from particular families to facilitate optimization and sampling---in
contrast to the present scheme, which approximates random variables
directly.
The present scheme has closer links to implicit filtering methods
\cite{chorin:2009:isf, chorin:2010:ipf} for sequential data
assimilation. These methods rely on a weighted-sample representation
of each probability distribution, but use maps to transport particles
to high-probability regions of the posterior. Implicit filtering
effectively uses \textit{many} maps, though, one for each particle and
assimilation step. Unlike the present scheme, maps are not constructed
explicitly; rather one evaluates the action of each map on a single
particle. The role of optimal transport in data assimilation has also
been raised in \cite{reich:2011:ads}.

This paper is organized as follows: In Section~\ref{sec:formulation}  we present the basic formulation and several alternative versions of the optimization problem that yields the map (Sections~\ref{subsec:optimaltransport} and
\ref{subsec:triangularformulation}). Methods for solving these optimization problems are presented in Section~\ref{sec:solalg}. Section~\ref{sec:results} presents a range of numerical results. First, we demonstrate our inferential approach on  a linear
problem where the posterior has known analytical form (Section~\ref{subsec:linearG}). Then we perform parameter inference in
several nonlinear ODE systems with strongly non-Gaussian posteriors,
and where the posterior differs markedly from the prior
(Sections~\ref{subsec:reactionkinetics} and
\ref{subsec:toggle}). Finally, we demonstrate the approach on
high-dimensional inverse problems involving the estimation of
spatially heterogeneous coefficients in elliptic PDEs
(Section~\ref{sec:results:subsec:PDE}).

\section{Formulation}
\label{sec:formulation}

\subsection{Bayesian framework}

We begin by describing the Bayesian inference framework and
setting notation for the subsequent developments.
Consider the task of inferring model parameters $x$ from observations
$d$. For simplicity, we will let both the parameters and observations
be real-valued and finite-dimensional. In the Bayesian setting, the
parameters and observations are treated as random variables. Let
$(\Omega,\CF,\PP)$ be a probability space, where $\Omega$ is a sample
space, $\CF$ is a $\sigma$-field, and $\PP$ is a probability measure
on $(\Omega, \CF)$. Then the model parameters $X: \Omega \rightarrow
\mathbb{R}^{n}$ are associated with a \textit{prior} measure $\mu_0$ on
$\mathbb{R}^{n}$, such that $\mu_0(A) = \PP\left (X^{-1}\left ( A \right
  ) \right ) $ for $A \in \mathbb{R}^{n}$. We then define $p(x) = d
\mu_0 / d x$ to be the density of $X$ with respect to Lebesgue
measure. For the present purposes, we assume that such a density
always exists. We then define the expectation with respect to the prior measure as $\mathbb{E}\left[g(X)\right] = \int g(x) d\mu_0(x) = \int g(x) p(x) d x$, for any  Borel-measurable function $g$ on $ \mathbb{R}^{n}$.
The observational data are described by a random variable $d$ taking
values in $\mathbb{R}^m$. Inference requires that we define a
probabilistic model for the data; in terms of probability densities,
this model is simply $p(d | x )$. Viewed as a function of $x$, this
conditional density defines the \textit{likelihood function} $L(x; d) \equiv p(
d | x )$. Bayes' rule then yields the \textit{posterior} probability density
$q(x) \equiv p(x | d)$:
\begin{equation}
q(x) = \frac{ L(x; d) \, p(x) } {\beta} \label{eq:posterior}
\end{equation}
where $\beta$ is the evidence or marginal likelihood $\beta = \int
L(x; d) p(x) \, dx$. This posterior density $q$ is associated with a
posterior measure $\mu$ on $\mathbb{R}^{n}$, such that $d \mu(z) =
q(z) dz$ and the likelihood function is proportional to the
Radon-Nikodym derivative of $\mu$ with respect to $\mu_0$, $L(z) \propto
d \mu / d \mu_0$ \cite{stuart:2010:ipa}.

In inverse problems \cite{kaipio:2004:bip, tarantola:2005:ipt, stuart:2010:ipa}, the
likelihood frequently incorporates a deterministic function $h$ mapping  
the parameters to some idealized observations. This function $h:
\mathbb{R}^n \rightarrow \mathbb{R}^m$ is termed the \textit{forward
  model}. The discrepancy between the idealized model predictions and
the actual data is often assumed to be additive: $d = h(x) +
\varepsilon$, where $\varepsilon$ is a random variable. A common
further assumption is that $\varepsilon$ is Gaussian, zero-mean, and
independent of the model parameters, i.e., $\varepsilon \sim N(0,
\Sigma)$, leading to the following form for the likelihood function:
\begin{equation}
  L(x; d) = \frac{1}{ (2 \pi)^{m/2} \left ( \det \Sigma \right )^{1/2} }
    \exp \left( -\frac{1}{2} \left \| h(x) - d \right \|^2_{\Sigma} \right)  
\label{eq:linearLH}
\end{equation}
where $\| u \|_A := \| A^{-\frac{1}{2}} u \|$, for any positive symmetric matrix $A$.

While the Bayesian formulation of inverse problems commonly leads to
likelihood functions that take the form of (\ref{eq:linearLH}), we
emphasize that this particular form is not an assumption or
requirement of the map-based inference method described below.

\subsection{Inference with a map}

\begin{figure}[htb]
 \centering
 \includegraphics[width=4.8in]{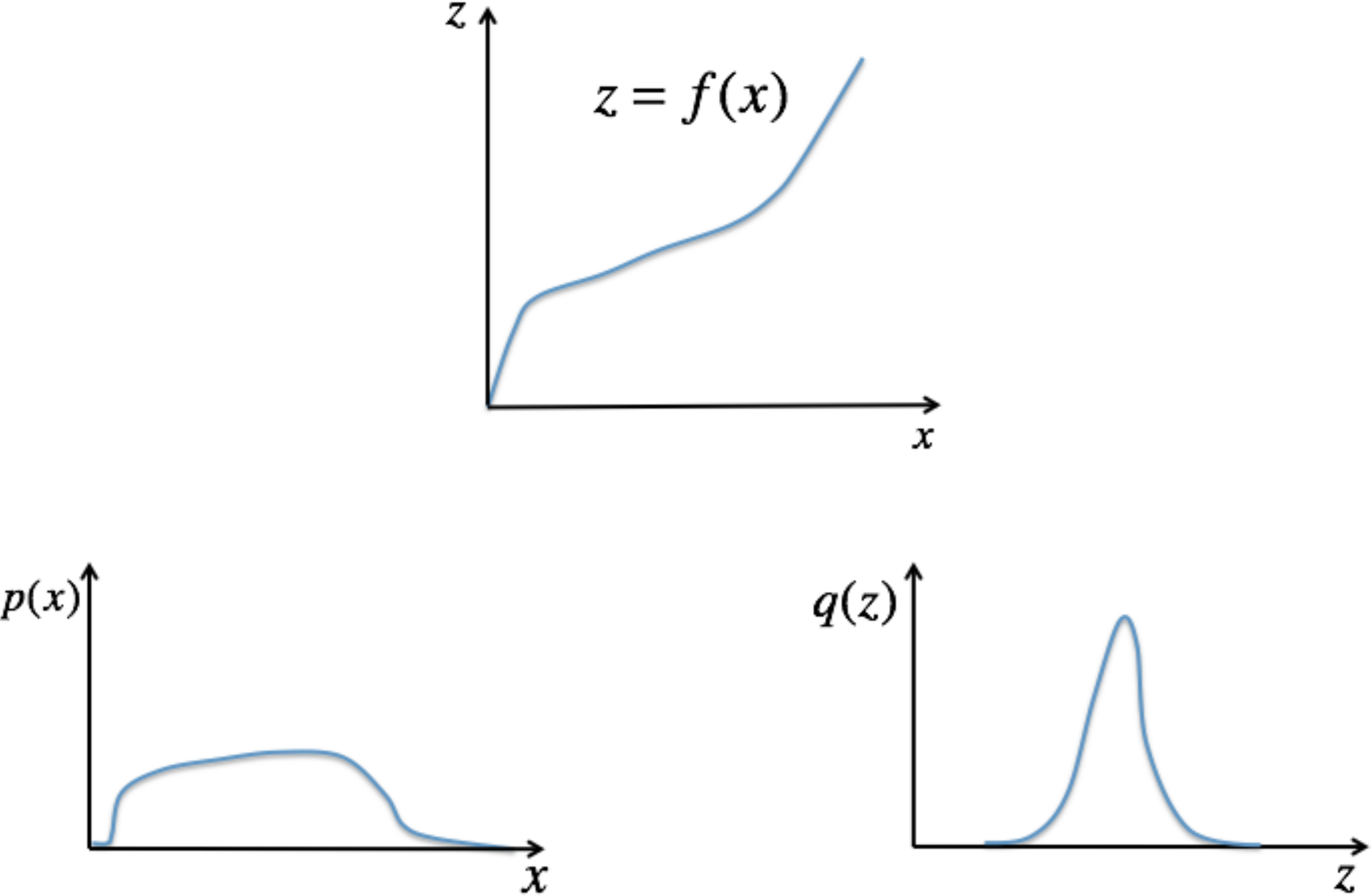}
 \caption{Schematic of a function $f$ that pushes forward the prior
   probability distribution (represented by density $p$) to the
   posterior probability distribution (represented by density $q$).
   The objective of our algorithm is to efficiently compute such a function
   in high dimensional spaces.}
\label{fig:measuretransform}
\end{figure}

The core idea of our approach is to find a map that pushes forward the
prior measure to the posterior measure. In other words, suppose that
$X$ is a random variable distributed according to the prior and that
$Z$ is a random variable distributed according to the posterior. Then
we seek a map $f: \mathbb{R}^{n} \rightarrow \mathbb{R}^{n}$ that
satisfies the following constraint, where equality is \textit{in
  distribution}:
\begin{equation}
Z = f(X), \ \mathrm{with} \ X \sim \mu_0, \, Z \sim \mu.
\label{eq:map}
\end{equation}
Equivalently, we seek a map $f$ which pushes forward $\mu_0$ to $\mu$, i.e., for which $\mu =
f_{\sharp}\mu_0$. A schematic of such a map is given in
Figure~\ref{fig:measuretransform}. The map necessarily depends on the
data and the form of the likelihood function. (In the case of
(\ref{eq:linearLH}), the map would thus depend on the data $d$, the
forward model $h$, and the distribution of the observational noise
$\varepsilon$.) If the posterior distribution were equal to the prior
distribution, an identity map would suffice; otherwise more complicated
functions are necessary. Note that, for clarity, we adhere to the
strict association of a random variable with a distribution. Thus the
prior and posterior distributions are associated with
\textit{distinct} random variables ($X$ and $Z$, respectively)
representing distinct states of knowledge about the same set of parameters.
This view is slightly different than the usual notion of a single
random variable for which we ``update'' our belief.

\begin{figure}[htb]
 \centering
 \includegraphics[width=4.8in]{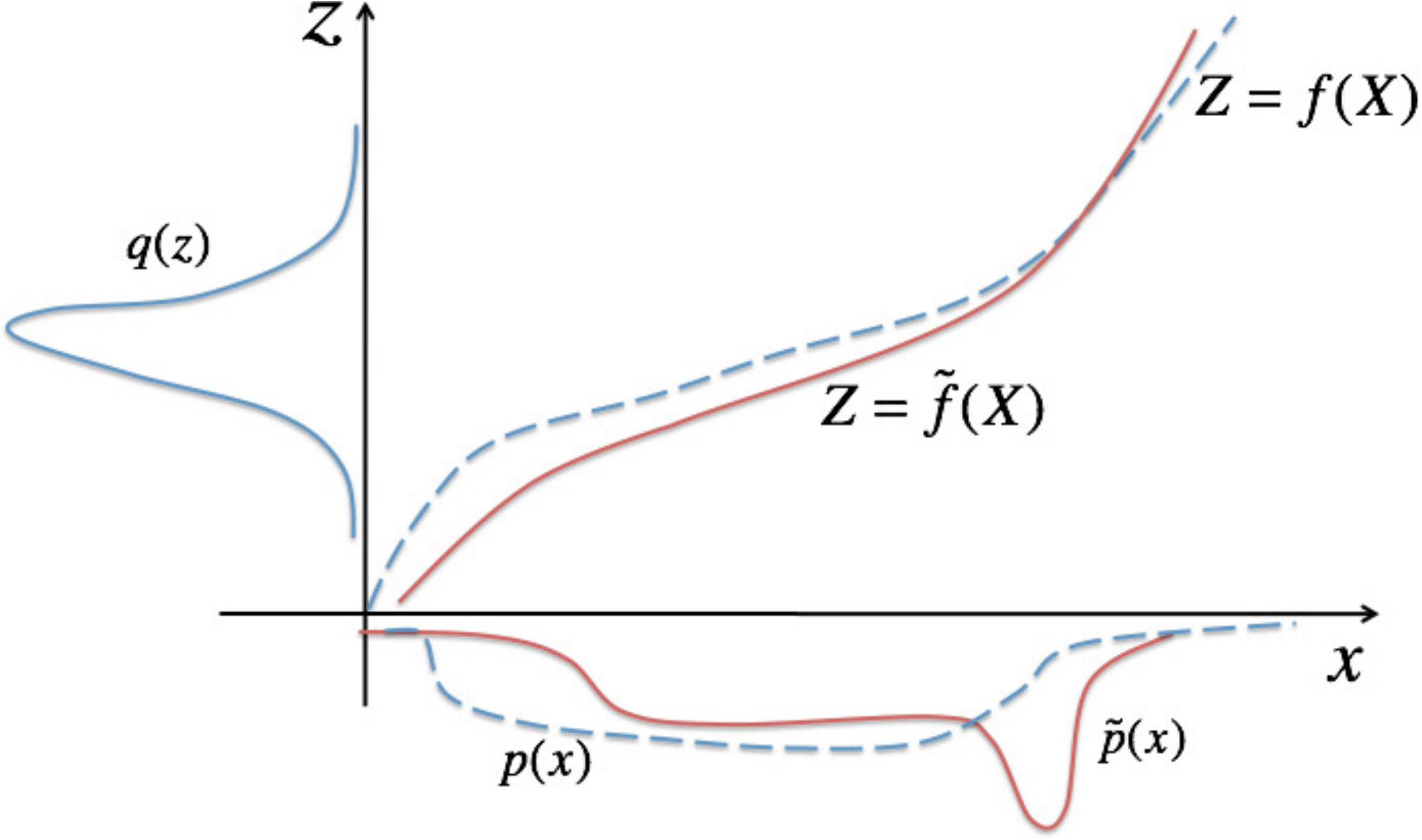}
 \caption{Pictorial representation of the optimization scheme. A
   candidate map $f$ transforms the posterior density $q$ into an approximation
   $\tilde{p}$ of the true prior density $p$. The map is adjusted to
   minimize the distance between $\tilde{p}$ and $p$; when this
   distance is brought below a prescribed threshold, optimization
   iteration termainate and one has obtained the desired map.}
\label{fig:mainideacartoon}
\end{figure}

Assuming that a map satisfying (\ref{eq:map}) exists and is
monotone,\footnote{A monotone (non-decreasing) map $f(x)$ on
  $\mathbb{R}^n$ is one for which $\left ( x_2 - x_1 \right )^T \left
    (f(x_2) - f(x_1) \right ) \geq 0 $ for every pair of distinct
  points $x_1, x_2 \in \RR^n$. 
  Issues of existence and monotonicity will be addressed in the next
  two subsections.} the measure of $Z$, represented by the posterior
probability density $q$, can be transformed into a probability
density $\pt$ for the input random variable $X$:
\begin{equation}
\pt(x) = q \left ( f(x) \right ) \left | \det{D_x f} \right | 
= \frac{ L \left ( f(x); d \right ) p \left ( f(x) \right) }
{\beta} \left | \det D_x f \right |  \label{eq:transformedposterior}
\end{equation}
where $D_x f = \partial f / \partial x$ is the Jacobian of the
map. But we already have a density for $X$, namely the prior density
$p$. This in turn defines an alternate criterion for an inferential
map: the transformed density $\pt$ should be equal 
to 
$p$, as depicted in Figure~\ref{fig:mainideacartoon}:
\begin{equation}
\pt(x) = p(x).
\end{equation} 
The Bayesian inference problem can thus be cast in the following
equivalent form:
\begin{problem}
  Find a monotone map $f(x)$ such that\footnote{To simplify notation we use $[L\circ f](x)$ to denote $L(f(x))$. We also drop the explicit dependence of $L$ on the data $d$.}
  \begin{equation}
    \frac{ \left[L\circ f\right](x) \left[p\circ f\right](x) }{\beta} 
    \left | \det D_x f \right |  = p(x) \label{eq:exactprobformulation}.
  \end{equation}
\end{problem}
 In many cases (e.g., when the likelihood function and the prior
density contain exponentials) the following expression is
more appropriate for computation than (\ref{eq:exactprobformulation}):
\begin{equation}
\log{\left( L \circ f \right)} + \log{\left( p \circ f \right)} -
\log{\beta} + \log \left | \det{D_x f} \right | - \log p = 0 .
\end{equation}
We emphasize that despite the appearance of the posterior normalizing constant $\beta$ in the problem above, our final formulation will never require knowledge of $\beta$. Indeed, the value of $\beta$ will emerge as a by-product of identifying the map. 

Instead of aiming for exact equality in
(\ref{eq:exactprobformulation}), we will define an optimization
problem that seeks to minimize the discrepancy between the prior
density $p$ and the approximate and map-dependent prior density
$\pt$. This discrepancy can be expressed, for instance, as the
Kullback-Leibler divergence or the Hellinger distance from $p$ to
$\pt$. We will begin by rewriting these discrepancies in a more
convenient form. For the Hellinger distance $Q(p, \pt)$ we have:
\begin{eqnarray}
  Q\left (p, \pt \right) & = &  \frac{1}{2} \int \left(\sqrt{p(x)} -
    \sqrt{\pt(x)} \right)^2 \, dx  \nonumber \\
  &=& \frac{1}{2} \int{ p(x) + \pt(x) - 2  \sqrt{p(x) \pt(x)  } \, dx } \nonumber \\
  &=& \int{ 1 - \sqrt{p(x) \pt(x)  } \, dx } \nonumber\\
  &=& 1 -  \int{ p(x)  \sqrt{\frac{ \pt(x) }{p(x)}  } \, dx } \nonumber \\
  &=& 1 -  \mathbb{E} \left[ \sqrt{\frac{ \pt(X) }{p(X)} } \right] . 
\label{eq:hellinger}
\end{eqnarray}
Similarly, the Kullback-Leibler (KL) divergence from $p$ to $\pt$ can be
rewritten as follows:
\begin{equation}
D_{\mathrm{KL}} \left ( p || \pt \right ) 
= \int{ p(x) \log \frac{p(x)}{\pt(x)} \, dx }  
= -\mathbb{E} \left[ \log{  \frac{\pt(X) }{p(X)} } \right] .
\label{eq:kldivergence}
\end{equation}
Both the square root in~(\ref{eq:hellinger}) and the
$\log$ in~(\ref{eq:kldivergence}) are concave functions $\Phi$,
and Jensen's inequality provides that
\begin{equation}
\mathbb{E}\left[ \Phi \left (\frac{\pt } {p}\right)  \right]  \le \Phi
\left (\mathbb{E}\left[  \frac{\pt } {p } \right] \right ) .
\end{equation}
To minimize either Hellinger distance (\ref{eq:hellinger}) or KL
divergence (\ref{eq:kldivergence}) we would therefore like to achieve
equality above, but equality holds if and only if the
ratio $\pt/{p}$ is constant in $x$ \cite{mcshane:1937:ji}. Consistent with
(\ref{eq:exactprobformulation}), the value of this constant should be
unity.  Re-arranging equation~(\ref{eq:transformedposterior}), we
obtain the following expression for $\beta$
\begin{equation}
  \beta = \frac{\left[L \circ f\right](x) \left[ p \circ f\right](x) }{ p(x)}
  \left | \det{D_x f} \right | .
 \label{eq:getbeta}
\end{equation}
Taking the logarithm of this expression we obtain
\begin{equation}
T(x; f ) := \log{\left( L \circ f \right)} + \log{\left( p \circ f \right)}
+ \log \left | \det{D_x f} \right | - \log p = \log\beta .
\label{eq:constantT}
\end{equation}
Equation~(\ref{eq:constantT}) is the cornerstone of our analysis. It
states that $T(x; f)$ should be constant over the support of the
prior. Setting $T$ to be a constant suggests the following problem
formulation:
\begin{problem}
\label{prob:equivalence}
Find $f^{\ast}(x)$ such that
\begin{equation}
  f^\ast = \arg \min_{f} {\V\left[T(X; f) \right]}
 \label{eq:reform0}   
\end{equation}
where $X \sim \mu_0$.
\end{problem}
Note that this formulation is equivalent to minimizing either of the
discrepancy measures between $\pt$ and $p$ considered above (and
others as well). Moreover, if $f$ is allowed to lie within a
sufficiently rich function space, we know exactly what minimum value
of $\V \left[T \right]$ can be attained: zero. As emphasized earlier, $f^\ast$ is computed without any knowledge of  the evidence $\beta$, since the latter does not appear in $T$.
However, $\beta$ can be evaluated as
\begin{equation}
\beta = \exp{\left(\mathbb{E}\left[T(X)\right] \right)} \ . 
\end{equation}
%
%
An alternative optimization problem can obtained by observing that
\begin{equation*}
  D_{\mathrm{KL}} \left ( p || \pt \right )  = 
  -\mathbb{E} \left[ \log{  \frac{\pt(X) }{p(X)} } \right] =  
  \log\beta - \mathbb{E} \left[ T(X) \right] .
\end{equation*}
Since $\log\beta$ is a constant, minimizing $D_{\mathrm{KL}} \left ( p
  || \pt \right )$ is equivalent to seeking:
\begin{equation}
f^\ast = \arg \max_{f} {\mathbb{E}\left[T(X; f) \right]} . 
 \label{eq:reform1} 
\end{equation}
In other words, we maximize the numerical estimate of the
evidence. Unlike Problem~\ref{prob:equivalence}, this formulation
does not have a known optimal value of the objective function, but
$\Var[T]$ can nonetheless be monitored to assess convergence.
%


%

\subsection{The optimal transport formulation}
\label{subsec:optimaltransport}

The multivariate map $f: \RR^{n} \rightarrow \RR^{n}$ takes the
following general form:
\begin{equation}
z_i = f_i(x_1, x_2, \ldots, x_n), \qquad i=1\ldots n
\end{equation}
where $f_i$ are the components of $f$.
Using the classic measure transform literature \cite{brenier:1991:pfa, mccann:1995:eau,
  caffarelli:1992:tro, gangbo:1996:tgo, villani:2009:oto} one can show that a map pushing $\mu_0$ forward to
$\mu$ exists under relatively weak conditions, but is not unique. To
guarantee uniqueness, extra constraints or penalties must be
enforced. Two types of constraints are employed in this paper. The
first, discussed in this section, is an optimal transport
constraint. It can be shown \cite{mccann:1995:eau}, under remarkably general
conditions, that the optimization problem
\begin{eqnarray}
\min_f {\mathbb{E}\left[ \left \| X-f(X) \right \|^2 \right ] }
\nonumber \\
\mathrm{subject \ to} \ \mu = f_{\sharp}\mu_0  \label{eq:optimaltransport}
\end{eqnarray}
has a unique solution $f^\ast$ and that this solution is
\textit{monotone}. Conditions for existence and uniqueness essentially
require that the measure $\mu_0$ (here, the prior) not contain any
atoms. We restate this result more precisely as follows:
\begin{theorem}
\label{tm:optimaltransport}
(After \cite{mccann:1995:eau, villani:2009:oto}) Let $\mu_0$ and $\mu$ be Borel
probability measures on $\mathbb{R}^n$ with $\mu_0$ vanishing on subsets
of $\mathbb{R}^n$ having Hausdorff dimension less than or equal to $n-1$. Then the
optimization problem (\ref{eq:optimaltransport}) has a solution $f$
that is uniquely determined $\mu_0$-almost everywhere. This map is the
gradient of a convex function and is therefore monotone.
\end{theorem}
For a detailed proof of this theorem see, e.g., \cite{villani:2009:oto}.

To be sure, our true objective is not to find the optimal transport of
Theorem~\ref{tm:optimaltransport} per se. Instead we want to add
enough regularity to Problem~\ref{prob:equivalence} such that we can
find \textit{a} monotone map satisfying $f_{\sharp} \mu_0 \approx \mu$ (or
equivalently $\pt \approx p $). To this end, we use the optimal
transport problem~(\ref{eq:optimaltransport}) to motivate the
following penalized objective:
\begin{equation}
\min_f \left \{ D_{\mathrm{KL}}\left( p || \pt \right) +
\lambda \mathbb{E}\left[ \left \| X-f(X) \right \|^2 \right ]  \right \}.
\label{eq:formOTkl}
\end{equation}
By analogy with Problem~\ref{prob:equivalence}, (\ref{eq:formOTkl}) can be replaced by:
\begin{problem}
\label{prob:OT}
Find $f$ 
to solve the following optimization problem
\begin{equation}
\min_{f} \left \{ \V \left[T(X) \right] + \lambda \mathbb{E}\left[ \left \|
    X-f(X) \right \|^2 \right ] \right \} .
\label{eq:formOTvar}
\end{equation}
\end{problem}
Equations~(\ref{eq:formOTkl}) and~(\ref{eq:formOTvar}) should be
understood as follows. The first term enforces the measure
transformation from prior to posterior,\footnote{See \ref{app:KLversusVar} for a discussion of the relative
magnitudes of $\Var[T]$ and the KL divergence.} while the second term is a
penalty inspired by the optimal transport formulation to promote
regularity and monotonicity in $f$.  
The magnitude of the penalty term is controlled by the multiplier
$\lambda$. In the optimization scheme to be detailed in
Section~\ref{sec:solalg}, this multiplier is chosen adaptively, such that the
magnitude of the penalty term is a prescribed fraction (e.g., $1/10$)
of the magnitude of $\Var\left [T \right ]$ or $ D_{\mathrm{KL}}\left(
  p || \pt \right)$ at the start of each cycle.
The stopping criterion for the optimization scheme evaluates only the
first term of (\ref{eq:formOTkl}) or (\ref{eq:formOTvar}), since this
is the term that enforces the accuracy of the posterior
representation.
We also emphasize that in every example tested below (see
Section~\ref{sec:results}), the value of $\lambda$ was not a critical
parameter in the optimization procedure. In fact $\lambda$ can be
assigned a zero value when $\V\left[T(X) \right]$ is sufficiently
small.

One could also, of course, revert to solving the constrained
optimization problem~(\ref{eq:optimaltransport}) and thus find the
true $L^2$-optimal transport.  
We do not pursue such a scheme here, however, because our core interest is in satisfying the measure transformation condition, not in finding the optimal transport per se.

\subsection{Triangular formulation}
\label{subsec:triangularformulation}

An alternative method for guaranteeing a unique solution is to enforce
a ``triangular'' structure for the map. This construction is inspired
by the Rosenblatt transformation \cite{rosenblatt:1952:roa, knothe:1957:ctt}, where now
the $i$th component of $f$ can depend only on the first $i$ inputs:
\begin{equation}
z_i = f_i(x_1, \ldots, x_i) . \label{eq:rosenblattform}
\end{equation}
In this formulation, the regularity of the problem is provided by the
triangular structure and no additional penalty term is required to
ensure uniqueness. Indeed, \cite{carlier:2010:fkt} shows that there
exists a \textit{unique monotone} map $f$ of the form
(\ref{eq:rosenblattform}) satisfying $\mu = f_{\sharp}
\mu_0$.\footnote{This result \cite{carlier:2010:fkt} places conditions
  on $\mu_0$ essentially requiring that the marginals of $\mu_0$ have no
  atoms. A sufficient condition is that $\mu_0$ be absolutely continuous
  with respect to Lebesgue measure. Also, just as in
  Theorem~\ref{tm:optimaltransport}, uniqueness of the map holds
  $\mu_0$-almost everywhere.} This map is known as the Knothe-Rosenblatt
re-arrangement \cite{carlier:2010:fkt,villani:2009:oto}.


The Rosenblatt re-arrangement is typically computed via an iterative
procedure \cite{villani:2009:oto} that involves evaluating and
inverting a series of marginalized conditional cumulative distribution
functions. This procedure is computationally very intensive, since it
involves computing a large number of high-dimensional
integrals. Instead of explicitly computing the Rosenblatt
transformation in this way, we propose to solve the following problem:
\begin{problem}
\label{prob:RT}
Find 
\begin{equation}
f^\ast = \arg\min_{f}  \V \left[T(X) \right],
\label{eq:formRTvar}
\end{equation}
where $f$ is subject to the structure~(\ref{eq:rosenblattform}).
\end{problem}
One could modify the optimization problem by replacing $\V \left[T(X) \right]$ with $ D_{\mathrm{KL}}\left(
  p || \pt \right)$ in~(\ref{eq:formRTvar}).
We thus propose to
find a Rosenblatt-type re-arrangement all-at-once rather than
component by component, and with a stopping criterion that involves
the magnitude of the entire objective function being smaller than a
prescribed threshold. We must acknowledge, however, that the current
theory \cite{carlier:2010:fkt} does not preclude the existence of
non-monotone measure-preserving maps that have the triangular
structure. Thus the question of whether minimizing $\Var \left [T(X)
\right ]$ subject to (\ref{eq:rosenblattform}) \textit{guarantees}
monotonicity of the map remains open. In numerical tests on a wide
variety of inference problems, however, the solutions we obtain using
the triangular constraint are consistently monotone. Monotonicity is easily
verified by examining $\det D_x f$; it should have the same sign
($\mu_0$-a.e.)  over the support of the prior.


Note that one can exploit the triangular structure in order to express
the determinant of the Jacobian of the map as a product of the
diagonal terms:
\begin{equation}
\det D_x f = \det \frac{\partial f}{ \partial x} 
= \prod\limits_{i=1}^n \frac{\partial f_i(x_1, \ldots, x_i) }{\partial x_i} .
\end{equation}

\section{Solution Algorithm}
\label{sec:solalg}

Problem~\ref{prob:OT} and its triangular variant are stochastic
optimization problems, in that they involve expectations with respect
to the prior distribution. They are also infinite-dimensional, in that
$f$ (in principle) may be an element of an infinite-dimensional
function space. Moreover, they may involve likelihood functions $L$
that are defined implicitly through the solution of forward problems
$h$ and that are computationally intensive. We will describe algorithms
for the solution of these problems, taking into account all of the above
challenges. The algorithms involve sample approximations of the prior
expectation, a flexible parameterization of $f$, and efficient
gradient-based optimization. Later we describe a continuation-type
method for solving a sequence of simpler optimization problems to
yield a ``cascade'' or sequence of maps.

\subsection{Monte Carlo sampling}

It is in general complicated or impossible to compute moments of
$T(X)$ using analytical formulae. Instead, we use Monte Carlo sampling
to approximate the expectation operator
\begin{equation}
\mathbb{E}\left[T(X) \right] \approx \frac{1}{N_s} \sum_{i=1}^{N_s} T \left
  ( x^{(i)} \right )
\end{equation}
where $x^{(i)} \sim \mu_0$.  Notice that the expectation is taken with
respect to the prior, which is assumed to be a distribution from which
one can easily generate independent samples. (Extensions to the method
may address situations where prior sampling is difficult; see
Section~\ref{s:concs}.) The number of samples $N_s$ is chosen
adaptively over the course of the optimization algorithm, as described
in Section~\ref{s:algstruct}. Moreover, the set ${ x^{(i)} }$ is
``refreshed'' with a new batch of i.i.d.\ samples between cycles. In
this sense, the present scheme bears strong similarities to the SAA
(sample average approximation) approach to stochastic programming
\cite{nemirovski:2009:rsa}.

%




\subsection{Polynomial approximation of $f$}
\label{subsec:polyapp}
As noted above, the optimization problems posed in
Section~\ref{subsec:optimaltransport}
and~\ref{subsec:triangularformulation} are infinite-dimensional. To
make them more computationally feasible, the map $f(x)$ is approximated
using a finite set of basis functions. In a high dimensional inference
problem, the use of localized basis functions is in general not
efficient; instead, a global basis is more suitable. In our
implementation, we represent $f$ using multivariate polynomials
orthogonal with respect to the prior measure; in other words, we write
the map as a polynomial chaos expansion~\cite{wiener:1938:thc,
  ghanem:2003:sfe, maître:2010:smf, xiu:2002:twa}. Some additional
intuition for polynomial maps is as follows. An identity map
$f:\mathbb{R}^n \rightarrow \mathbb{R}^n$ yields a posterior that is
equal to the prior. An affine map containing  a diagonal linear transformation, i.e., where $f_i$ depends only on $x_i$, allows changes of location and scale in each component of the
parameter vector, from prior to posterior, but preserves the form of
each distribution. An affine map containing a general linear
transformation (with each $f_i$ depending on all components of $x$) allows
the introduction of new correlations in the posterior. Quadratic,
cubic, and higher-degree maps allow even more complex distributional
changes from prior to posterior.

We write the polynomial expansion of $f$ as:
\begin{equation}
f(x) = \sum_{\mathbf{i} \in \mathcal{J}} g_{\mathbf{i}} \, \psi_{\mathbf{i}}(x)
\end{equation}
where $\mathbf{i} = (i_1, i_2, ..., i_n) \in \mathbb{N}^n$ is a
multi-index, $g_{\mathbf{i}} \in \mathbb{R}^n$ are the expansion
coefficients, and $\psi_{\mathbf{i}}$ are $n$-variate polynomials. We
write each of these polynomials as
\begin{equation}
\psi_{\mathbf{i}}(x) = \prod_{j=1}^n \varphi_{i_j}(x_j)
\end{equation}
where $\varphi_{i_j}$ is a univariate polynomial of order $i_j$ in
the variable $x_j$, orthogonal with respect to the distribution of
$x_j$. For simplicity, we assume here that the prior can be described,
perhaps after some transformation, by a set of independent random
variables. That said, orthogonal polynomial expansions for
\textit{dependent} random variables with arbitrary probability measure
can certainly be constructed~\cite{soize:2004:psw}. The set of multi-indices
${\mathcal J}$ describes the truncation of the expansion. For example
$\mathcal{J} = \{ \mathbf{i}: \left | \mathbf{i} \right |_1 \leq n_0 \}$
corresponds to a total-order expansion of degree $n_0$, containing
\begin{equation}
K = \card\left (\mathcal{J} \right ) = \left(\begin{array}{c} n+n_0 \\
    n_0 \end{array}\right) =  \frac{(n+n_0)!}{n! \, n_0!}
\end{equation}
terms.

The orthogonal polynomial representation of the map offers several
advantages, besides being convenient and flexible (by choice of
$\mathcal{J}$). First, moments of the posterior distribution---i.e.,
moments of $f(X)$, with $X \sim \mu_0$---can be evaluated analytically,
without the need for further sampling or additional forward model
solves. For example, the posterior covariance is a weighted sum of
outer products of the coefficients: $\Cov(Z) = \sum_{\mathbf{i}}
g_{\mathbf{i}}^{ } g_{\mathbf{i}}^T \, \mathbb{E} \left [ \psi_{\mathbf{i}}^2
\right ]$. Higher moments can be evaluated using known formulas for
the expectations of products of orthogonal polynomials
\cite{maître:2010:smf,rupert:2007:aao}.
Second, obtaining a polynomial chaos expansion of the posterior
facilitates efficient propagation of data-conditioned uncertainty
through subsequent computational models. This is a key step in
posterior prediction and in sequential data assimilation (e.g.,
filtering). With a polynomial chaos representation in hand, a vast
array of stochastic Galerkin and stochastic collocation methods can be
employed for uncertainty propagation.

Treating each coefficient $g_{\mathbf{i}} \in \mathbb{R}^n$ as a
column vector, we assemble the set $\{ g_{\mathbf{i}} \}_{\mathbf{i}
  \in \mathcal{J}}$ into a matrix $F^T$ of size $n \times K$, where $K$
denotes $\card\left (\mathcal{J} \right )$:
\begin{equation}
F^T = \left[\begin{array}{cccc}g_1 & g_2 & \ldots &
    g_K \end{array}\right] .
\end{equation}
In the optimal transport formulation all $n \times K$ coefficients are
considered the optimization variables, whereas in the triangular
formulation some of these coefficients are fixed to zero. The map $f$
is then then represented as
\begin{equation}
f(x) = F^T \Psi(x)
\label{e:pcmap}
\end{equation}
where $\Psi(x)$ is a column vector containing every basis polynomial
$\psi_{\mathbf{i}}$, $\mathbf{i} \in \mathcal{J}$. 

\subsection{Newton's method and nonlinear least squares}

In our implementation we use two alternative algorithms to solve the
optimization problem: Newton's method and nonlinear least squares. In
the standard Newton's method, we compute both the first and second
derivatives of the objective function. It is then necessary to compute
the derivatives of $T(x)$ with respect to the optimization variables
$F$:
\begin{eqnarray*}
T(x; F) = \log\left({L \circ f}\right) + \log\left({p \circ f }\right)
+ \log{ \left | \det{D_x f} \right |} - \log p \\
\frac{\partial T}{\partial F} = \left( \frac{1}{L \circ f}\frac{dL(f)}{df} +
  \frac{1}{p \circ f}\frac{dp(f)}{df}  \right) \frac{\partial f}{\partial F} + D_x \Psi  \left( D_x f \right)^{-1} 
\end{eqnarray*}
where ${\partial f}/{\partial F} = \left[ I_n \otimes \Psi(z)
\right] $. The last term is obtained as follows:
\begin{eqnarray}
D_x f = D_x \left(F^T \Psi\right)  &=& F^T D_x \Psi \nonumber\\
\frac{\partial}{\partial F_{ij}} \left ( \log \left | \det D_x f
  \right | \right ) &=&  \trace( \left( D_x f \right)^{-1} E_{ji} D_x \Psi ) \nonumber\\
&=&  \trace( \left( D_x f \right)^{-1} e_j \left[ D_x \Psi \right] (i, :) ) \nonumber\\[8pt]
&=&  \left[ D_x \Psi \right] (i, :) \left( D_x f \right)^{-1} e_j  \nonumber\\
\frac{\partial}{\partial F}  \left ( \log \left | \det D_x f \right | \right ) &=&  D_x \Psi  \left( D_x f \right)^{-1} 
\end{eqnarray}
where $E_{ji}$ is a matrix of all zeros except for a single 1 at the
location $(j,i)$ and $e_j$ is a vector of zeros except for a 1 in
row $j$.

The second derivatives of $T$ can be obtained in a similar fashion. In the
case of a Gaussian prior and Gaussian additive noise, explicit
expressions for the first and second derivatives are given
in~\ref{app:additivegaussian}. Note that to compute any
derivatives of the likelihood function, derivatives of the forward
model are required. Using modern adjoint techniques, gradients and Hessian-vector products can typically be computed at the cost of $O(1)$ forward solves, independent of the dimension of the parameter space~\cite{martin:2011:asn}. In many
problems, however, the forward model and its derivatives can be
accurately approximated using surrogate
models~\cite{marzouk:2009:dra}.

The second derivatives of the log-determinant can be calculated explicitly as:
\begin{eqnarray}
\frac{\partial^2}{\partial F \partial F_{ij}} \left ( \log \left | \det D_x
  f \right | \right ) &=&  -D_x \Psi  \left( D_x f \right)^{-1} \left(  E_{ji} D_x  \Psi \right)  \left( D_x f \right)^{-1}  \nonumber \\
&=&  -D_x \Psi  \left( D_x f \right)^{-1} \left(  e_{j}  \left[D_x\Psi\right] (i,:) \right)  \left( D_x f \right)^{-1}  \nonumber \\[8pt]
&=&  - \left(  D_x \Psi  \left( D_x f \right)^{-1}  e_{j} \right) \left( \left[D_x\Psi\right] (i,:)   \left( D_x f \right)^{-1}\right) . 
\end{eqnarray}
From the previous relations it is clear that in order to compute the
Hessian of the log-determinant, we must first compute the matrix $D_x\Psi \left( D_x f
\right)^{-1}$ and then use the outer product of appropriate column and
row vectors to assemble the desired Hessian. 

Alternatively, one can use a nonlinear least squares (NLS) method to
solve the problem. This method is particularly suited to the
triangular formulation of Section~\ref{subsec:triangularformulation},
which does not have a penalty term.  To cast the optimization
objective as a nonlinear least squares problem, it is observed that
the zero variance condition $\Var \left[T(X)\right] = 0$ is equivalent
to $T \left (x^{(i)} \right ) = \mbox{constant}$, for $x^{(i)} \sim
p$. In other words, $T$ is constant in an $L^2$ sense. This leads to
the set of equations
\begin{equation}
T \left (x^{(i)}; F \right ) = \frac{1}{N_s}\sum\limits_{j=1}^{N_s} T
\left (x^{(j)}; F \right ) .
\label{eq:nls}
\end{equation}
If the number of optimization parameters $\ell < nK$ in $F$ is less than $N_s$,
then (\ref{eq:nls}) is an overdetermined set of nonlinear equations
in $F$, which can be solved via
\begin{eqnarray}
M \Delta F &=& b \nonumber\\
F^{k+1} &=& F^{k} - \Delta F 
\end{eqnarray}
where the rectangular matrix $M \in \mathbb{R}^{N_s\times \ell}$ and the column vector
$b\in R^{N_s}$ are:
\begin{eqnarray}
M(i, :) &=& \frac{\partial T \left (x^{(i)} \right )}{\partial F} -  \frac{1}{N_s}\sum\limits_{j=1}^{N_s} \frac{\partial T \left (x^{(j)} \right )}{\partial F}  \nonumber\\
b(i) &=& T \left ( x^{(i)}; F \right ) - \frac{1}{N_s}
\sum\limits_{j=1}^{N_s} T \left (x^{(j)} \right ) . 
\end{eqnarray}
Here, the derivative of a scalar with respect to a vector is assumed to be a row vector.

We have observed that in certain cases, e.g., weakly multi-modal
posteriors, convergence of the algorithm from arbitrary initial maps is more reliable if the objective function is modified to become
\begin{equation}
\min_f \left \{ \V\left[T(X)\right] + \lambda \mathbb{E}\left[ \left(
      T(X)-\tilde{\beta} \right)^2 \right] \right \} ,
\end{equation}
where $\tilde{\beta}$ is an estimate of the evidence computed from the previous iteration.
In this case, the nonlinear least squares formulation proceeds from the following set of equations:
\begin{equation}
T \left (x^{(i)}; F \right) = 
\frac{1}{1+\lambda} \left( \frac{1}{N_s}\sum\limits_{j=1}^{N_s} T
\left (x^{(j)}; F \right )  + \lambda \tilde{\beta} \right), \ \ i = 1 \ldots N_s.
\end{equation}

In general, we initialize our algorithm using the identity map, $f(x)
= x$. Alternatively, if the likelihood contains additive Gaussian
noise and the prior is Gaussian, we linearize the forward model and
compute the resulting map analytically. This approximate linear map is
then used to initialize the algorithm.


\subsection{Algorithm structure}
\label{s:algstruct}

We solve the optimization problem in stages, by iterating over the
order of the polynomial representation of $f(x)$. We typically start
the algorithm with only linear basis functions in $\Psi(x)$. After
computing the best linear map, the total order of the polynomial basis
is increased and a new corresponding map is computed---allowing
coefficients of \textit{all} the polynomial terms to be adjusted, not
just those of the highest degree. This process is repeated until
reaching a stopping criterion $\V\left[T(X)\right] < \delta$, where
$\delta$ is a preset threshold. The samples used to approximate the
prior expectation or variance are renewed each time the order of the
expansion is increased. The number of samples $N_s$ is chosen such
that approximations of the mean or variance obtained with successive
sample batches lie within some predetermined relative tolerance
$\alpha$ of each other, typically 5\%.

Algorithm~\ref{alg:mainalgorithm} summarizes our complete
approach. Notice that the multiplier $\lambda$ is used only when
computing the penalized map of Problem~\ref{prob:OT}. The value of
this multiplier is adjusted each time the order of the expansion is
increased; it is chosen such that the penalty term is 10\% of the
total value of the objective function at the start of the current
stage. Thus $\lambda$ decreases as $f$ converges towards the desired
measure transformation.
  
\begin{algorithm}[htb]
\label{alg:mainalgorithm}
\KwData{Likelihood function $L$, observed data, prior density $p$,
  stopping criterion $\delta$, sample tolerance $\alpha$} 
\KwResult{Map $f(x)$ }
$f(x) = x$, $i=0$, and $\lambda=1$\;
$n_0 = 1$\; 
\While{$\Var\left[T(X)\right] > \delta$}{
      i = i + 1\;
	Generate $N_s$ i.i.d.\ samples of the prior random variable\;
	\If{$i > 1$}{
		Compute $\Var[T(X)]$\;
		\If{$ \left | \( \Var[T(X)]/V^{\ast} \)  - 1 \right | > \alpha$}{
			Double the number of samples $N_s \leftarrow 2 N_s$\;
		}
	}
	\KwSolve{the optimization problem for $f(x)$ up to desired order
        $n_0$, stopping when $\Delta F$ is smaller than a threshold}
	Get current $\Var[T(X)]$ and store it in  $V^{\ast} = \Var[T(X)]$\;
	Increase order of expansion: $n_0 \leftarrow n_0 + 2$\;
	If applicable compute a new $\lambda = 0.1\, V^{\ast}/\mathbb{E}\left[\|X-f(X)\|^2 \right]$
}
\caption{Structure of optimization algorithm for computing the map.}
\end{algorithm}



\subsection{Composite map}
\label{subsec:composite}

In certain problems, the map from prior to posterior can be accurately
approximated only with very high-degree polynomials. Unfortunately,
the number of coefficients in a multivariate polynomial expansion
grows very rapidly with degree, leading to a more challenging
optimization problem.

To overcome this difficulty, we can use a \textit{composition} of maps
to efficiently approximate a high-degree transformation from prior to
posterior. That is, instead of mapping the prior to the posterior with
a single map, we introduce a sequence of intermediate distributions
that evolve gradually from prior to posterior. With this sequence of
distributions, we introduce a corresponding sequence of maps.
Intuitively, this approach ``relaxes'' the measure transformation
problem; if the input and output distributions are similar to each
other, then the presumably the map can be well approximated using
low-degree polynomials. Letting the number of steps approach infinity, one could imagine a continuous transformation from prior to posterior, for instance as proposed in~\cite{reich:2011:ads}.

\begin{figure}[htb]
 \centering
 \includegraphics[width=4.8in]{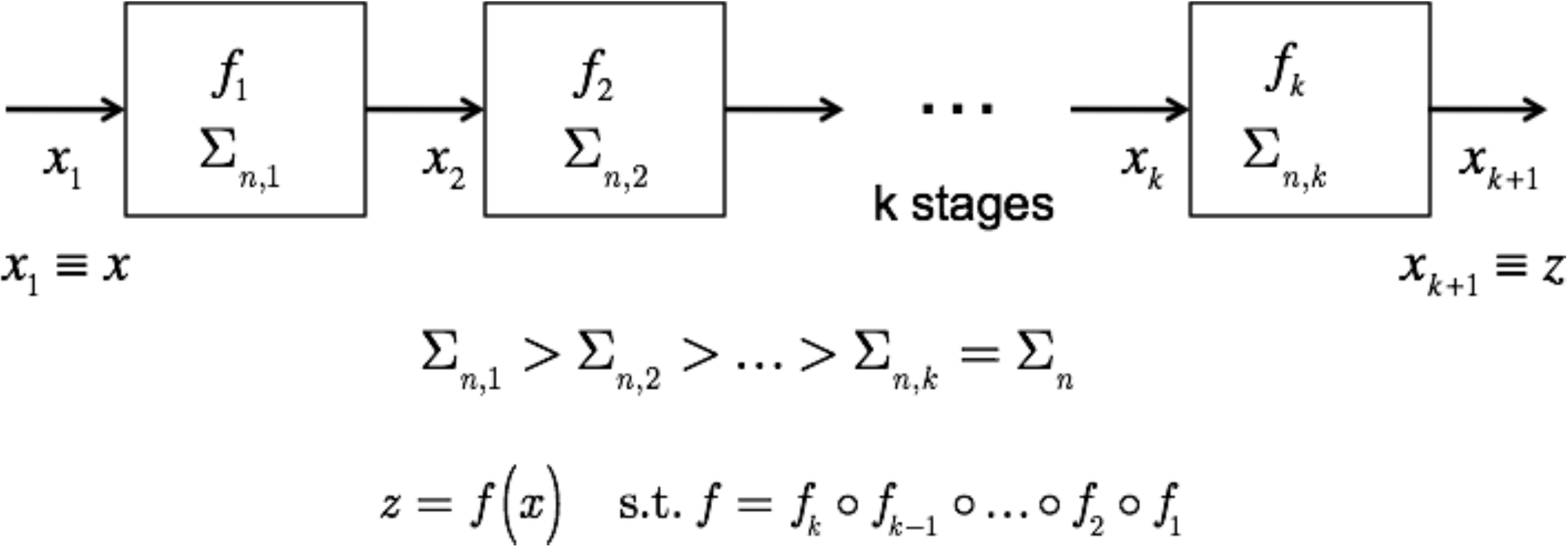}
 \caption{A pictorial representation of the composite-map algorithm.
   Successive stages represent a gradual transition from prior to
   posterior, imposed, e.g., via a sequence of noise levels, or by iterating
   through the data set or the fidelity of the forward model. 
   }
\label{fig:cascade}
\end{figure}

Here we focus on finite sequences of intermediate distributions. There are many ways that one could create such a sequence: iterating over the accuracy of the forward model,
gradually introducing components of the data vector $d$, or iterating
over the scale of the noise covariance in the likelihood function. We
choose the latter tactic, beginning with a large covariance
$\Sigma^{1}$ and introducing $k$ stages such that the variance of
the last stage is the true $\Sigma$ of the original inference problem:
\begin{equation}
\Sigma^{1} > 
\Sigma^{2} > 
\cdots > 
\Sigma^{k}  = \Sigma 
\end{equation}
For a given noise variance $\Sigma^i$,  a composite map
\begin{equation}
\phi^i \equiv  f^i \circ f^{i-1} \circ \cdots \circ f^1
\end{equation}
is computed such that it maps the prior distribution $\mu_0$ (with 
density $p$) to the \textit{intermediate} posterior distribution
$\mu_i$ (with density $q^i$) corresponding to $i$th-stage noise level.
Then we can minimize, e.g., 
\begin{equation}
  \min_{F^i}{D_{\mathrm{KL}}\left(p \, ||  \left ( q^i \circ \phi^i
      \right ) \left | \det D_x \phi^i \right | \right)}
\label{e:intermed}
\end{equation}
with a regularization term added above as needed. Here only the
parameters $F^i$ of the $i$th map $f^i$ are optimization variables!
Coefficients of the preceding maps are fixed, having been determined
in the previous stages' optimization problems. See
Figure~\ref{fig:cascade} for a schematic. The resulting map $f =
\phi^k \equiv f^k \circ f^{k-1} \circ \cdots \circ f^1 $ will have been
computed with significantly less effort (and fewer degrees of freedom)
than a total-order polynomial of the same maximum degree.

One advantage of the sequential construction above is that the
accuracy of an intermediate map, transforming the prior to any of the
intermediate distributions $q^i$, does not affect the accuracy of the
final map. Because the prior density $p$ always appears
in~(\ref{e:intermed}), the final stage can be be computed with tight
error tolerances to yield the appropriate measure transformation.
Intermediate maps may be computed with looser tolerances, as they
serve in a sense only to find distributions that are closer in shape
to the actual posterior. Any error in an intermediate $f$ can be
compensated for by the final $f^k$. In our implementation, we have
observed that this scheme is robust and that it allows for simple
error control.

Alternatively, one could think of computing the map one stage
at a time---i.e., at each step computing a transformation $f^i$ from
$q^{i-1}$ to $q^i$ by minimizing
 \begin{equation}
\min_{F^i}{D_{\mathrm{KL}}\left( q^{i-1} \, || \left ( q^i \circ f^i
      \right ) \left | \det D_x f^i \right | \right)}.
\end{equation}
The final map is again $f = \phi^k \equiv f^k \circ f^{k-1} \circ
\cdots \circ f^1$. A possible advantage of this second construction is
that one needs to consider only a single map at a time. Each stage
must be solved exactly, however, as the construction depends on
satisfying $\mu_{i} = \phi_{\sharp}^i \mu_0$ for all $i$. If $\phi^i$ does not push
forward $\mu_0$ to $\mu_i$, then any error incurred in the first $i$
stages of the sequence will remain and corrupt the final stage. Also,
because the error tolerances in the intermediate stages cannot be
relaxed, the difficulty of computing any of the intermediate maps
could conceivably be as great as that of computing the full map. For these
reasons, we prefer the first construction (\ref{e:intermed}) and will
use it when demonstrating composite maps in the numerical examples
(Section~\ref{subsec:toggle}) below.

\section{Results}
\label{sec:results}

We now demonstrate map-based inference on a series of example
problems, ranging from parameter estimation in linear-Gaussian models
(allowing verification against a known analytical solution) to
nonlinear ODEs and high-dimensional PDEs.

\subsection{Linear-Gaussian model}
\label{subsec:linearG}
Here we demonstrate the accuracy and convergence of our method on over- and under-determined linear-Gaussian models. We evaluate both the optimal transport formulation and the triangular formulation.

Consider a linear forward model $h(x) = A x$, where the parameters $x$
are endowed with a Gaussian prior, $x \sim N(0, \Sigma_P)$, and additive
measurement noise $\varepsilon$ yields the data:
\begin{equation}
d = h(x) + \varepsilon =  A x + \varepsilon \, .
\end{equation}
If the noise is further assumed to be Gaussian, $\varepsilon \sim N(0,
\Sigma_N)$, then the posterior is Gaussian, and the map $f$ is linear and
available in closed form:
\begin{equation}
  f(x) = z_0 + Z_1 x 
\end{equation}
with
\begin{eqnarray}
  Z_1 \Sigma_P Z_1^T & = &   \Sigma_{\mathrm{post}} =  \left(A^T \Sigma_N^{-1} A
  + \Sigma_P^{-1} \right)^{-1} \nonumber \\
  z_0 & = & \mu_{\mathrm{post}} = \Sigma A^T \Sigma_N^{-1} d  .
\label{e:linearmap}
\end{eqnarray}
In addition, the evidence can be computed in closed form:
\begin{equation}
  \beta = \exp{\left( -\frac{1}{2} \left ( d^T \Sigma_N^{-1} d - \mu_{\mathrm{post}}^T
     \Sigma_{\mathrm{post}}^{-1} \mu_{\mathrm{post}} \right )   \right) } \sqrt{ \left
   | \det \Sigma_N \right |} .
\end{equation}
A detailed derivation of the previous relations is given
in~\ref{app:lineargaussian}.  

We first test the two optimization formulations (the penalized
optimal transport objective of
Section~\ref{subsec:optimaltransport} and the triangular construction
of Section~\ref{subsec:triangularformulation}) on an over-determined inference
 problem with 10 parameters and 16 observations. The forward model,
represented by $A \in \mathbb{R}^{16 \times 10}$, is randomly
generated with entries of $A$ independently sampled from a standard
normal distribution. We use an identity prior covariance
$\Sigma_P = I$ and put $\Sigma_N = \sigma^2 I$ with
$\sigma = 0.06$.

With both formulations  (Problem~\ref{prob:OT} and Problem~\ref{prob:RT}), we observe convergence to a map matching
(\ref{e:linearmap}) after 15 iterations. The evidence $\beta$
also converges to its analytical value, to within machine
precision. With the optimal transport formulation, we observe that our
$Z_1$ differs from the symmetric matrix square root of the posterior
covariance by roughly 5\%. More precisely, the Frobenius norm $\| \cdot \|_F$ of the
difference, $ \| Z_1 - \Sigma_{\mathrm{post}}^{1/2} \|_F$, is roughly
5\% of $ \| \Sigma_{\mathrm{post}}^{1/2} \|_F$. When the triangular
construction is employed, the Frobenius norm of $Z_1 - L$ is
less than $10^{-6} \| L \|_F$, where $L$ is the lower-triangular
Cholesky factor of $\Sigma_{\mathrm{post}}$, i.e., $\Sigma_{\mathrm{post}} =
L L^T$.
With the optimal transport formulation, we have further observed that
if the map $f$ is forced to remain symmetric and if the penalty factor
$\lambda$ is gradually reduced to zero, then $Z_1$ matches
$\Sigma_{\mathrm{post}}^{1/2}$ with a relative error of less than
10$^{-6}$.

\begin{figure}[htb]
 \centering
 \includegraphics[width=4.8in]{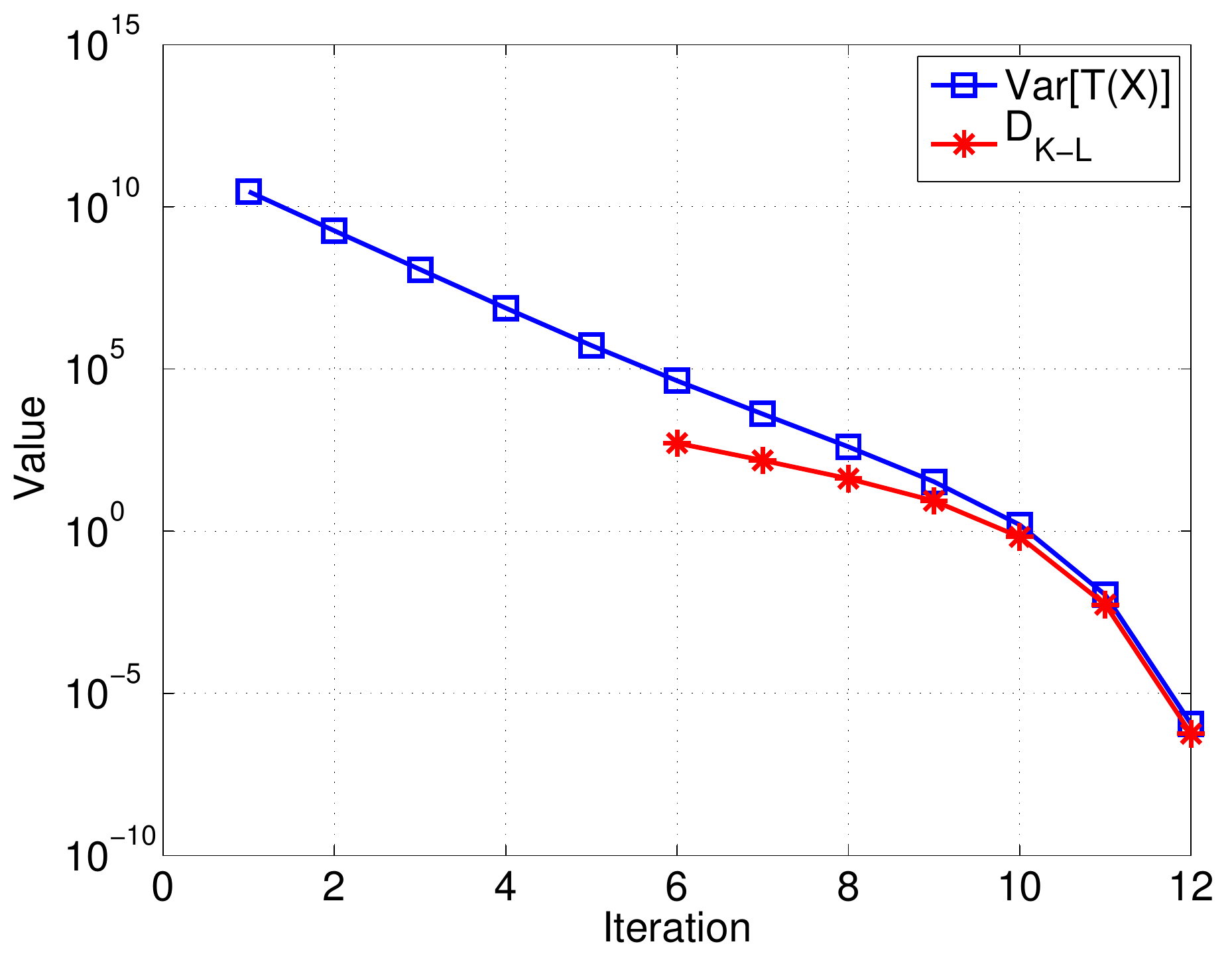}
 \caption{Linear-Gaussian problem: Variance and Kullback-Leibler
   divergence versus iteration number. The magnitude of both
   quantities converges to machine precision after 12 iterations.}
\label{fig:linearVarKLmapversusanalytic100param}
\end{figure}

\begin{figure}[htb]
 \centering
 \includegraphics[width=4.8in]{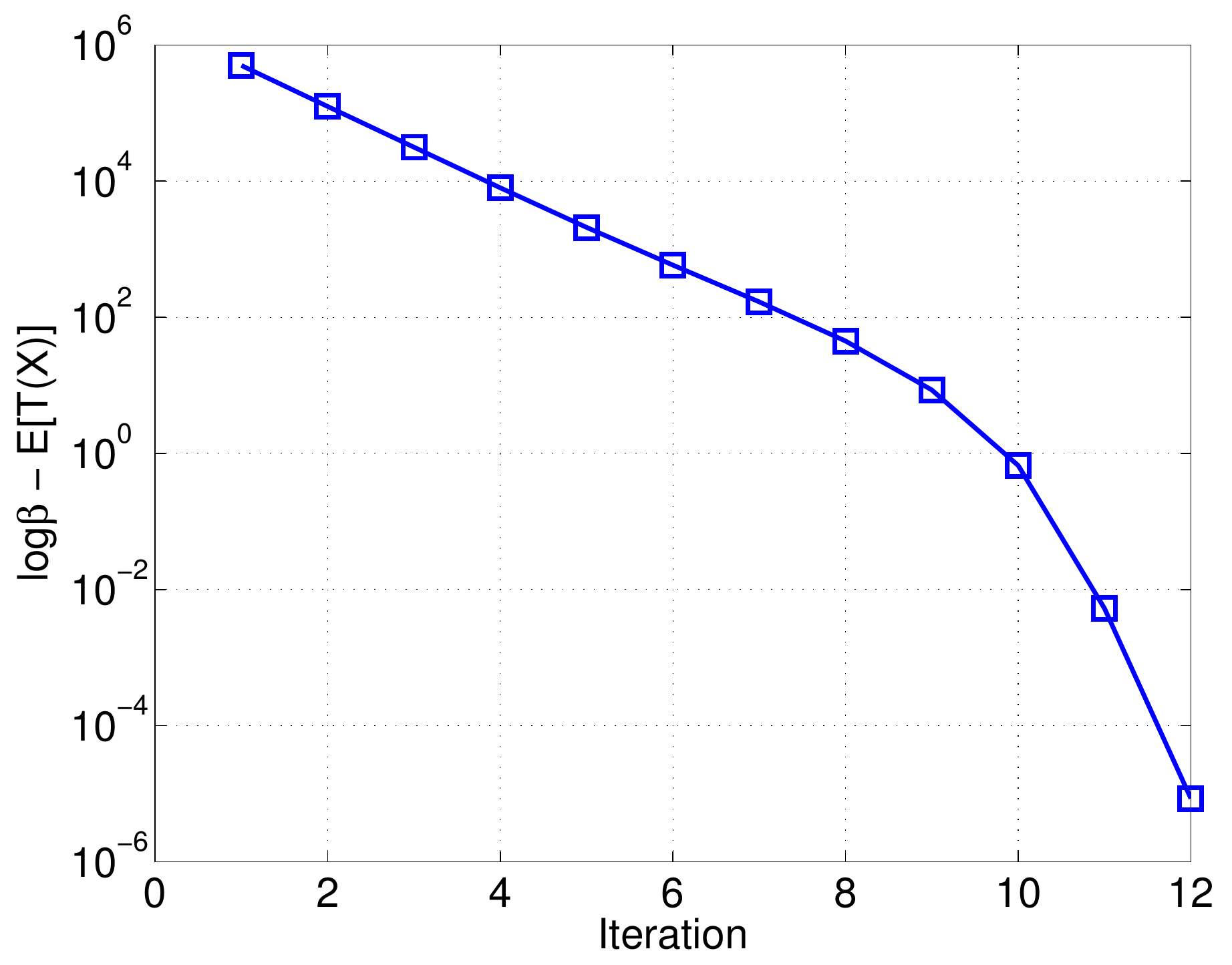}
 \caption{Linear-Gaussian problem: Difference between the exact
   evidence $\beta$ and the evidence $\mathbb{E}[T(X)]$ computed with
   the map algorithm. The evidence converges to the exact
   value in 12 iterations.}
\label{fig:linearEvidence100param}
\end{figure}

Next we consider a larger but under-determined linear-Gaussian  problem with 100 parameters
and 8 observations, again randomly generating $A$. Using the
triangular construction, we compute the map. Convergence plots of $\Var
\left [T \right ]$ and the Kullback-Leibler divergence
$D_{\mathrm{KL}} \left ( p || \pt \right )$ are given in
Figure~\ref{fig:linearVarKLmapversusanalytic100param}, while
convergence of the evidence $\beta$ is shown in
Figure~\ref{fig:linearEvidence100param}. The optimization iterations
begin with the identity map, and the variance and KL divergence
decrease rapidly to zero as the correct map is identified.
(Note that KL divergence is not shown in the first five iterations of
Figure~\ref{fig:linearVarKLmapversusanalytic100param}, because the
sample-estimated divergence is infinity.)
Near convergence, it is observed that the value of the
Kullback-Leibler divergence is approximately half the variance of
$T(X)$. This relation is proven in~\ref{app:KLversusVar}.


\subsection{Reaction kinetics}
\label{subsec:reactionkinetics}
This example demonstrates map-based inference on a nonlinear problem that yields a non-Gaussian posterior, with a sharp lower bound and strong correlations. The problem is two-dimensional, enabling direct visualization of the map. We examine the impact of truncating the polynomial order on convergence and monotonicity of the map. 

The objective of the problem is to infer the forward and reverse rates of reaction in the
chemical system $A \stackrel[k_2]{k_1}{\leftrightharpoons} B$~\cite{haario:2006:dra}. The
governing equations are as follows, with $u$ representing the
concentration of component $A$ and $v$ the concentration of component
$B$:
\begin{eqnarray}
\frac{du}{dt} & = & -k_1 u + k_2 v \nonumber\\
\frac{dv}{dt} & = &  k_1 u - k_2 v \, .
\end{eqnarray}
The initial condition is fixed at $u(0) = 1$ and $v(0) = 0$. The rate
parameters $k_1$ and $k_2$ are endowed with independent Gaussian prior
distributions, $k_1 \sim N(2, 200^2)$ and $k_2 \sim N(4, 200^2)$. The
``true'' parameter values are set to $k_1=2$, $k_2=4$, and synthetic
data consisting of noisy observations of $u$ at times $t = 2$, 4, 5,
8, and 10 are generated. Observational noise is assumed to be i.i.d.\
Gaussian. Because the observations occur relatively late in the
dynamics of the ODE initial value problem, the only information which
can be inferred is the ratio of $k_1$ and $k_2$---i.e., the
equilibrium constant of the system.

We compute a map using the algorithm detailed in
Section~\ref{sec:solalg}. The map is represented using a total-order
polynomial expansion. We begin with a linear map and progressively
increase its order; at the end of iterations with a 5th-order map ($n_0=
5$), the Kullback-Leibler divergence $D_{\mathrm{KL}} \left (
  p || \pt \right )$ is less than 10$^{-3}$. The algorithm requires a
total of 30 inner-loop optimization steps to reach this level of
accuracy. 

\begin{figure}[htb]
\centering
\subfigure[Samples from the prior distribution.]
{\includegraphics[width=4.7in]{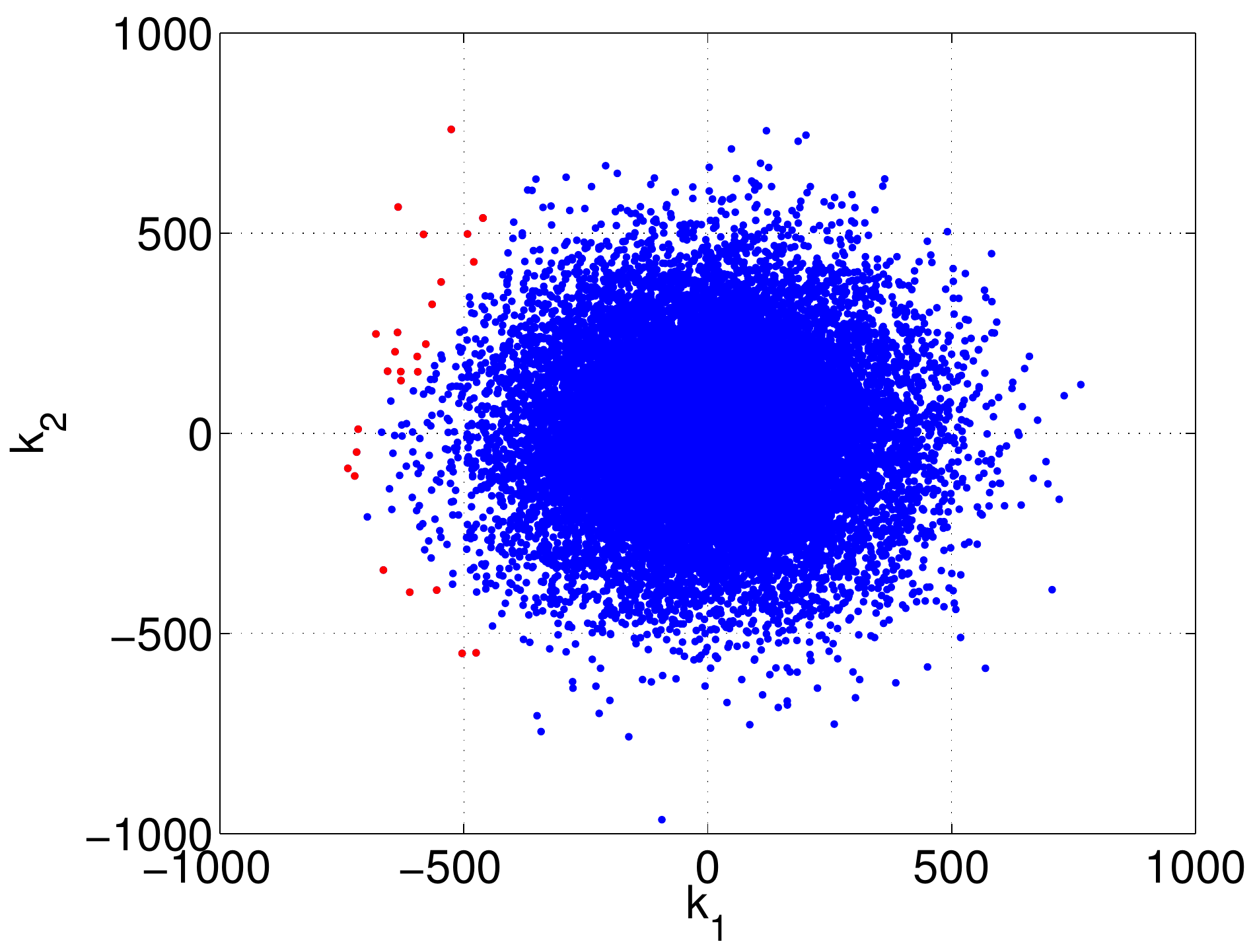}
\label{fig:priorsamps}}
\subfigure[Corresponding samples of the posterior distribution.]
{\includegraphics[width=4.7in]{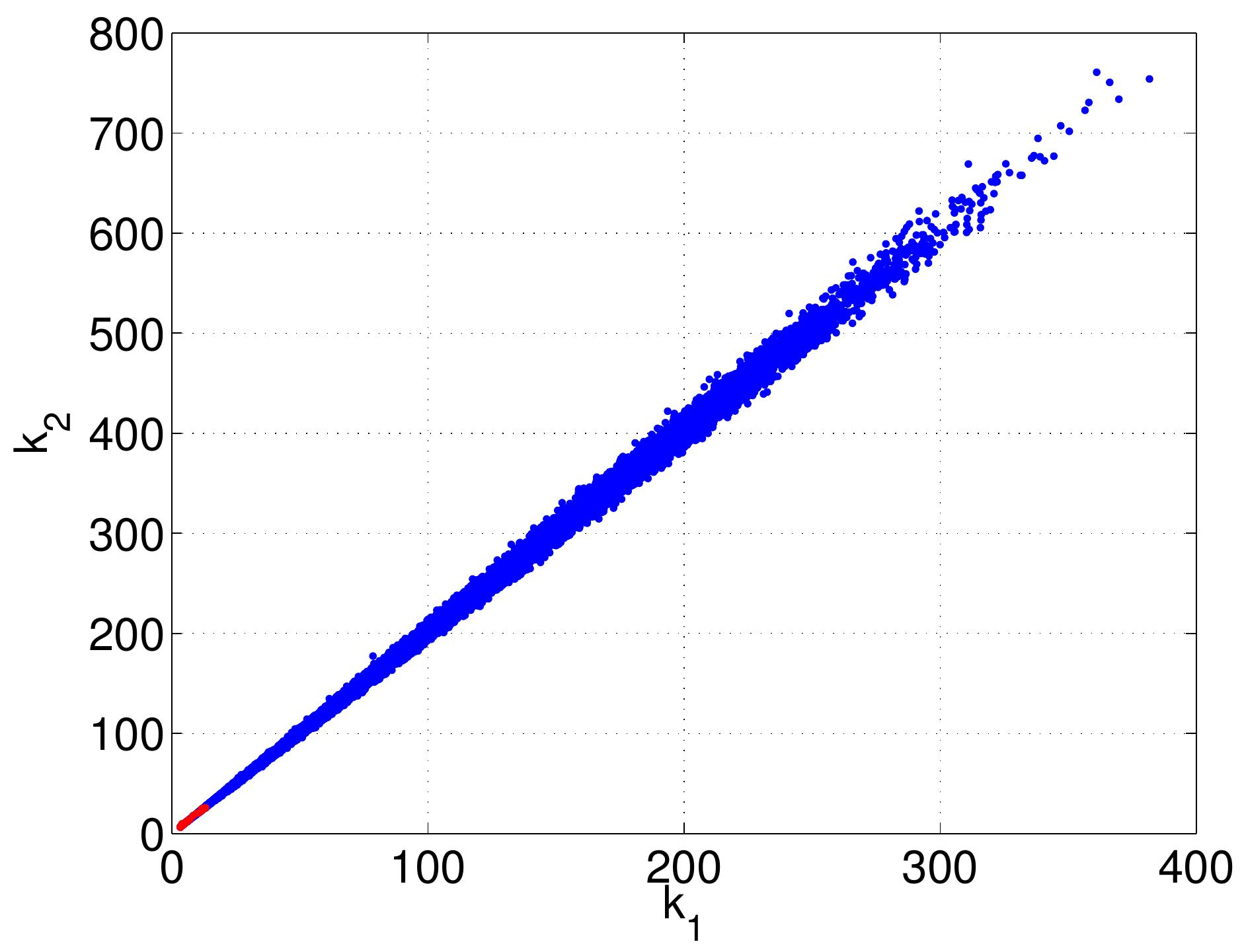}
\label{fig:postsamps}}
\caption{Reaction kinetics problem: samples from the prior and
  posterior distributions. Red dots represent samples at which the
  determinant of the Jacobian of the map is negative. The determinant
  is positive for 9992 out of 10000 samples. The posterior density is
  concentrated along a line of slope 2, which is the ratio $k_2/k_1$ }
\label{fig:twodrk_priorposteriorsamples}
\end{figure}

Figure~\ref{fig:twodrk_priorposteriorsamples} shows $10^4$ samples
from both the prior and posterior distributions. As expected, the
posterior samples concentrate around the line $k_2 = 2 k_1$. Also,
posterior samples are localized in the upper left quadrant of the
$k_1$-$k_2$ plane, which corresponds to stable trajectories of the
ODE. In both Figures~\ref{fig:priorsamps} and~\ref{fig:postsamps},
sample points at which the determinant of the Jacobian $D_x f$ is
negative are shown in red. Elsewhere (blue points), the Jacobian
determinant is positive. This sign change corresponds to regions of
the parameter space where monotonicity of the map is lost; we note
that all of these points are relegated to the \textit{tails} of the
prior distribution. Only 0.08\% of the samples have negative
determinant.

\begin{figure}[htb]
\centering
\subfigure[First component of the map.]
{\includegraphics[width=4.7in]{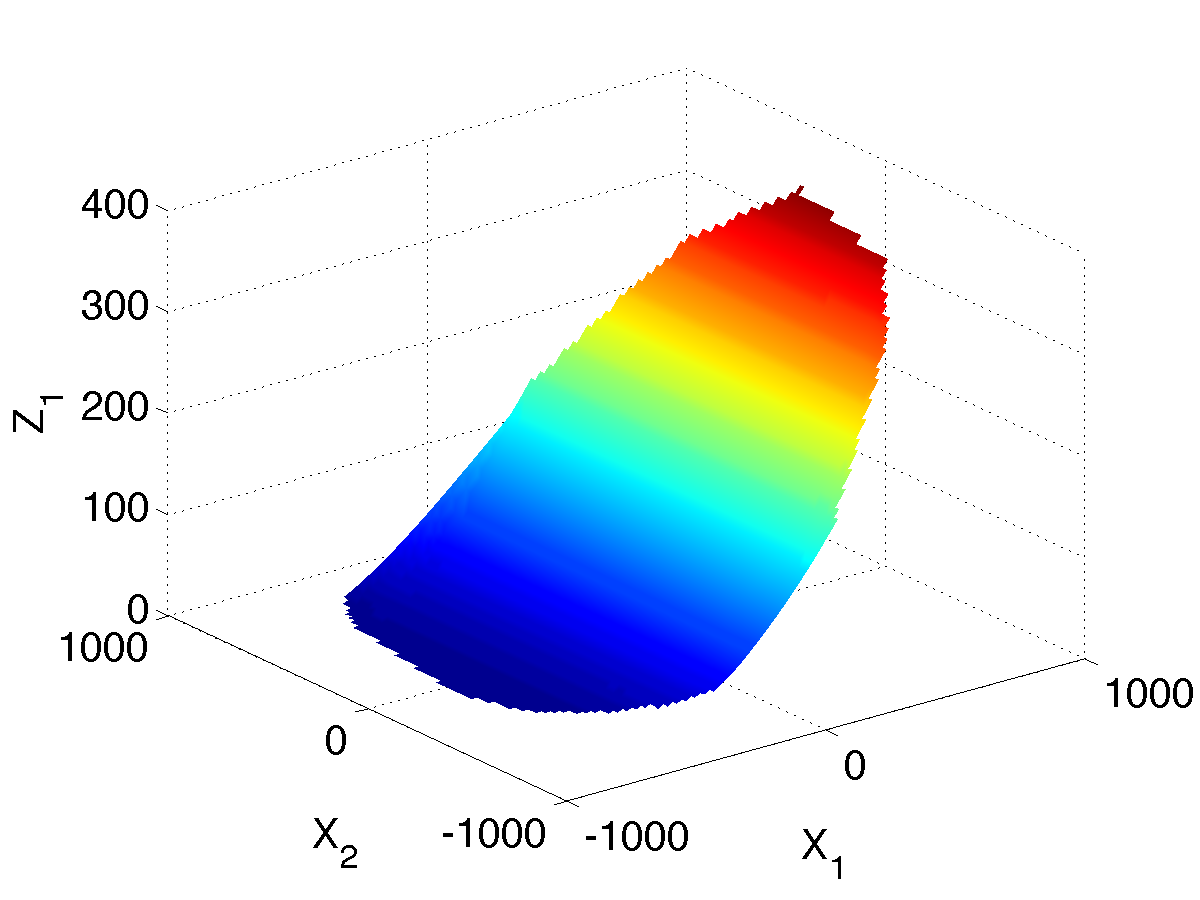}
\label{fig:twodrk_map_1}
}
\subfigure[Second component of the map.]
{\includegraphics[width=4.7in]{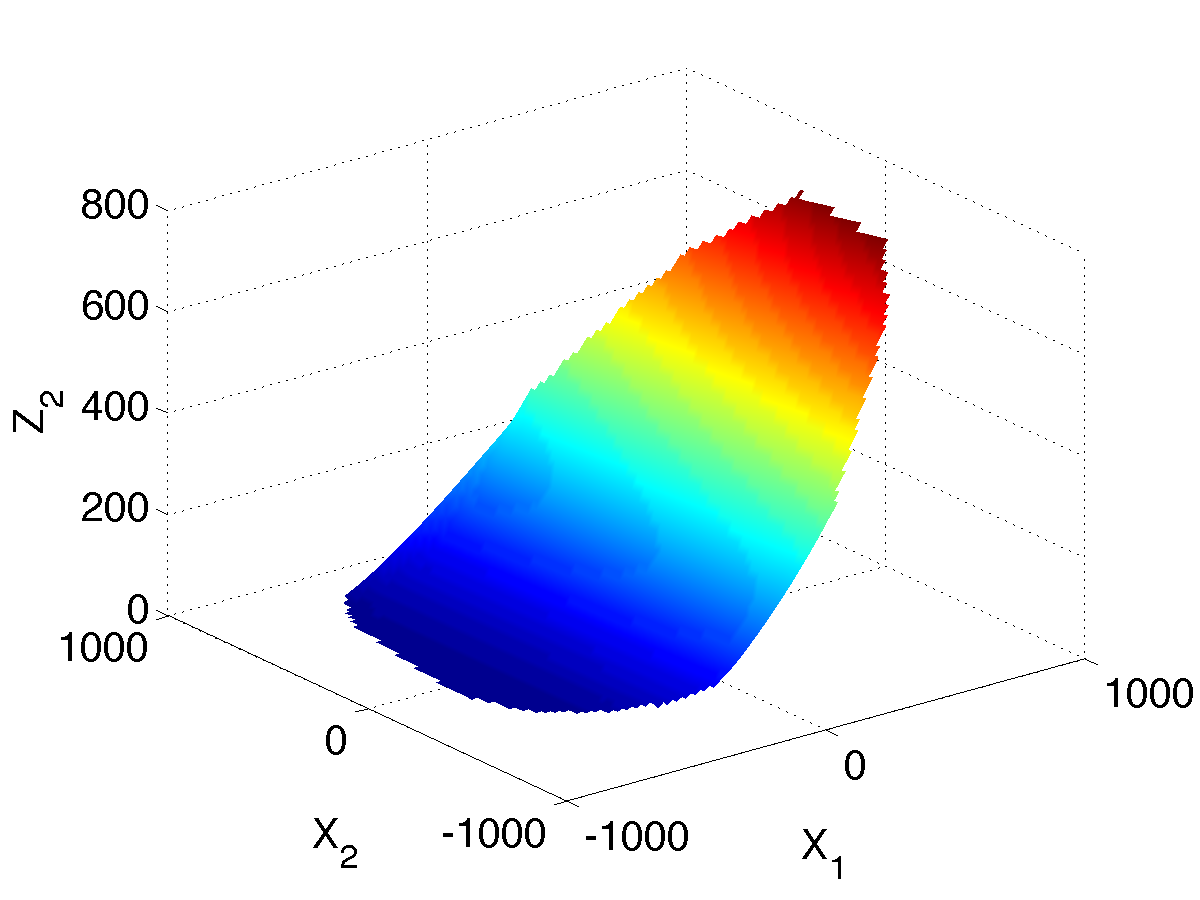}
\label{fig:twodrk_map_2}
}
\caption{Reaction kinetics problem: two-dimensional transformation $f$
  from $X \sim \mu_0$ (the prior random variable) to $Z \sim \mu$ (the
  posterior random variable). Due to the strong correlation between
  $z_1$ and $z_2$, both figures look almost identical up to a
  multiplicative factor of two. }
\label{fig:twodrk_map} 
\end{figure}

Figures~\ref{fig:twodrk_map_1} and~\ref{fig:twodrk_map_2} show
components of the map itself, via surface plots of $f_1(k_1, k_2)$ and
$f_2(k_1, k_2)$. The two components of the map are nearly identical up
to a scaling factor of two, which is to be expected, as the posterior
distribution contains a strong correlation between the parameters.

\subsection{Genetic toggle switch}
\label{subsec:toggle}
In this example we demonstrate the composite map on a six-dimensional  nonlinear parameter inference problem using real experimental data. We compare the accuracy of our results to those obtained with a standard MCMC algorithm. 

The example involves the dynamics of a
genetic ``toggle switch''~\cite{gardner:2000:coa}. The toggle switch
consists of two repressible promotors arranged in a mutually
inhibitory network: promoter 1 transcribes a repressor for promoter 2,
while promoter 2 transcribes a repressor for promoter 1. Either
repressor may be induced by an external chemical or thermal signal.
Genetic circuits of this form have been implemented on {\em
  E.\ coli} plasmids, and the following ODE 
model has been proposed \cite{gardner:2000:coa}:
\begin{eqnarray}
\frac{du}{dt} &=& \frac{\alpha_1}{1+v^\beta} - u \nonumber\\
\frac{dv}{dt} &=& \frac{\alpha_2}{1+w^\gamma} - v \nonumber\\
w &=& u \left(1 + \frac{\mathrm{[IPTG]}}{\kappa}\right)^{-\eta}
\end{eqnarray}
Here $u$ is the concentration of the first repressor and $v$ is the
concentration of the second repressor. $\mathrm{[IPTG]}$ is the
concentration of IPTG, the chemical compound that induces the switch.
At low values of $\mathrm{[IPTG]}$, the switch is in the `low' state,
reflected in low values of $v$; conversely, high values of
$\mathrm{[IPTG]}$ lead to strong expression of $v$. As in
\cite{marzouk:2009:asc}, we would like to infer the six model
parameters $\alpha_1$, $\alpha_2$, $\beta$, $\gamma$, $\eta$ and
$\kappa$. To this end, we employ actual experimental
data\footnote{Data are courtesy of Dr.\
  T.\ S.\ Gardner.} consisting of normalized steady-state values of
$v$ at selected IPTG concentrations, spanning the `low' and `high'
sides of the switch: $\mathrm{[IPTG]} \in \left\{10^{-3}, 0.6, 1, 3, 6, 10\right\} \times 10^{-3}$.

At steady state, the model yields the following relations
\begin{eqnarray*}
u &=& \frac{\alpha_1}{1+v^\beta}  \\
v &=& \frac{\alpha_2}{1+w^\gamma} \\
w &=& u\left(1 + \frac{\mathrm{[IPTG]}}{\kappa}\right)^{-\eta} 
\end{eqnarray*}
which can be expressed as an implicit function $v = g(v, \xi)$, where
$\xi=\left(\alpha_1, \alpha_2, \beta, \gamma, \eta, \kappa \right)$.
The elements of $\xi$ are endowed with independent uniform prior
distributions, centered at the nominal values $\xi_0$ suggested
in~\cite{gardner:2000:coa}:
\begin{equation*}
\xi_0 = \left ( 156.25, 15.6, 2.5, 1, 2.0015, 2.9618\times
  10^{-5} \right ) . 
\end{equation*}
In other words, we have
\begin{equation}
\xi_i = \xi_{0,i} \left(1 + \sigma_i \theta_i \right)
\label{eq:thetatoxi}
\end{equation}
where $\theta$ is a vector of uniformly-distributed random variables, $\theta_i \sim
U(-1,1)$, and entries of $\sigma$ are $(0.20, 0.15, 0.15, 0.15, 0.30,
  0.20 )$. As detailed in \cite{marzouk:2009:asc}, the
observational error is assumed Gaussian and zero-mean, with a
standard deviation that depends on whether the expression level is low
or high. This simplified error model is consistent with experimental
observations.

The use of uniform priors adds an extra constraint on the map: since
the prior support is a unit hypercube (with appropriate scaling), the
range of the map must be an improper subset of the unit hypercube,
just as the support of the posterior is contained within the
support of the prior. This constraint is difficult to satisfy when the
map is approximated using global polynomials. To circumvent this
difficulty, we first map the uniform random variables $\theta$ to
independent standard normal random variables $x \sim N(0, I)$ using
the error function:
\begin{equation}
\theta = \erf \left(x / \sqrt{2}\right) \sim U(-1, 1)
\label{eq:xtotheta}
\end{equation}
Computation of the map can now proceed using a Gaussian prior on the
input $x$. After computing the map $f(x)$, the posterior random
variable $\xi_{\mathrm{post}}$ is obtained through the same transformation:
\begin{equation}
  \xi_{\mathrm{post}} =  \xi_0 \left [ 1 + \sigma \erf \left({f(x)
    }/{\sqrt{2}} \right )  \right]
\end{equation}
All derivatives of $\xi$ with respect to $x$ are computed analytically
using~(\ref{eq:thetatoxi}) and (\ref{eq:xtotheta}), e.g.,
\begin{equation}
\frac{ d\xi}{d x} =  \xi_0  \sigma  \frac{\sqrt{2}}{\pi}
\exp{\left(-\frac{x^2}{2}\right)} .
\end{equation}

Using the implicit expression for $v$, we can compute the derivatives of
$v$ with respect to any model parameter $\xi_i$:
\begin{eqnarray}
\frac{\partial v}{\partial \xi_i} &=& \frac{\partial g}{\partial \xi_i} + \frac{\partial  g}{\partial v} \frac{\partial  v}{d \xi_i} \nonumber\\
\frac{\partial v}{\partial \xi_i} &=& \frac{\partial  g}{\partial
  \xi_i} \left(1 - \frac{\partial g}{\partial v} \right)^{-1} .
\end{eqnarray}
The second derivatives are:
\begin{eqnarray}
  \frac{\partial^2v}{\partial \xi_i\xi_j} &=& \frac{\partial g ^2}{\partial \xi_i \partial \xi_j} + \frac{\partial^2 g}{\partial \xi_i \partial v} \frac{\partial v}{\partial \xi_j} + \left( \frac{\partial^2 g}{\partial v^2} \frac{\partial  v}{d\xi_j} + \frac{\partial^2 g}{\partial v \partial \xi_j} \right) \frac{\partial v}{\partial \xi_i} +  \frac{\partial g}{\partial v} \frac{\partial^2 v}{\partial \xi_i \partial \xi_j} \nonumber\\
  \frac{\partial^2v}{\partial \xi_i\xi_j} &=& \left(1 - \frac{\partial g}{\partial v} \right)^{-1}\left( \frac{\partial^2 g}{\partial \xi_i \partial \xi_j} + \frac{\partial^2 g}{\partial \xi_i \partial v} \frac{\partial v}{\partial \xi_j} + \left( \frac{\partial^2 g}{\partial v^2} \frac{\partial  v}{d\xi_j} + \frac{\partial^2 g}{\partial v \partial \xi_j} \right) \frac{\partial v}{\partial \xi_i}  \right) .
\end{eqnarray}
To compute derivatives with respect to the transformed Gaussian random
variables $x$, the chain rule is applied:
\begin{eqnarray*}
  \frac{\partial v}{\partial x} &=& \frac{\partial v}{\partial \xi} \frac{d\xi}{dx} \nonumber\\
  \frac{\partial^2v}{\partial x^2} &=& \frac{\partial v}{\partial \xi} \frac{d^2\xi}{dx^2} + \left(\frac{d\xi}{dx}\right)^T  \frac{\partial^2v}{\partial \xi^2} \frac{d\xi}{dx} . %
\end{eqnarray*}

\begin{figure}[htb]
\centering
\subfigure[First stage: noise level $\Sigma^1 = 16 \Sigma$]
{
\includegraphics[width=2.5in]{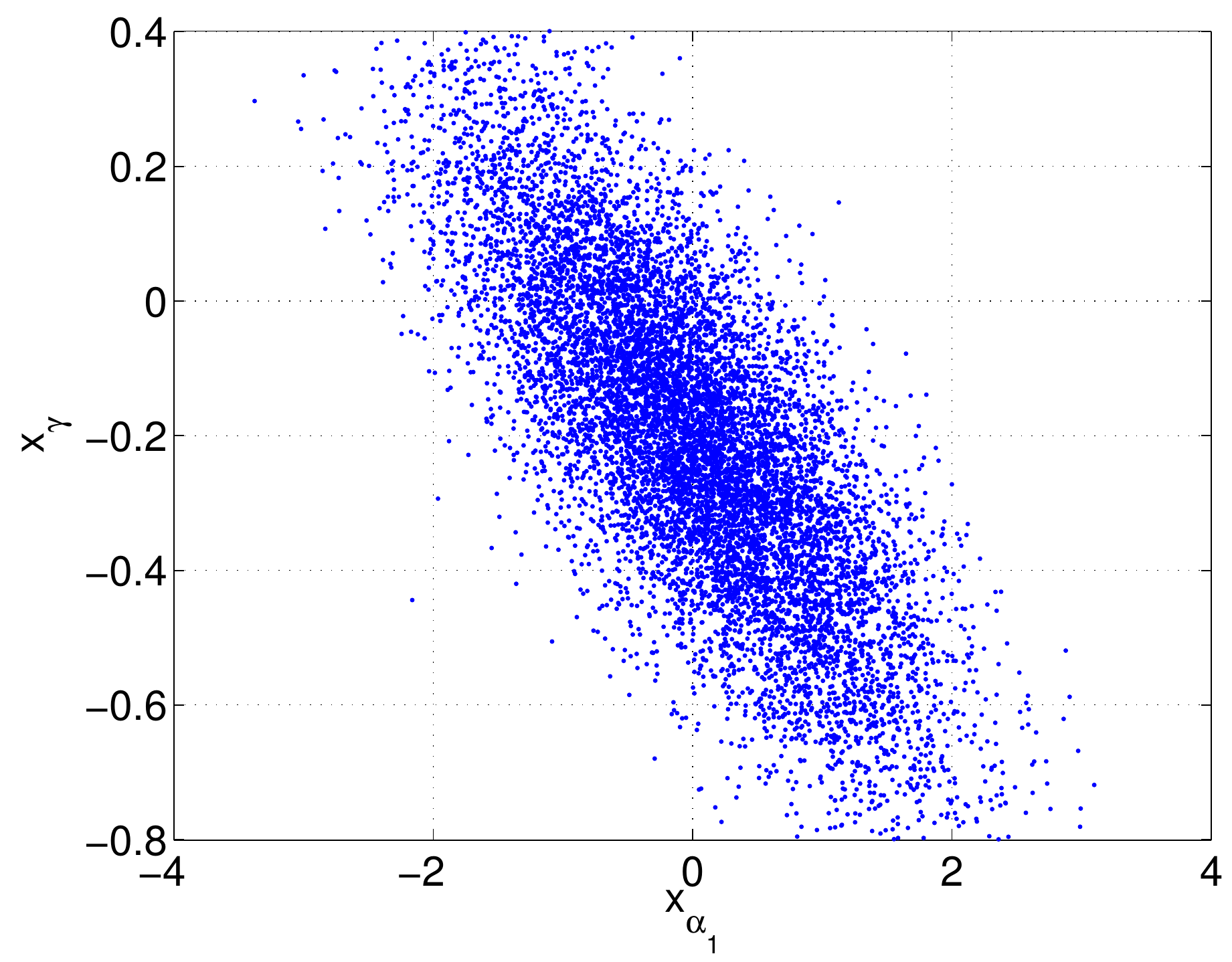}
\label{fig:toggle_scatterstage1}
}
\subfigure[Second stage: noise level $\Sigma^2 = 8 \Sigma$]
{\includegraphics[width=2.5in]{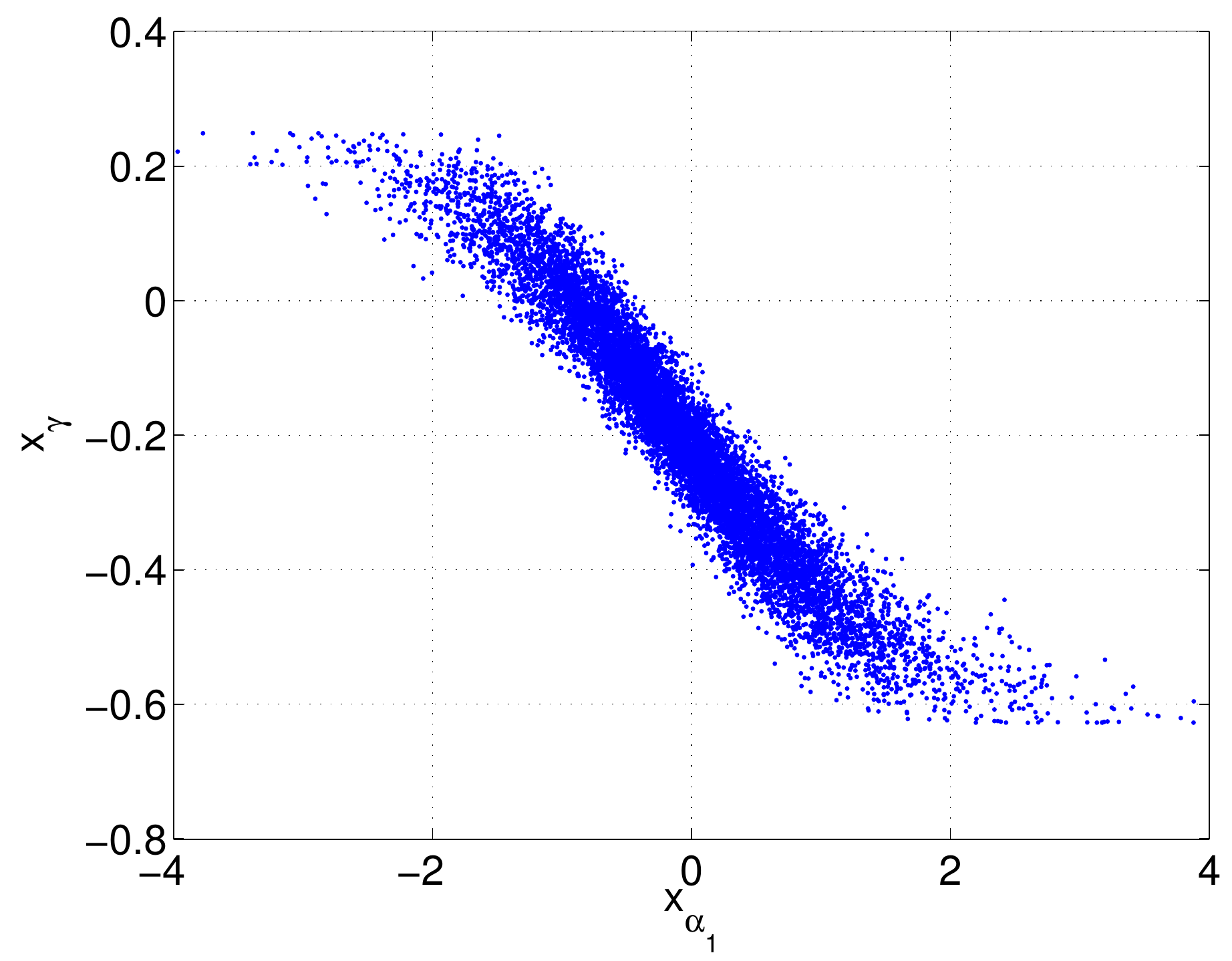}
\label{fig:toggle_scatterstage2}
}
\subfigure[Third stage: noise level $\Sigma^3 = 2 \Sigma$]
{\includegraphics[width=2.5in]{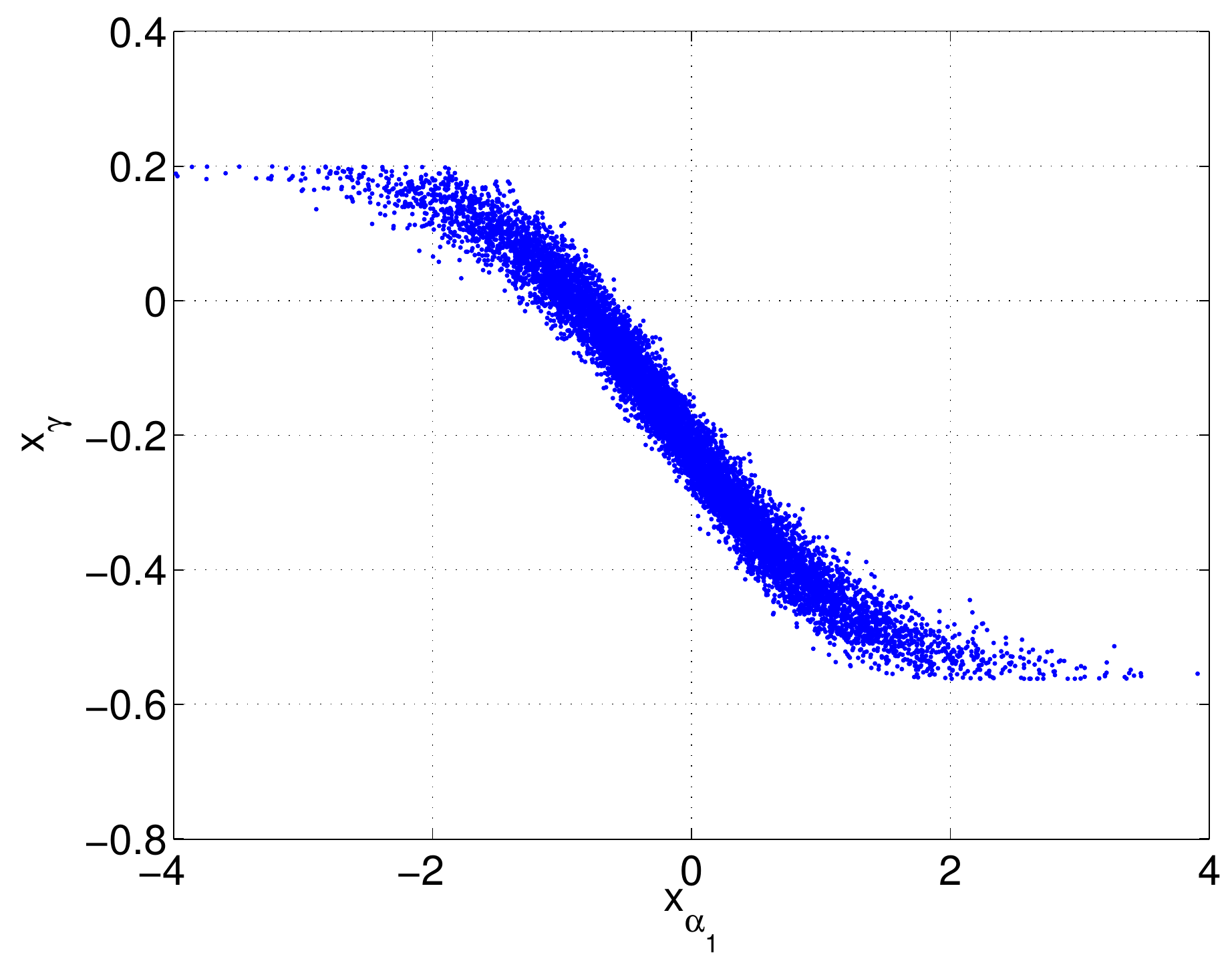}
\label{fig:toggle_scatterstage3}
}
\subfigure[Fourth stage: noise level $\Sigma^4 = \Sigma$]
{\includegraphics[width=2.5in]{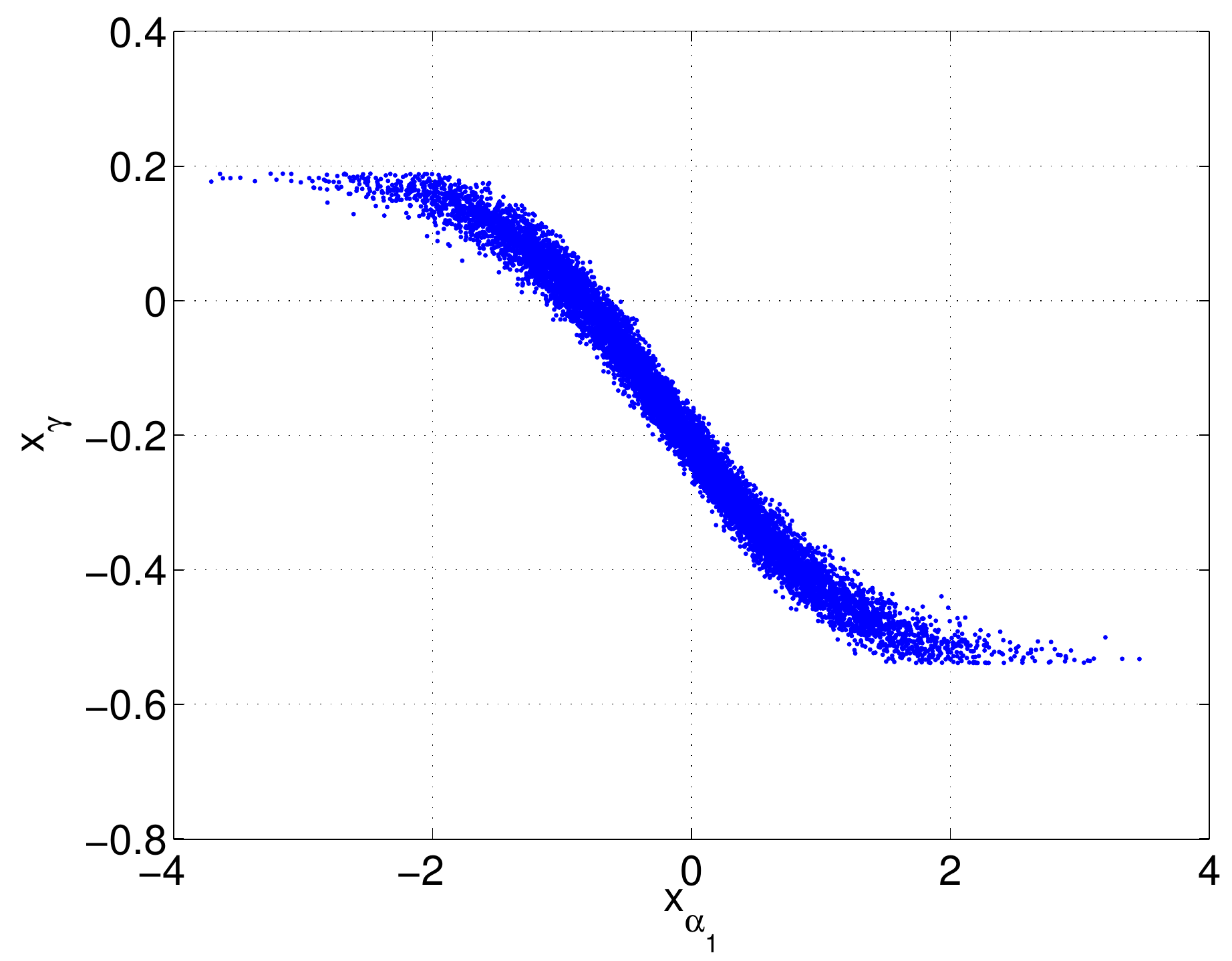}
\label{fig:toggle_scatterstage4}
}
\subfigure[MCMC samples]
{\includegraphics[width=2.5in]{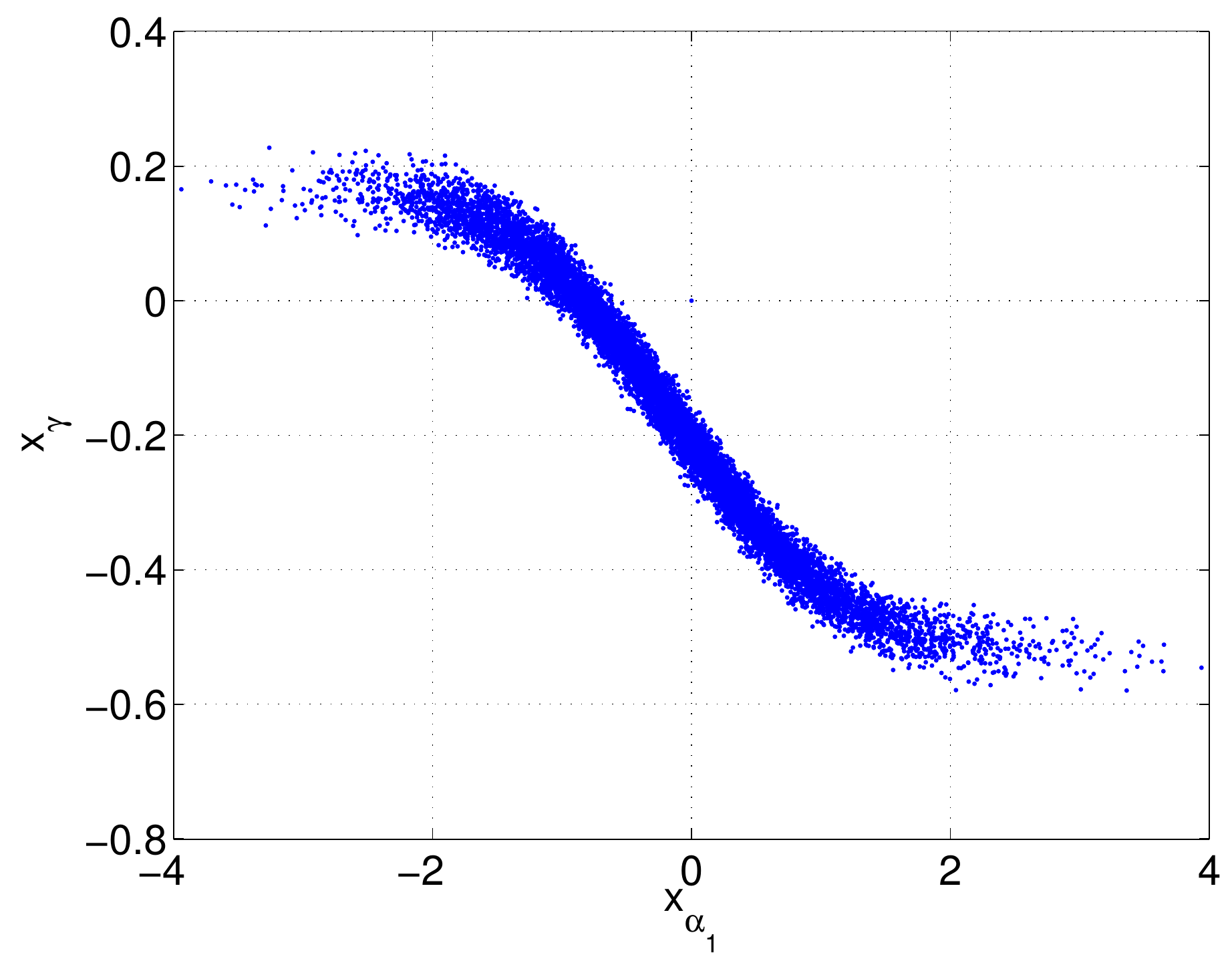}
\label{fig:toggle_scattermcmc}
}

\caption{Toggle switch problem: posterior samples of $x_{\alpha_1}$
  and $x_{\gamma}$ computed using four stages of the composite map
  algorithm. For comparison, we also show posterior samples obtained with
  adaptive MCMC.}
\label{fig:togglestages}
\end{figure}

\begin{figure}[htb]
 \centering
  \subfigure[Posterior pdf of $\alpha_1$.]
{
 \includegraphics[width=2.5in]{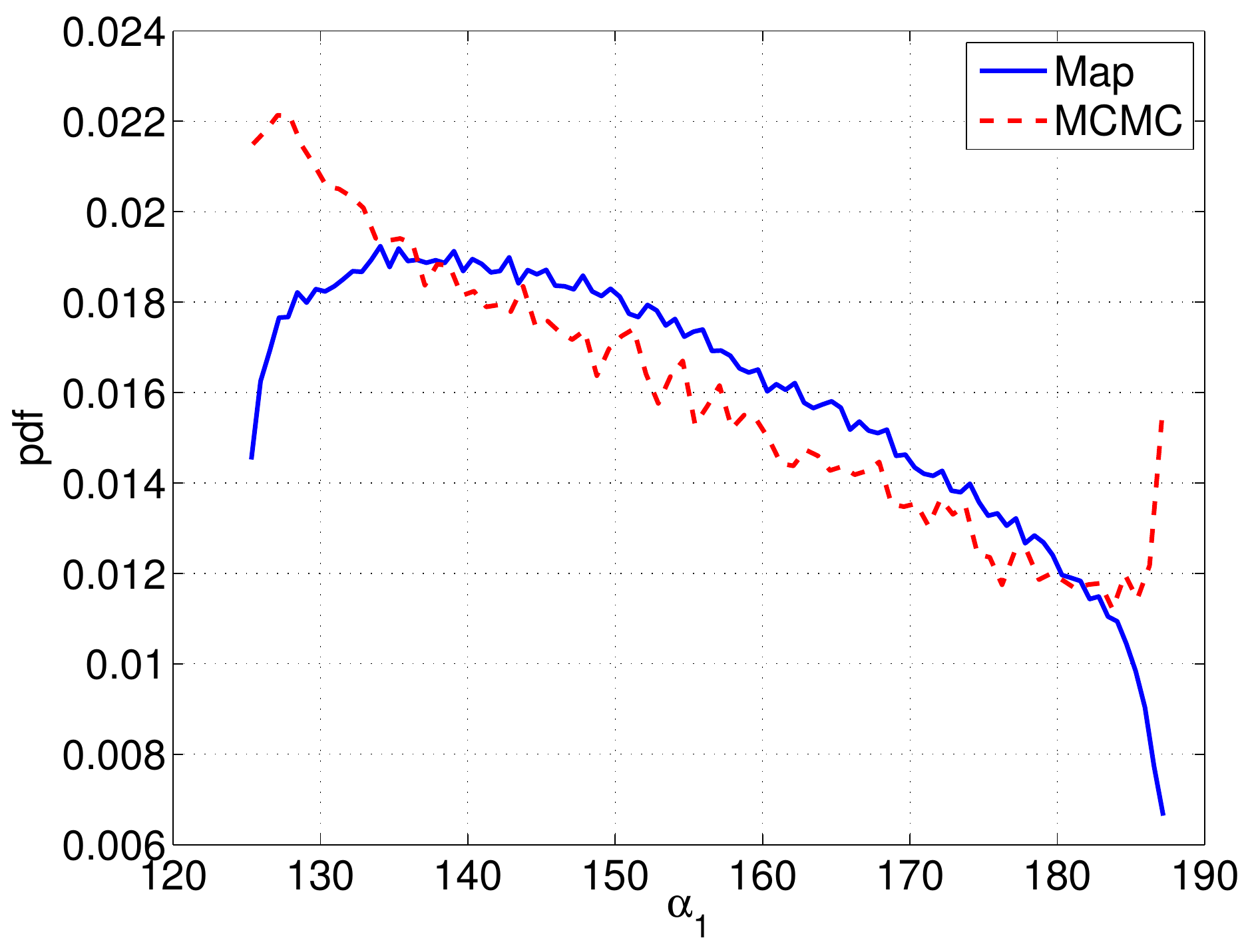}
\label{fig:toggle_pdfalpha1}
}
  \subfigure[Posterior pdf of $\alpha_2$.]
{
  \includegraphics[width=2.5in]{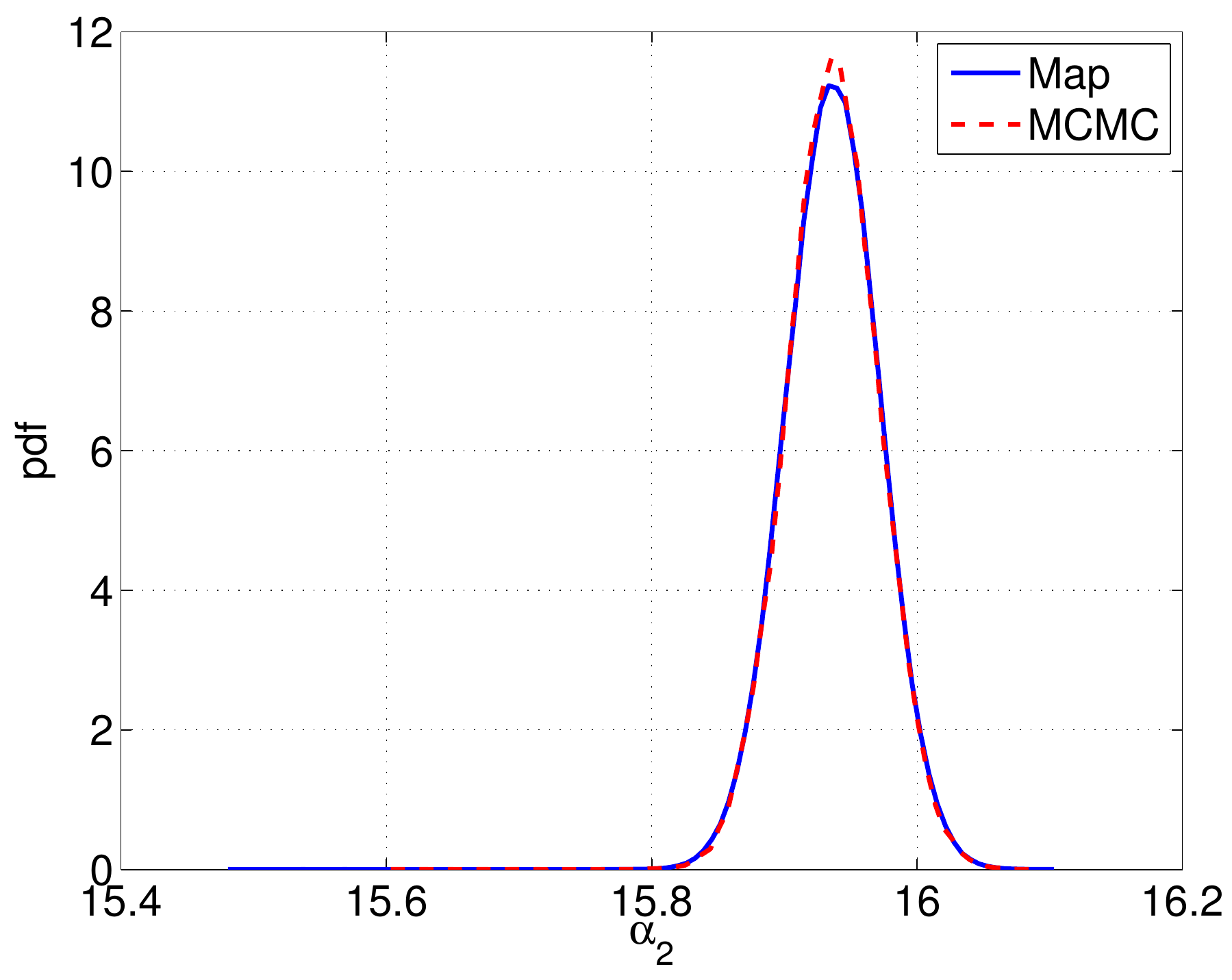}
\label{fig:toggle_pdfalpha2}
}
  \subfigure[Posterior pdf of $\beta$.]
{
 \includegraphics[width=2.5in]{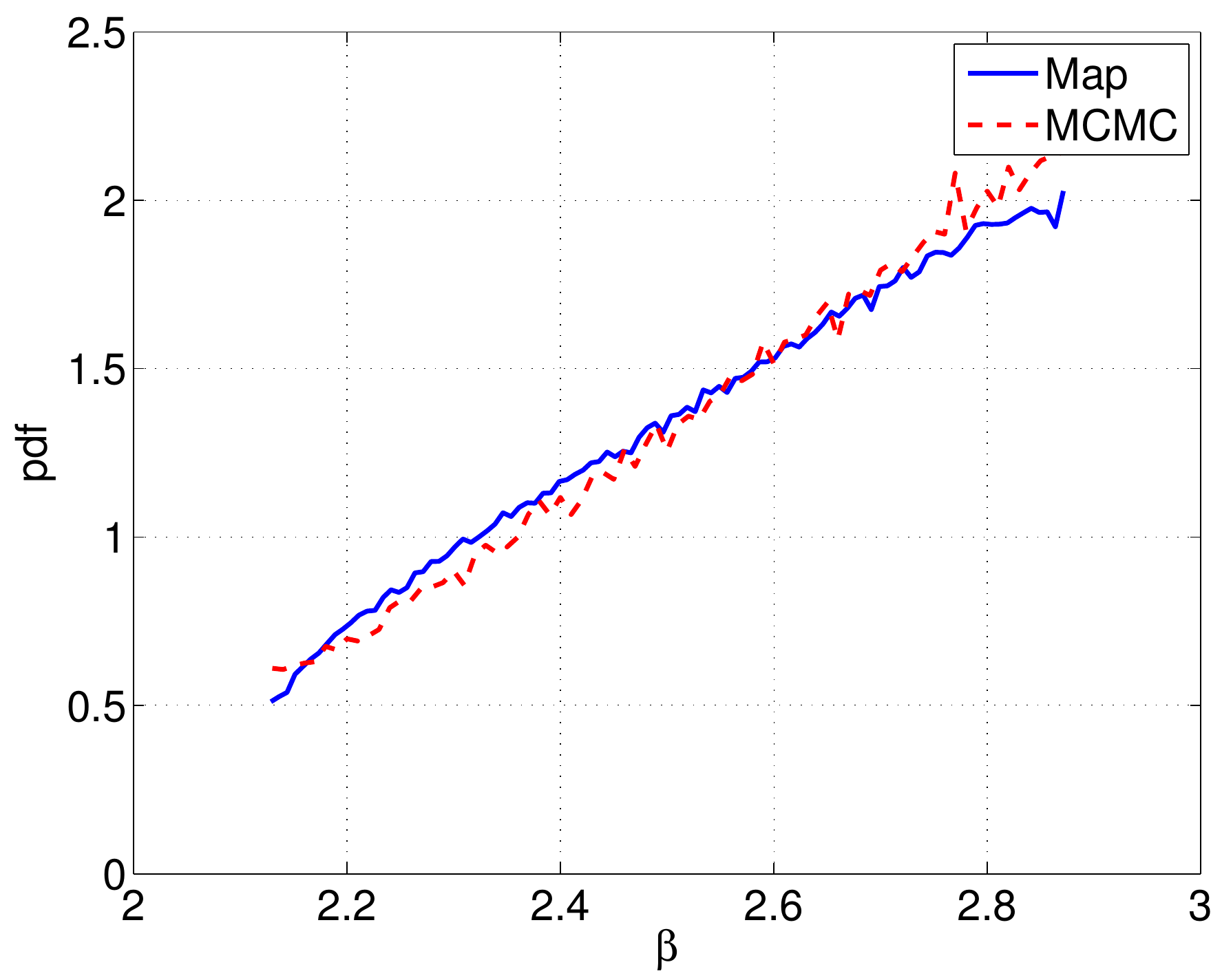}
\label{fig:toggle_pdfbeta}
}
  \subfigure[Posterior pdf of $\gamma$.]
{
 \includegraphics[width=2.5in]{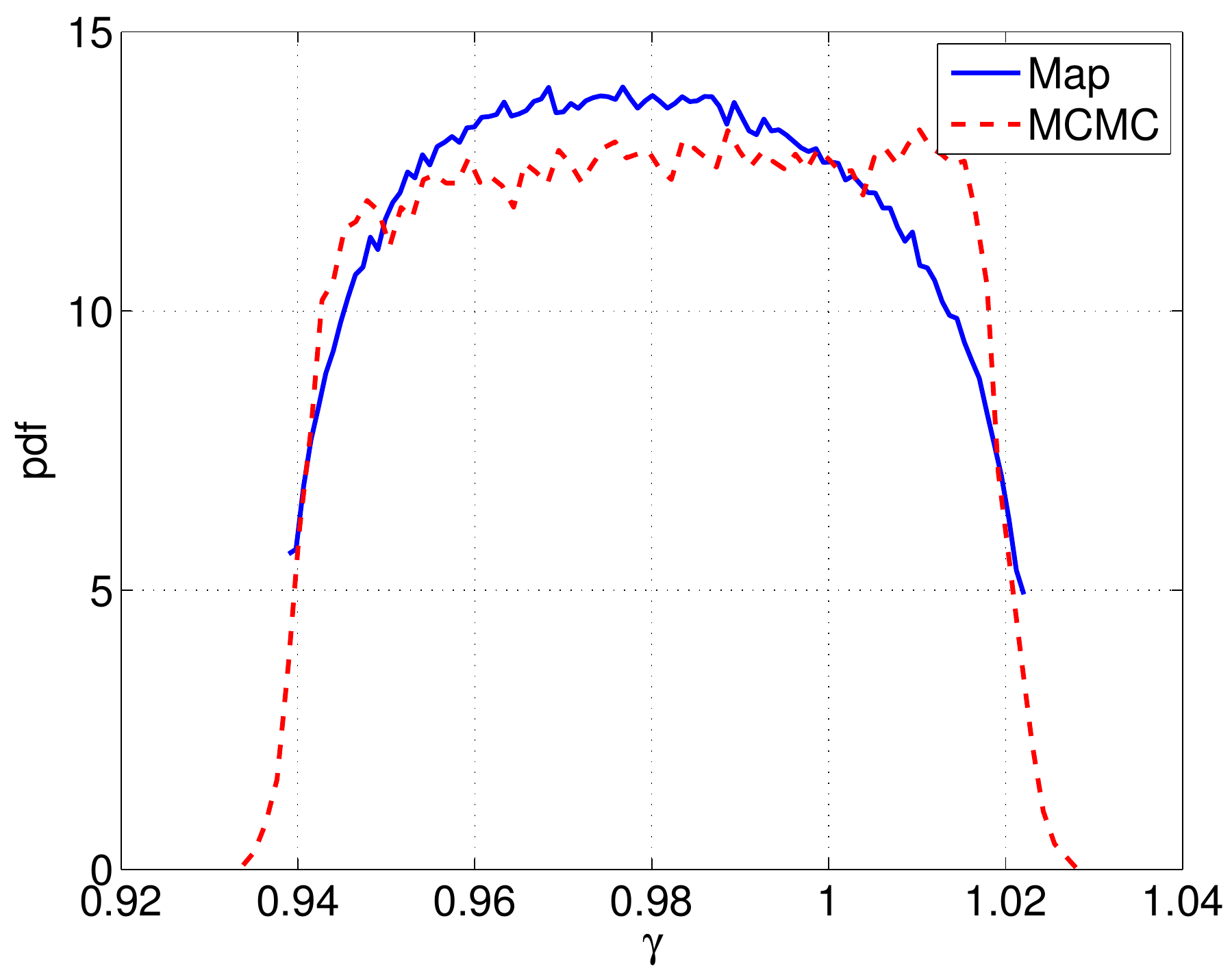}
\label{fig:toggle_pdfgamma}
}  
 \subfigure[Posterior pdf of $\eta$.]
{
 \includegraphics[width=2.5in]{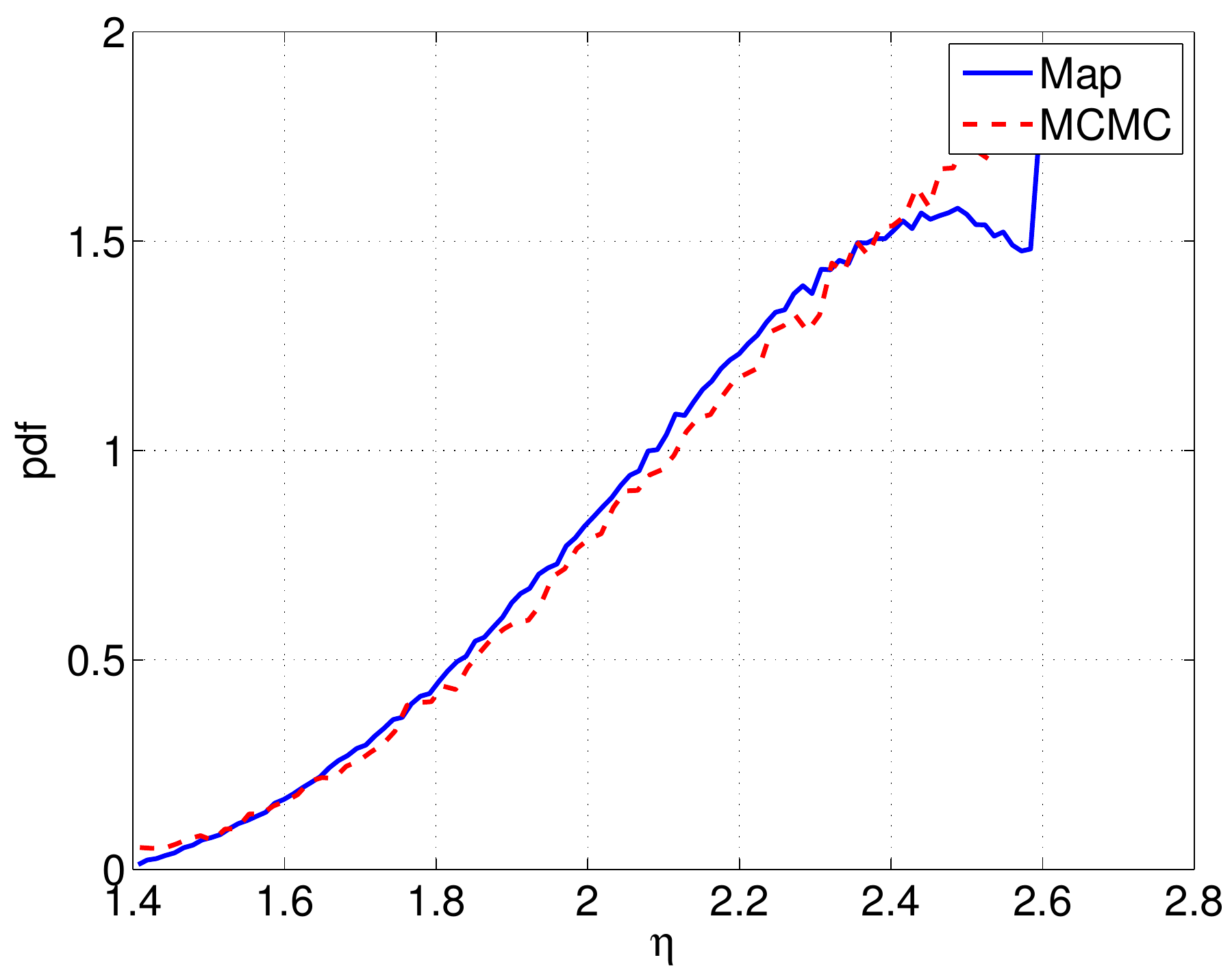}
\label{fig:toggle_pdfeta}
}
 \subfigure[Posterior pdf of $\kappa$.]
{
  \includegraphics[width=2.5in]{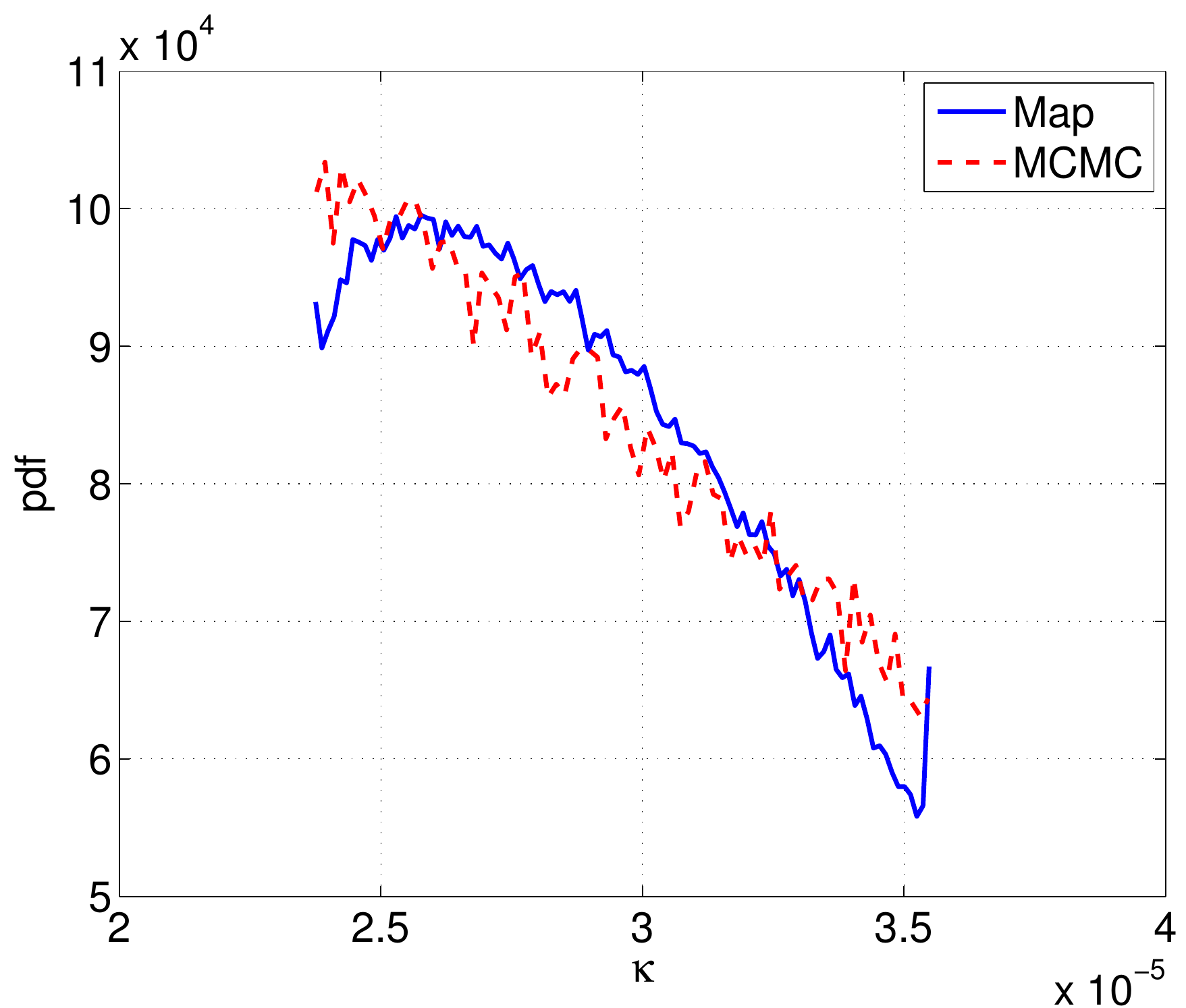}
\label{fig:toggle_pdfkappa}
}
\caption{Toggle switch problem: Marginal posterior probability density
  functions of each model parameter, computed using the map and
  compared with MCMC.}
\end{figure}

\begin{figure}[htb]
 \centering
 \includegraphics[width=4.8in]{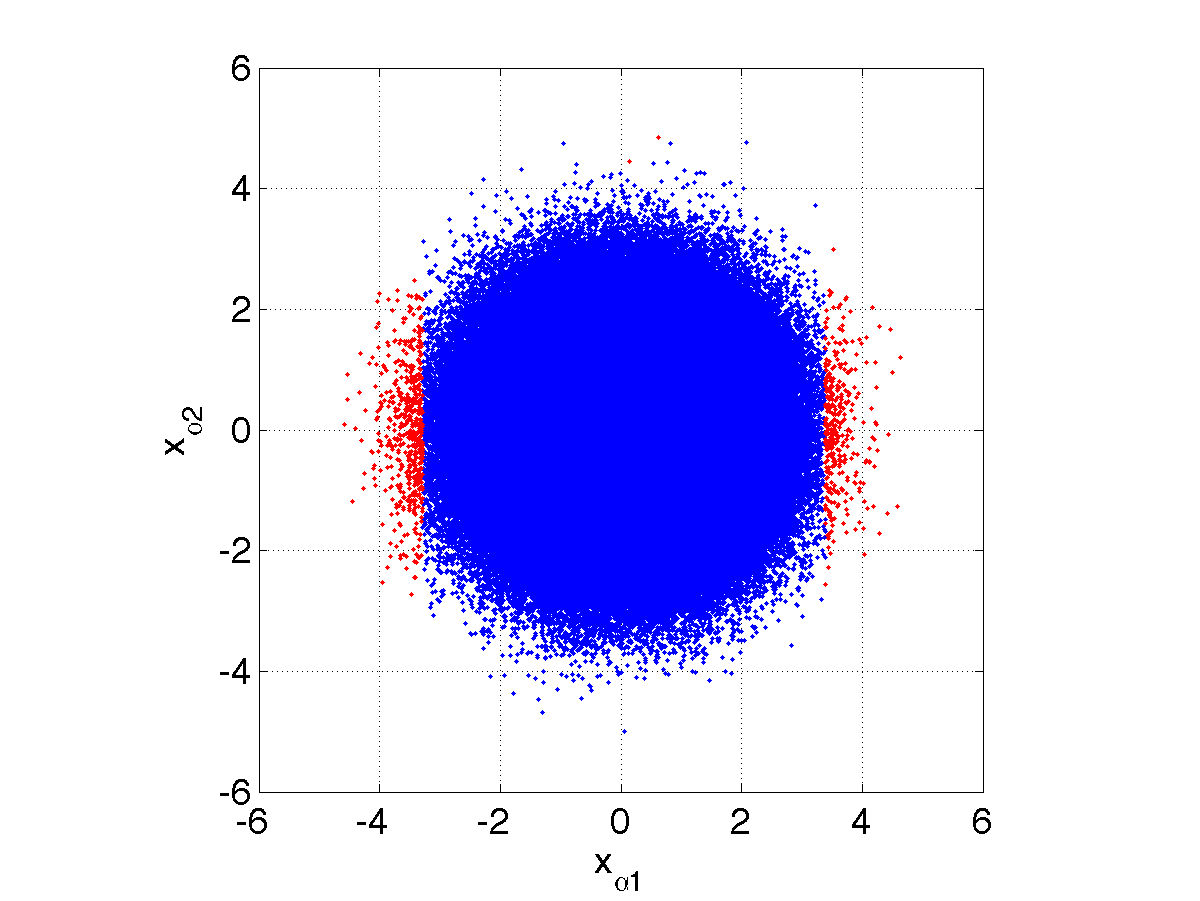}
 \caption{Toggle switch problem: colors represent the sign of the
   determinant of the Jacobian of $f$ (the final composite map). Only
   861 out of $10^6$ points, shown in red, have negative Jacobian
   determinant. These points lie in the low probability region of
   $x_{\alpha_1}$. }
\label{fig:toggle_jacobian}
\end{figure}

With these transformations in place, we turn to the numerical results
in Figures~\ref{fig:togglestages}--\ref{fig:toggle_jacobian}. In this
problem, a high-order map is required to accurately capture the
posterior distribution, and thus we employ the \textit{composite map}
of Section~\ref{subsec:composite}. In particular, we compute the
overall map using a four-stage cascade of third order maps, $f = f^1
\circ \cdots \circ f^4$. The distributions obtained after each stage
are shown in
Figures~\ref{fig:toggle_scatterstage1}--\ref{fig:toggle_scatterstage4},
using a scatter plot of output samples from each map ($\phi^1$,
$\phi^2$, etc). While the posterior is six-dimensional, these plots
focus on the pairwise marginal distribution of $\alpha_1$ and
$\gamma$; this is the most ``complex'' pairwise marginal and is
particularly sensitive to the accuracy of $f$.
These distributions are shown on the transformed Gaussian domain, and
thus they correspond to the first and fourth components of $x$:
$x_{\alpha_1}$ and $x_{\gamma}$. For comparison,
Figure~\ref{fig:toggle_scattermcmc} shows results obtained using a
long run of delayed-rejection adaptive Metropolis (DRAM)
MCMC~\cite{haario:2006:dra}, with 10$^6$ samples.

In Figures~\ref{fig:toggle_pdfalpha1}--\ref{fig:toggle_pdfkappa} we
show the marginal posterior probability density of each component of
$\xi$, obtained at the end of the four-map sequence. These densities
are compared with densities obtained using the long MCMC run; the
map-based results show good agreement with the latter.
Figure~\ref{fig:toggle_jacobian} shows samples from the joint prior of
$x_{\alpha_1}$ and $x_{\alpha_2}$. As specified above, these
parameters are independent and standard normal. The point of this
figure is not the distribution per se, but rather the color labeling,
which diagnoses the monotonicity of the map. The map is evaluated on
$10^6$ samples from the prior. Among these samples, only 861 have
Jacobian determinants that are negative; all the rest are positive. As
in the previous example, the few samples with negative Jacobian
determinant are concentrated in the tails of the distribution. Away
from the tails, the map is monotone.

\subsection{PDE-constrained inverse problems}
\label{sec:results:subsec:PDE}
Examples in this section focus on high-dimensional inverse problems involving partial
differential equations (PDEs).
In particular, our objective is to estimate a spatially
heterogeneous coefficient $\kappa$ appearing in an elliptic PDE, from noisy and limited observations of
the solution field~\cite{dashti:2012:uqa}. We solve the problem on one-dimensional (Section~\ref{sec:results:subsubsec:PDE1d}) and two-dimensional (Section~\ref{sec:results:subsubsec:PDE2d})
spatial domains. We employ the triangular parameterization of the map, performing quantitative comparisons of computational efficiency and accuracy with  MCMC  for a range of data sets and observational noise magnitudes.


\subsubsection{One-dimensional domain}
\label{sec:results:subsubsec:PDE1d}

Consider a linear elliptic PDE on the unit interval $\mathcal{S} = [0,1]$
\begin{equation}
\nabla \cdot \left ( \kappa(s) \nabla u \right ) = -f(s)
\label{e:1dpde}
\end{equation}
where $s \in \mathcal{S}$ is the spatial coordinate, $\nabla
\equiv \partial/\partial s$, and $f$ is a known source term. In the
subsurface context, this equation describes Darcy
flow 
through porous media, where $\kappa$ represents a permeability field
and $u$ is the pressure. We apply boundary conditions $\left
  . \pp{u}{s} \right |_{s=0} = 0$ and $\left . u \right |_{s=1} =
1$. The source term is given by:
\begin{equation}
f(s) = a \exp \left ( -\frac{1}{2b^2} \(s-s_m \)^2  \right )
\label{e:sourceterm}
\end{equation}
with $a = 100$, $s_m = 0.5$, and $b = 0.01$.

We place a log-normal prior on the permeability field $\kappa$, 
\begin{equation}
\log \[ \kappa(s, \omega) - \kappa_0 \] \sim \mathcal{GP} \( 0, C \) 
\label{e:lnprior}
\end{equation}
and let the covariance kernel  $c(s, s^\prime)$ of the Gaussian process have exponential form
\begin{equation}
c\(s, s^\prime \) = \sigma^2 \exp \( - \frac{ \left | s - s^\prime
  \right |}{L_c} \) .
 \end{equation} \footnote{Note that $c(s, s^\prime)$ is the Green's function of the differential operator 
\begin{equation}
-\frac{L_c}{2 \sigma^2} \frac{d^2}{d s^2} + \frac{1}{2 \sigma^2 L_c} 
\label{eq:diffop}
\end{equation} 
with appropriate boundary conditions~\cite{rasmussen:2006:gpf, Papaspiliopoulos:2012:neo}. Hence the inverse covariance operator $C^{-1}$ is explicitly represented by~(\ref{eq:diffop}).}
We use a prior standard deviation of
$\sigma = 1.5$ and a correlation length $L_c=0.5$, along with an
offset $\kappa_0 = 0.5$. Realizations of this spatial process are
rough; in fact, they are not mean-square differentiable.

The problem is spatially discretized using finite differences
on a uniform grid, with spacing $\Delta s = 1/100$. The log-normal
permeability field is stochastically discretized using the Karhunen-Lo\`{e}ve
expansion of the underlying Gaussian process:
\begin{equation}
\log{ \[ \kappa(s, \omega) -\kappa_0 \]} \approx \sum\limits_{i=1}^n \phi_i(s) \sqrt{\lambda_i} x_i(\omega)
\end{equation}
where $x_i \sim N(0, 1)$ are independent standard normal random
variables, and $\phi_i$, $\lambda_i$ are the eigenfunctions and
eigenvalues of the linear operator corresponding to the covariance
kernel: $\int_{\mathcal{S}} c(s, s^\prime) \phi_i(s) ds = \lambda_i
\phi(s^{\prime})$. We discretize the eigenfunctions on the same grid
used to discretize the PDE solution field $u$. To capture 99\% of the
spatially-integrated variance of the log-normal process, we retain $n=
66$ Karhunen-Lo\`{e}ve modes.

Noisy observations of $u$ are obtained at $m$ points in space. The
noise is additive and i.i.d.\ Gaussian, such that $d_j = u(s_j; x) +
\varepsilon_j$ with $\varepsilon_j \sim N(0, \sigma_n^2)$, $j=1\ldots
m$. The observational data is generated by choosing a ``true''
permeability field $\kappa$, solving the \textit{full} PDE model
(\ref{e:1dpde}) to obtain the corresponding pressure field $u(s)$,
then corrupting $u(s)$ at the $m$ measurement locations with
independent realizations of the noise. In the inference process, on
the other hand, we use the polynomial chaos expansion
(\ref{e:pcforward}) as the forward model. This discrepancy ensures that we do not
commit an ``inverse crime'' \cite{kaipio:2004:bip}.

Any Bayesian inference strategy, whether the map-based optimization
approach or MCMC, requires repeated evaluations of the forward
model. As suggested in~\cite{marzouk:2009:dra}, exploiting regularity
in the dependence of $u$ on the parameters $x$ can make these
evaluations more computationally tractable. We therefore approximate
$u(s; x)$ using a polynomial chaos expansion. We apply the iterative
algorithm developed in~\cite{moselhy:2011:aai}, for the solution of
high-dimensional stochastic PDEs, to obtain an approximation of $u$ in
the form:
\begin{equation}
u(s; x) = \sum\limits_{\mathbf{k} \in \mathcal{K}} u_{\mathbf{k}}(s) \psi_{\mathbf{k}}(x)
\label{e:pcforward}
\end{equation} 
where $u_{\mathbf{k}}$ are coefficients and $\psi_{\mathbf{k}}(x)$ are multivariate Hermite
polynomials, chosen adaptively within a very large basis set
$\mathcal{K}$.

Algorithm~\ref{alg:mainalgorithm} allows the expansion order of the
map $f(x)$ to be increased in stages; i.e., we begin by finding a
linear map, then a cubic map, and so on. Since inference problems
involving distributed parameters are typically high-dimensional (in
the current problem $f$ maps $\RR^{66}$ onto itself, for example),
writing $f$ as a total-order polynomial expansion in all $n$ of its
components will lead to a very large number of degrees of freedom,
most of which are not needed to achieve an accurate representation of
the posterior. Instead, we refine the polynomial description of the
map in a more finely-grained fashion.
Recall that $f(x) = F^T \Psi(x)$, where $\Psi$ is a vector of
orthogonal polynomials in $x_1 \ldots x_n$. In the structure of
Algorithm~\ref{alg:mainalgorithm}, decisions to expand the polynomial
description of the map are made outside of the inner optimization
loop, after checking whether $\Var[T]$ satisfies the desired threshold
$\delta$. Now, rather than raising the polynomial degree of the entire
map (e.g., $n_0 \leftarrow n_0 + 2$), we choose a subset of the inputs
$\{x_i: i \in \mathcal{I} \}$ and introduce higher-degree polynomial
terms in these variables only. The index set $\mathcal{I}$ initially
consists of $\{1, 2, \ldots, i_1 < n \}$. At the next iteration, if
the variance threshold is still not satisfied, $i_1$ is replaced with
a larger index $i_2$, where $i_1 < i_2 < n$. This process continues
until the largest element in $\mathcal{I}$ is $n$ or until $\Var[T]$
stagnates. Now $\mathcal{I}$ is reset and new polynomial terms of even
higher degree, involving only $x_1$ through $x_{i_1}$, are added to the
expansion. Then the index set is expanded once again. In any of these
iterations, adding terms to the expansion is equivalent to adding rows
to $\Psi(x)$ and to the matrix of polynomial coefficients $F$.

To make this process more concrete, consider what happens in the
present inference problems. We begin by solving for the polynomial
coefficients of a linear map (possibly subject to the triangular
constraint). The optimal linear map is not able to reduce $\Var[T]$
below the requisite threshold. Polynomial terms of total order $n_0=3$
in the first $i_1=10$ (for example) components of $x$ are then
introduced, and all the coefficients collected in $F$ are adjusted via
optimization. This refinement of $f$ is still not sufficient, so now
the order-3 expansion is extended to the first $i_2=20$ (again for example) components of
$x$. Extension to additional components of $x$ is observed to yield
little decrease in the minimal value of $\Var[T]$, so now the
polynomial space is enriched by adding terms of total order $n_0=5$ in
the first $i_1$ components of $x$. Iterations continue until
convergence, resulting an ``adapted'' map $f$ whose components are a
subset of a total-order expansion in $x_1 \ldots x_n$.

\begin{figure}[htb]
 \centering
 \subfigure[Case I: Posterior mean of $\log(\kappa- \kappa_0)$.]
 {
 \includegraphics[width=2.5in]{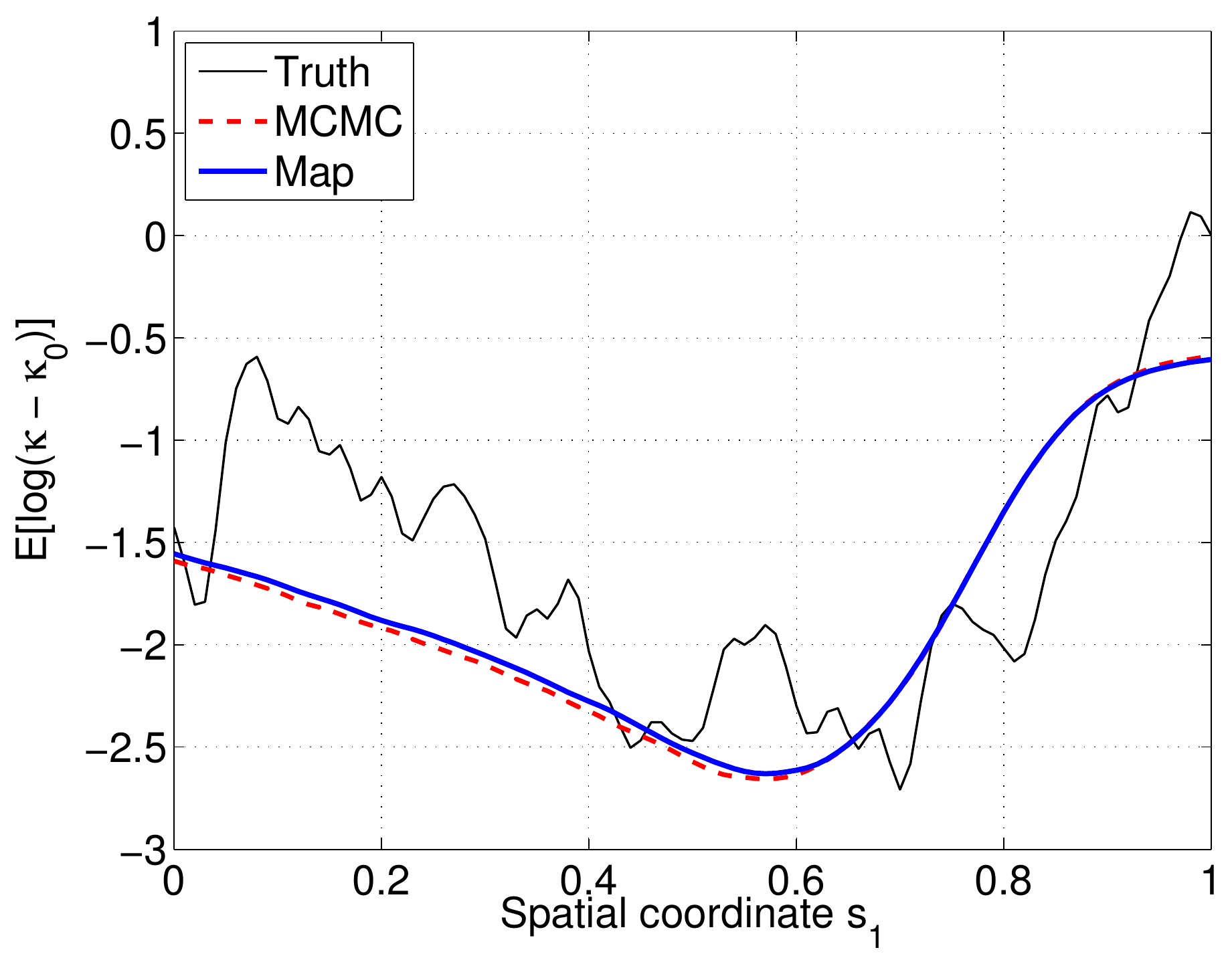}
\label{fig:onedpde_sdln_logN_mean}
}
 \subfigure[Case I: Posterior std of $\log(\kappa- \kappa_0)$]
 {
 \includegraphics[width=2.5in]{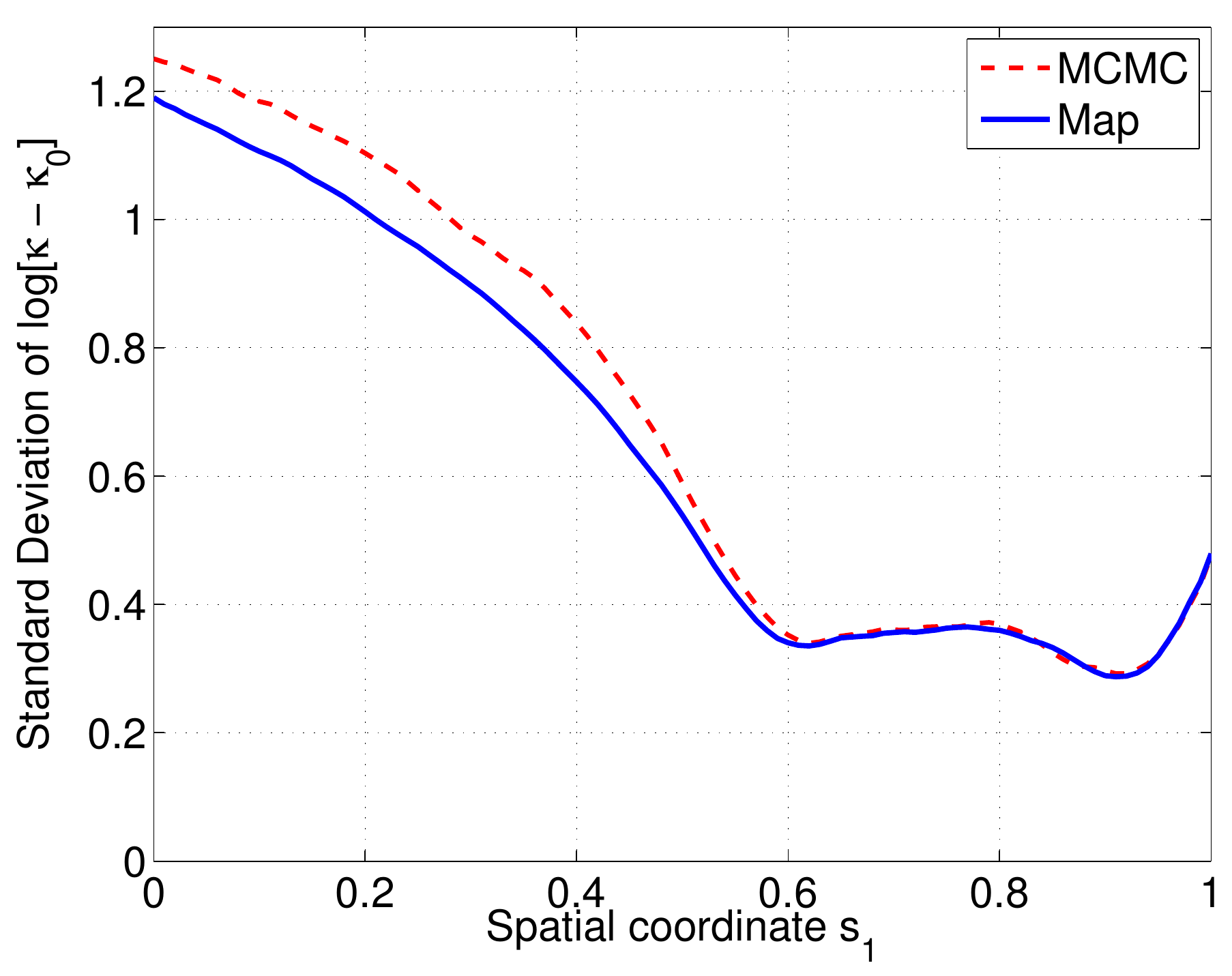}
\label{fig:onedpde_sdln_logN_std}
}
 \subfigure[Case II: Posterior mean of $\log(\kappa- \kappa_0)$]
 {
 \includegraphics[width=2.5in]{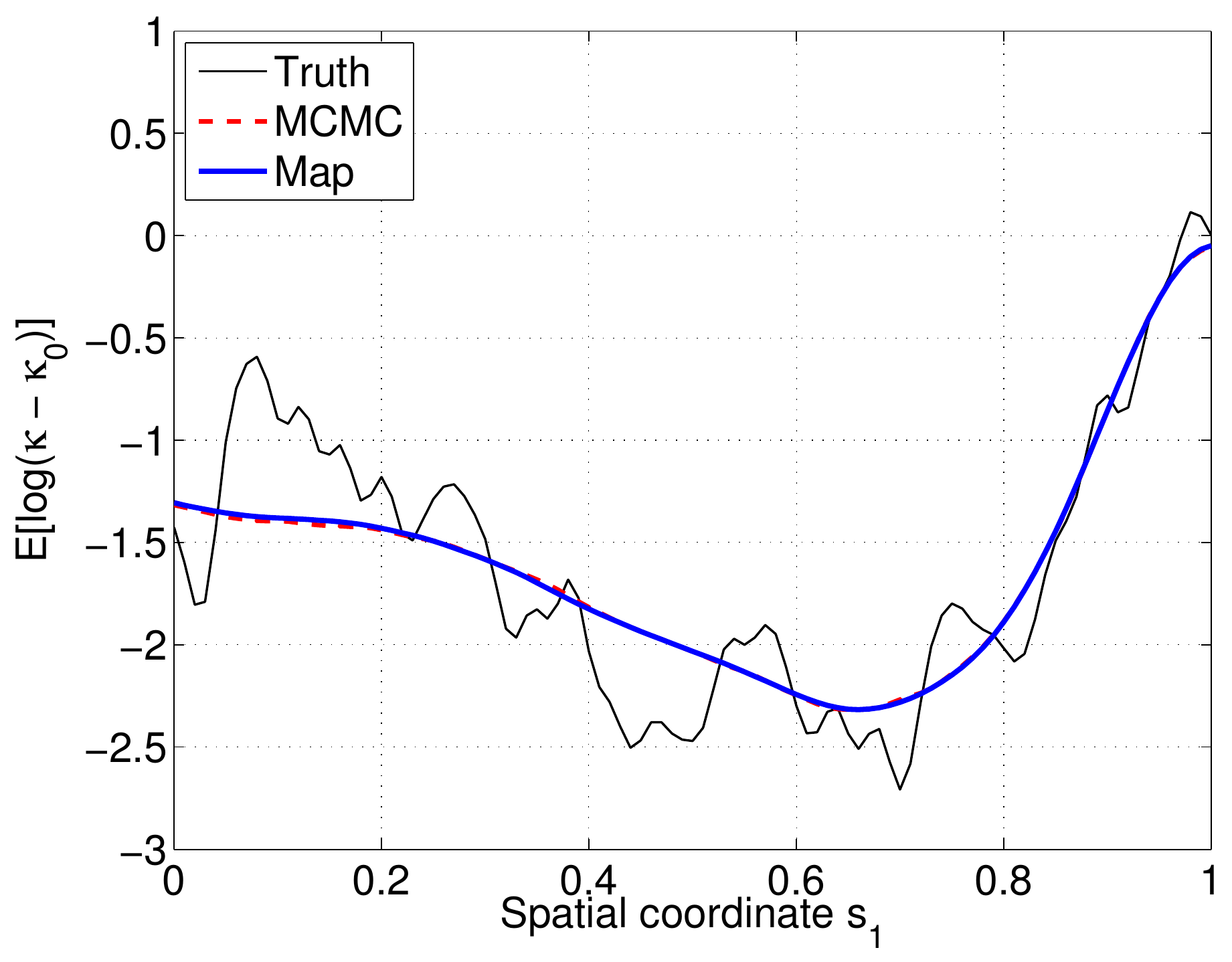}
\label{fig:onedpde_ldln_logN_mean}
}
 \subfigure[Case II: Posterior std of $\log(\kappa- \kappa_0)$]
 {
 \includegraphics[width=2.5in]{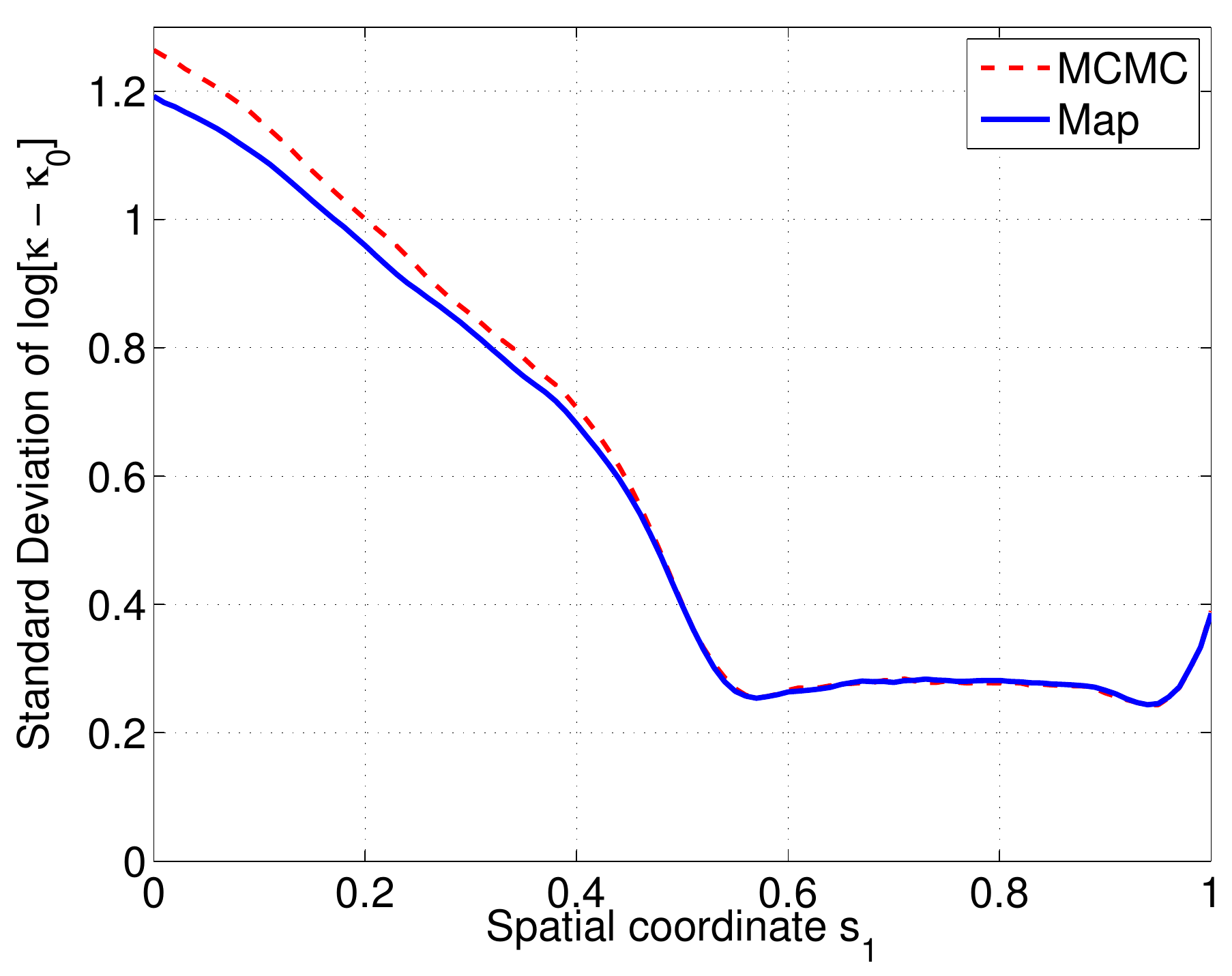}
\label{fig:onedpde_ldln_logN_std}
}
 \subfigure[Case III: Posterior mean of $\log(\kappa- \kappa_0)$]
 {
 \includegraphics[width=2.5in]{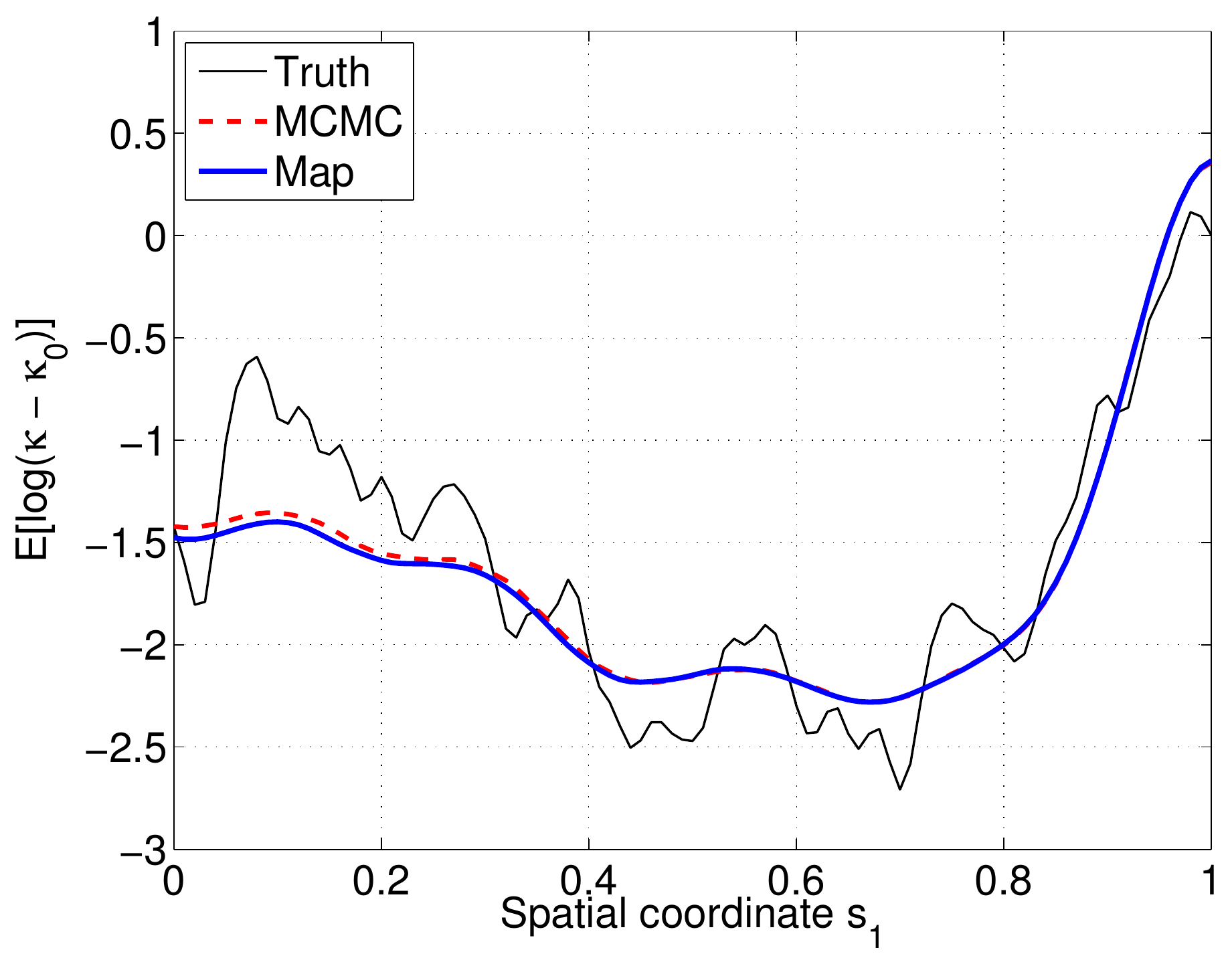}
\label{fig:onedpde_ldsn_logN_mean}
}
 \subfigure[Case III: Posterior std of $\log(\kappa- \kappa_0)$]
 {
 \includegraphics[width=2.5in]{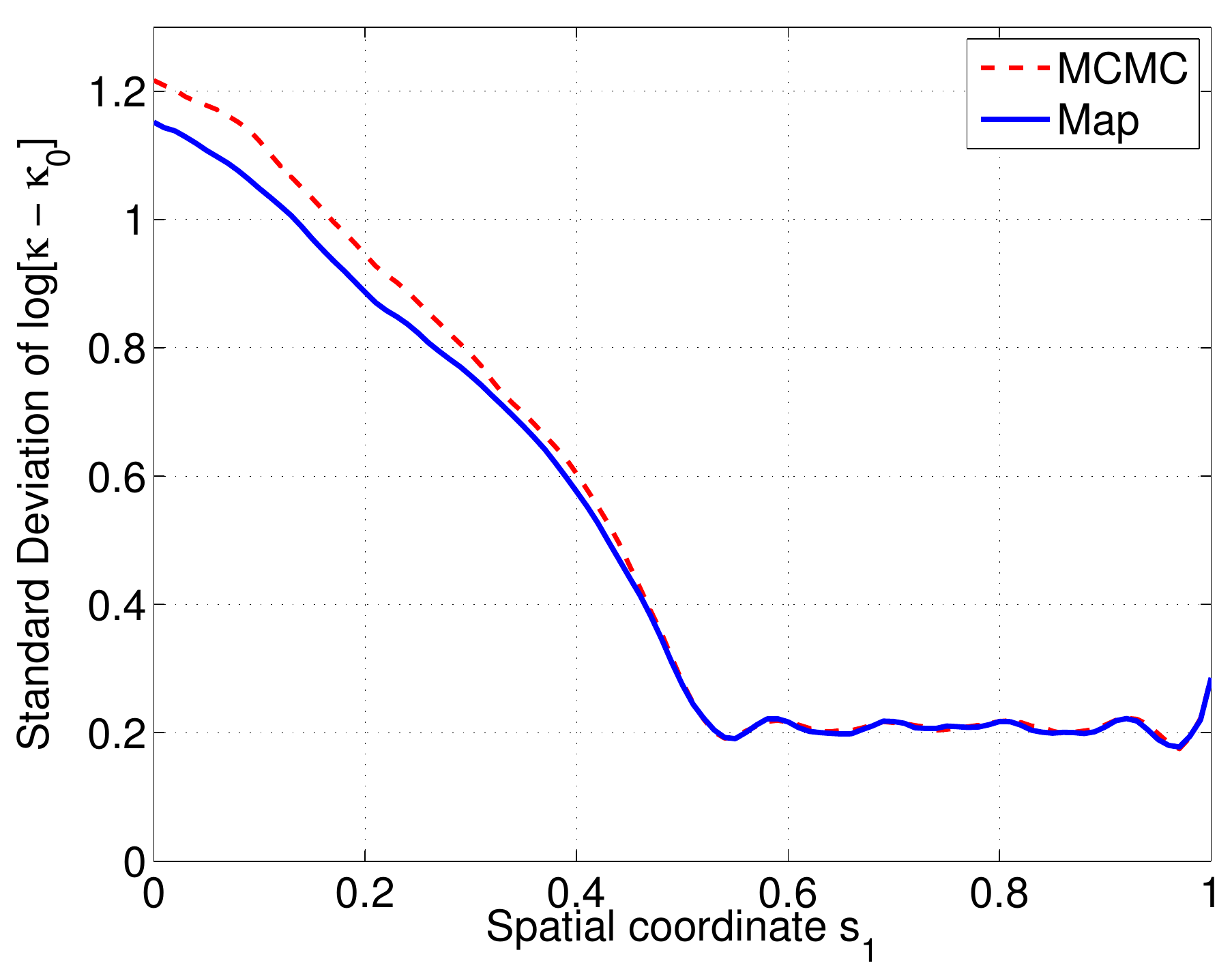}
\label{fig:onedpde_ldsn_logN_std}
}
\caption{One-dimensional elliptic PDE: results of the map algorithm
  compared to results of MCMC for three different data cases, detailed
  in the text. The cases are: (I) fewer data points and larger noise
  variance; (II) many data points and larger noise variance; (III)
  many data points and smaller noise variance. MCMC experiences
  significant convergence difficulties in the final case.}
\label{fig:onedpde_ldsn_logN_meanstd}
\end{figure}

\begin{figure}[htb]
 \centering
\subfigure[Case I: Map realizations]
 {
 \includegraphics[width=2.5in]{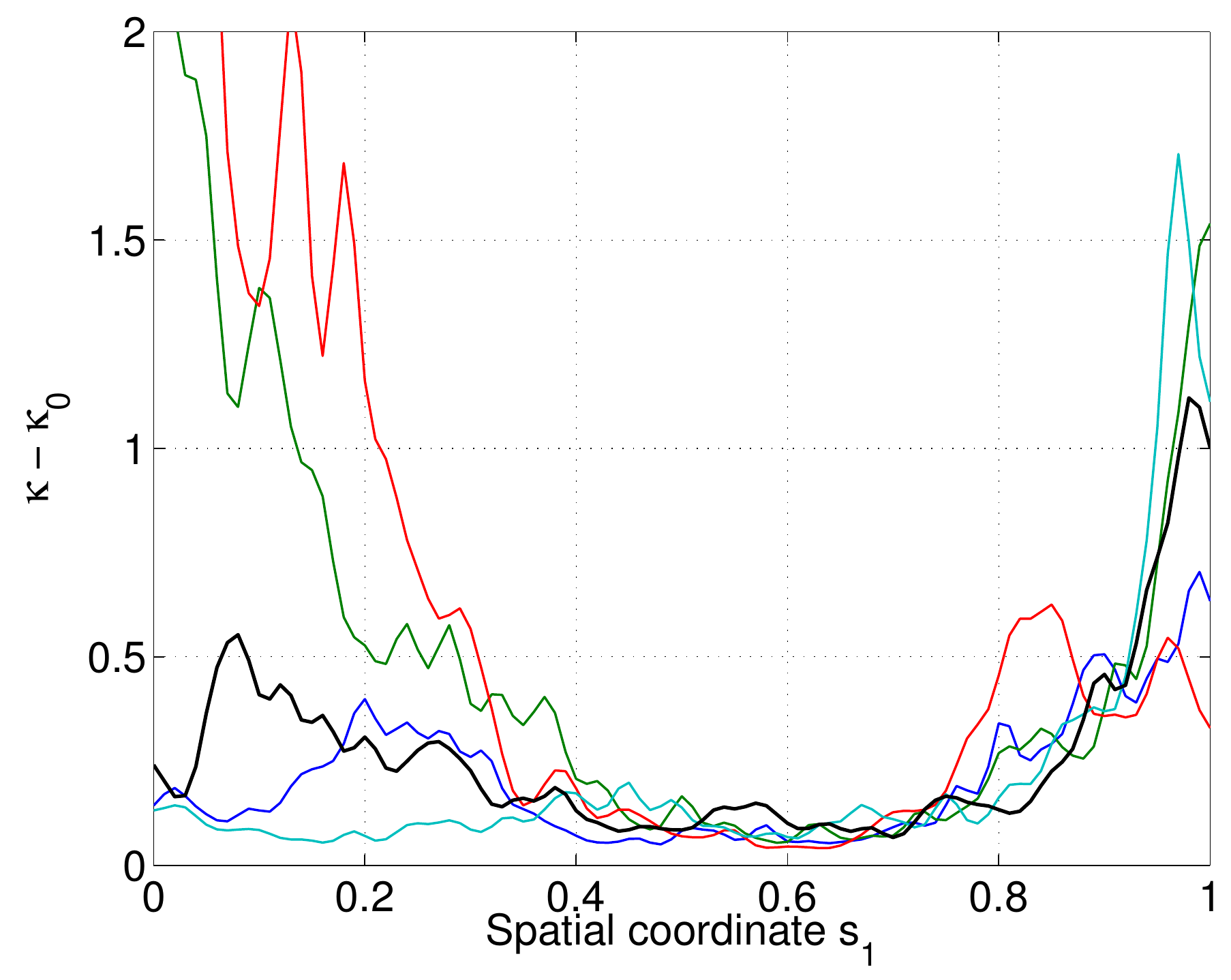}
\label{fig:onedpde_realizations_map1}
}
\subfigure[Case I: MCMC realizations]
 {
 \includegraphics[width=2.5in]{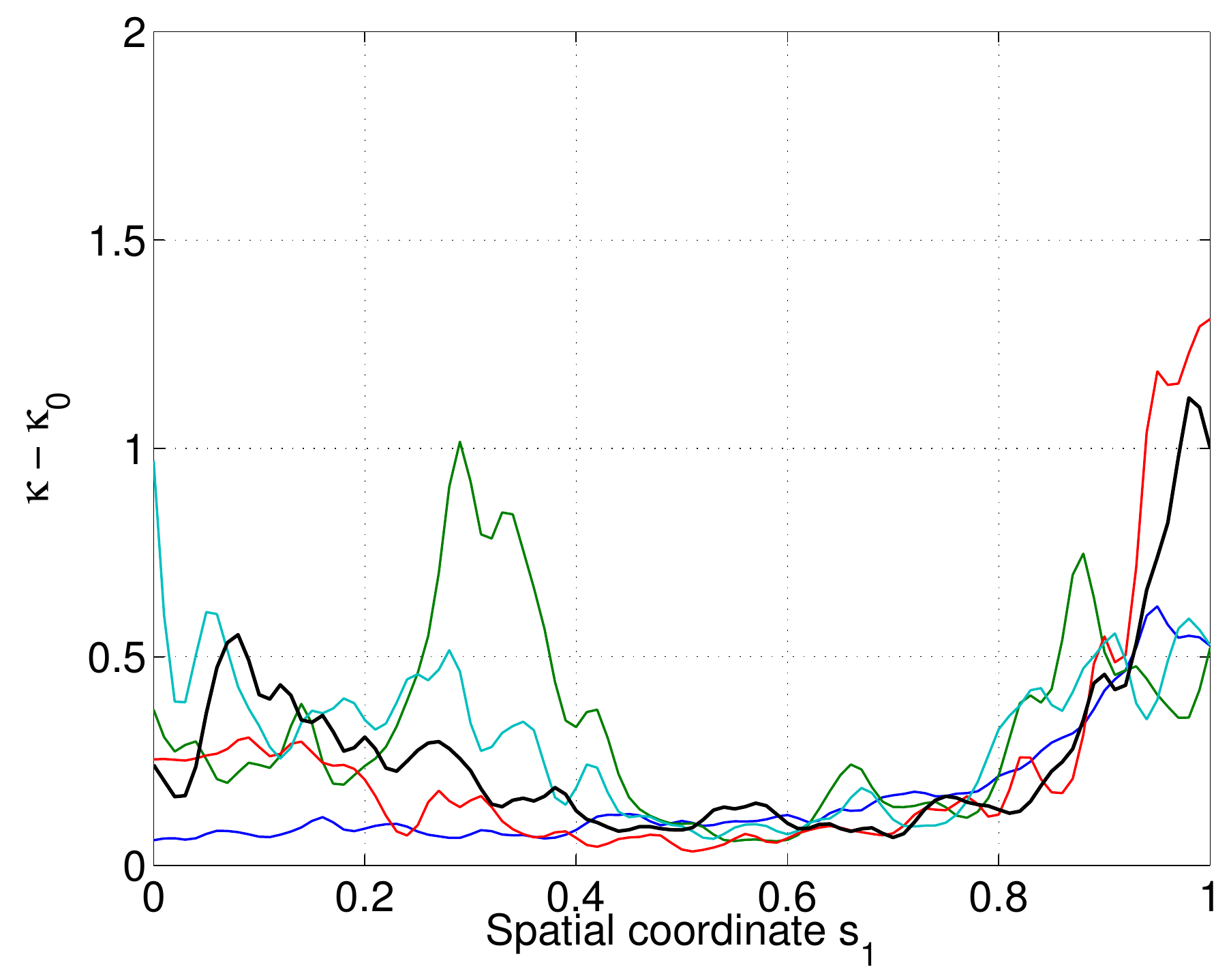}
\label{fig:onedpde_realizations_mcmc1}
}
\subfigure[Case III: Map realizations]
 {
 \includegraphics[width=2.5in]{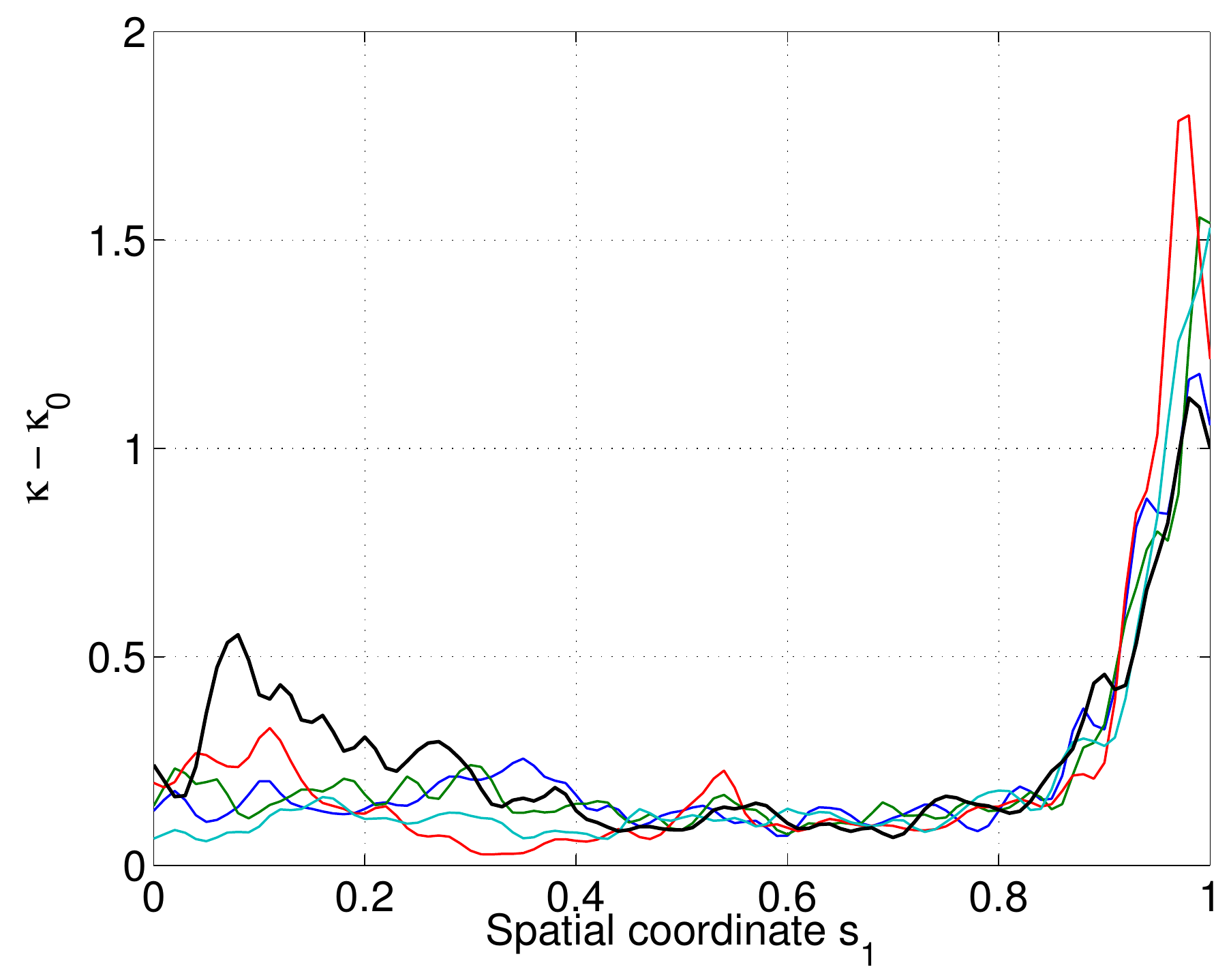}
\label{fig:onedpde_realizations_map3}
}
\subfigure[Case III: MCMC realizations]
 {
 \includegraphics[width=2.5in]{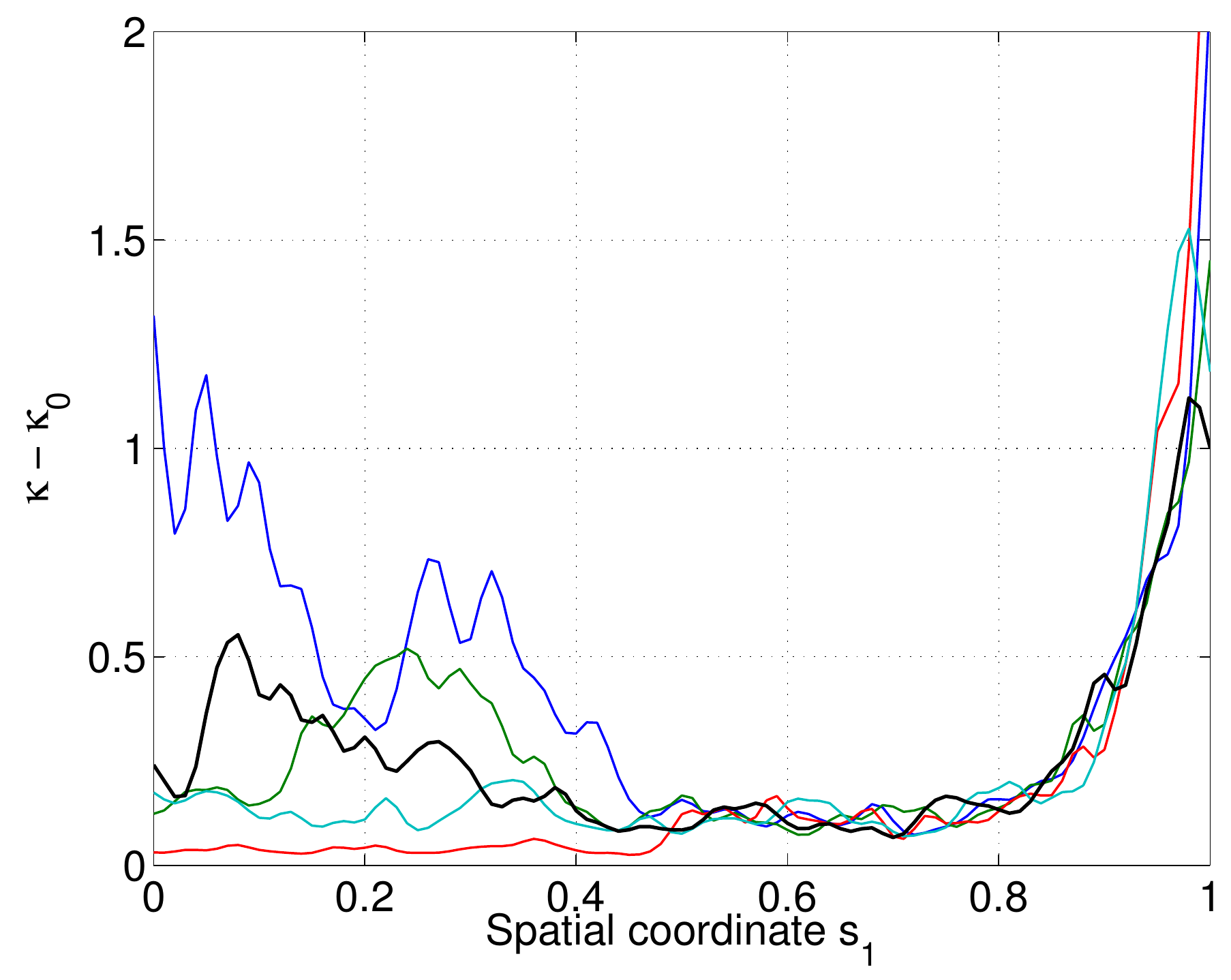}
\label{fig:onedpde_realizations_mcmc3}
}
\caption{One-dimensional elliptic PDE: four posterior realizations
  (colored lines) from Case I and Case III, computed with the map and
  with MCMC. The true permeability field is shown in black on all
  figures.}
\label{fig:onedpde_realizations}
\end{figure}

\begin{figure}[htb]
 \centering
\subfigure[Surface plot of the posterior covariance $C(s_1, s_2)$,
calculated with the map.]
 {
 \includegraphics[width=4.8in]{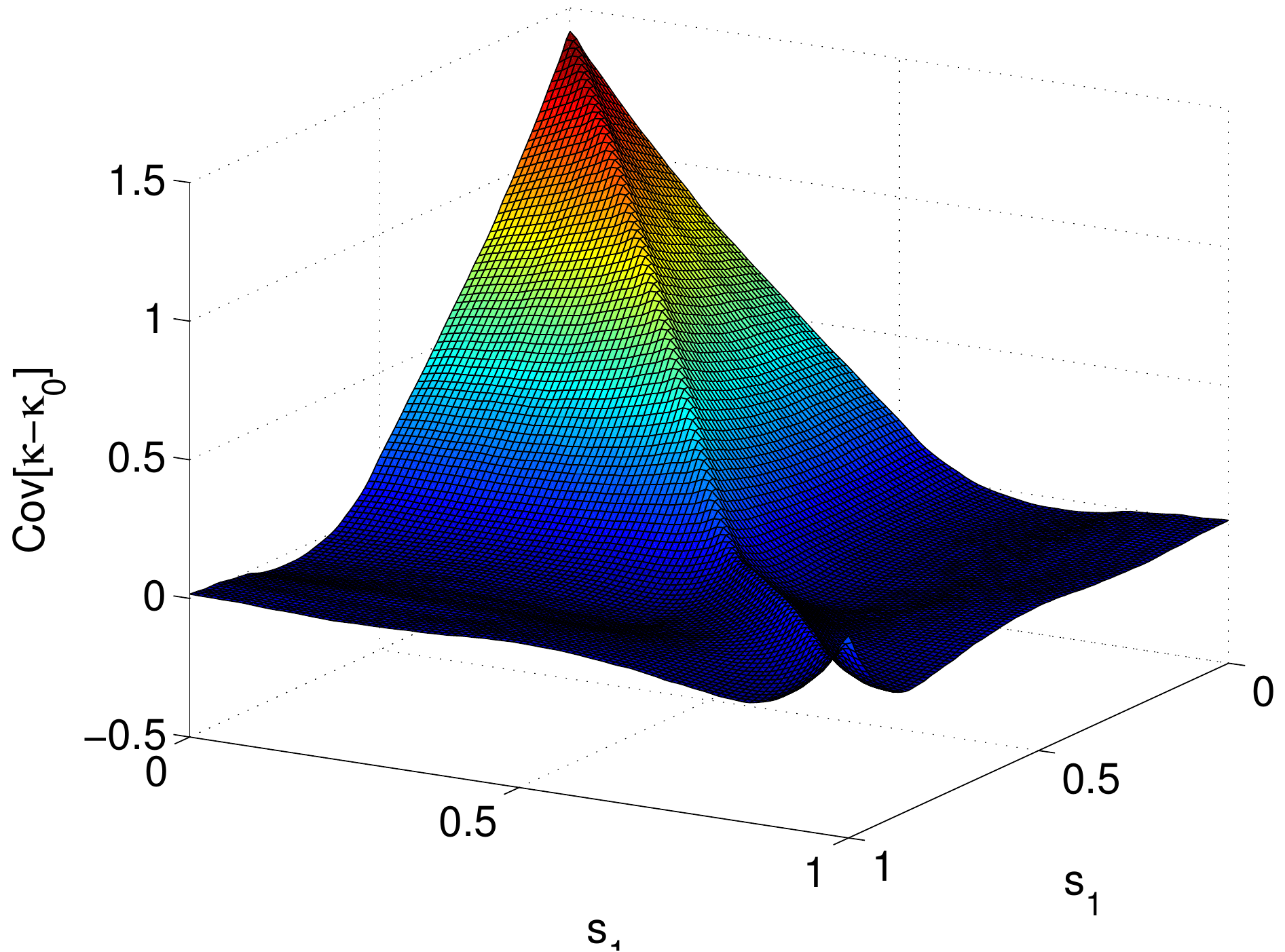}
\label{fig:onedpde_covsurf}
}
\subfigure[Map (thick contours) versus MCMC (thin contours).]
 {
 \includegraphics[width=4.8in]{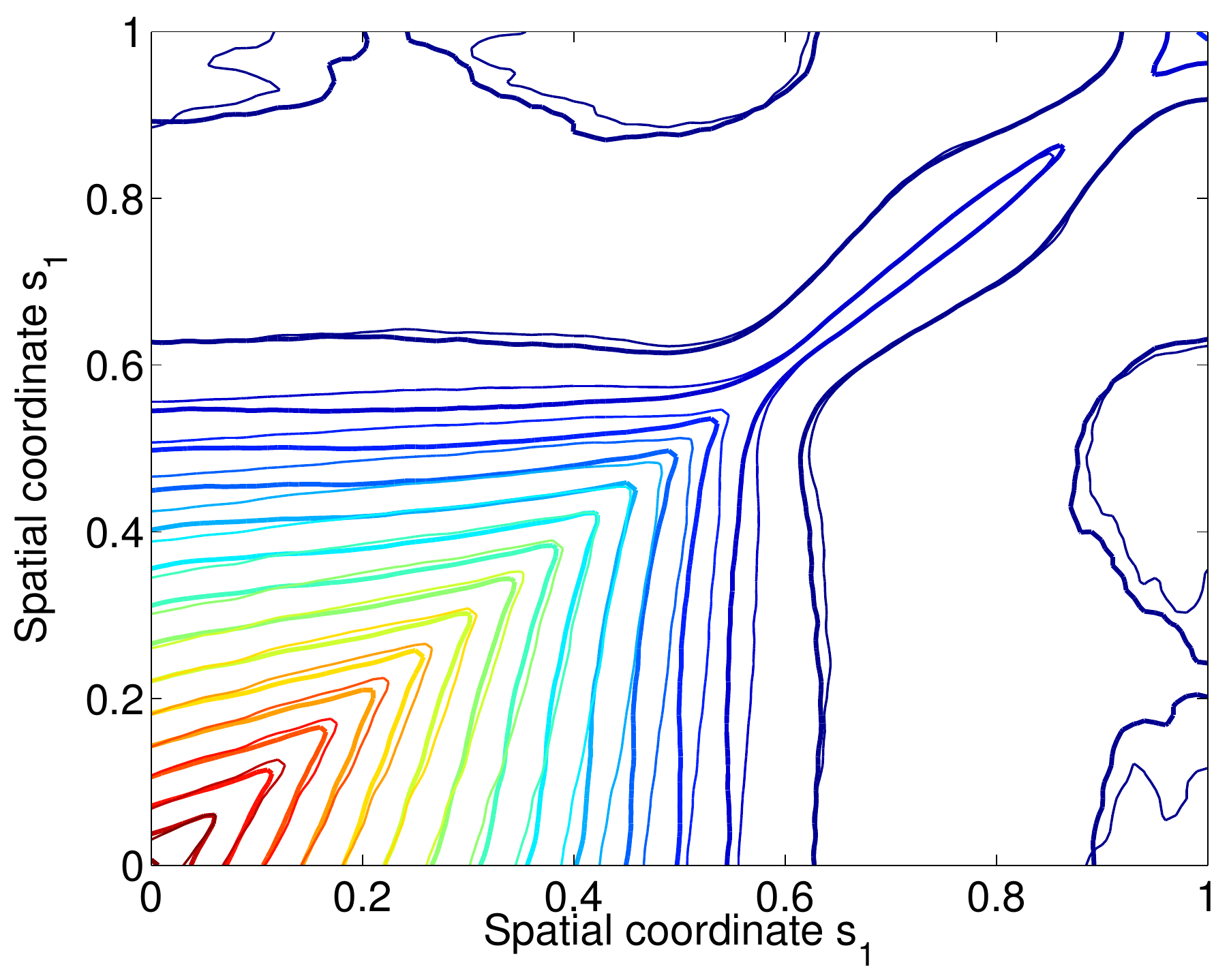}
\label{fig:onedpde_cov}
}
\caption{One-dimensional elliptic PDE: posterior covariance of the
  log-permeability field, Case I. The lower plot compares results obtained
  with the map against results obtained with MCMC, for a fixed set of
  contour levels.}
\label{fig:onedpdecov}
\end{figure}

Results of our algorithm are shown in
Figures~\ref{fig:onedpde_ldsn_logN_meanstd}--\ref{fig:onedpdecov},
where we also report comparisons with MCMC. We consider three cases,
corresponding to different numbers of observations $m$ and different
noise levels $\sigma_n^2$. In Case I, we have $m=31$ observations and
a noise standard deviation of $\sigma_n = 0.05$; in Case II, we
increase the number of data points to $m=101$ and retain $\sigma_n =
0.05$; in Case III we keep $m=101$ observations and reduce the noise
standard deviation to $\sigma_n = 0.01$. In all cases, the maximum
polynomial order of the map is 5 and the optimization routine is
terminated when $\Var[T] < \delta = 0.1$. We use the triangular
formulation and a single map (rather than a composite map) for these
problems.

Figures~\ref{fig:onedpde_sdln_logN_mean}--\ref{fig:onedpde_ldsn_logN_std}
plot the posterior mean and standard deviation of the log-permeability
field as computed with optimal maps and with MCMC. The ``truth''
log-permeability field, used to generate the data, is shown in
black. As expected in this ill-conditioned problem, only the smooth
part of the permeability field can be reconstructed. As the number of
data points increases and the noise level decreases, however, more
features of the permeability field can be recovered. The posterior
covariance is non-stationary, and we note that the posterior variance
is much smaller at the right boundary, where there is a Dirichlet boundary
condition, than at the left boundary, where a zero Neumann condition
was imposed. Overall, posterior uncertainty decreases from Case I to
Case III, reflecting additional information in the data. Good
agreement between the map and MCMC results is observed, though MCMC
yields a slightly larger standard deviation in the left half of the
domain.
The current MCMC results were obtained using $10^6$ samples of DRAM,
with the first $5\times 10^5$ samples discarded as burn-in. Simple
MCMC convergence diagnostics suggest that this is an adequate number
of samples, but it is not entirely clear which set of results is more
authoritative. Indeed, we observed that DRAM failed to converge in
almost half the attempted runs of Case III. (These runs were initiated
from the prior mean; differences among the attempted runs result from
randomness in the proposals.) On the other hand, the optimal map
algorithm reliably converges to the desired tolerance $\delta$.

%

Figure~\ref{fig:onedpde_realizations} shows posterior realizations for
Case I and Case III, as computed with the map and with MCMC. Note that
the realizations are different in character than the posterior mean;
they are significantly rougher, as one would expect given the exponential
covariance kernel in the Gaussian process prior. But the map and MCMC
results yield posterior realizations of similar character.

In Figure~\ref{fig:onedpdecov} we plot the posterior covariance of
$\log{(\kappa-\kappa_0)}$ from Case I. We choose this case because of
the apparent discrepancy in posterior standard deviations shown in
Figure~\ref{fig:onedpde_sdln_logN_std}. Figure~\ref{fig:onedpde_covsurf}
is a surface plot of the posterior covariance as computed using our
map algorithm. The exponential prior covariance has a discontinuity in
its derivative at the diagonal, smoothed slightly due to truncation at
a finite number of Karhunen-Lo\`{e}ve modes. This feature is preserved
in the posterior, reflecting the fact that posterior realizations are
also rough and that data are not informative at the smallest
scales. The overall scale of the covariance is reduced significantly
from prior to posterior, however.  While the prior covariance was
stationary with $\sigma^2=1.5$, the posterior shows smaller variance
throughout the spatial domain.

In Figure~\ref{fig:onedpde_cov} we compare the contours of posterior
covariance obtained with the map to those obtained using MCMC
algorithm. The 16 contour levels range uniformly from $-0.1$ to
1.4. Contours computed using the two methods are remarkably similar.
It should be emphasized that finding the map $f$ enables the posterior
covariance to be computed analytically.

\subsubsection{Two-dimensional domain}
\label{sec:results:subsubsec:PDE2d}

To explore the performance of our algorithm in a more challenging
setting, we solve the same inverse problem as in
Section~\ref{sec:results:subsubsec:PDE1d} but on a two-dimensional
spatial domain $\mathcal{S} = [0,1]^2$. The governing equation is
still (\ref{e:1dpde}), but now $s \equiv (s_1, s_2) \in \mathcal{S}$
and $\nabla \equiv (\partial/\partial s_1, \partial/\partial s_2)$. We
apply deterministic Dirichlet boundary conditions on all four sides of
$\mathcal{S}$, with $u(0,0)=0$, $u(0,1)=1$, $u(1,1)= - 1$, $u(1,0)=2$
and a linear variation on $\partial S$ between these corners. The
source term $f$ is of the form (\ref{e:sourceterm}), centered at $s_m
= (0.5, 0.5)$ and with width $b = \sqrt{10}/40$. The equation is discretized on a $41 \times 41$ grid.

Again we place a log-normal prior on the permeability field
(\ref{e:lnprior}) with $\kappa_0 = 1$, choosing an isotropic
exponential covariance kernel $c(s, s^{\prime})$ with $\sigma=1.25$
and $L_c = 0.5$. To capture 95\% of the spatially integrated variance
of the prior permeability field, the next two cases employ $n=39$
Karhunen-Lo\`{e}ve modes. 

Two different data configurations are
considered. The first (Case A) uses $m=121$ observations randomly
scattered on the spatial domain, with noise standard deviation
$\sigma_n=0.08$; the second (Case B) involves $m=234$ randomly scattered observations and
noise standard deviation $\sigma_n=0.04$. As in the one-dimensional
problem, a polynomial chaos approximation of $u(s, x)$ is used as the
forward model for inversion.

\begin{figure}[htb]
\centering
 \subfigure[Boxplot of Karhunen-Lo\`{e}ve mode weights, obtained with
 the map. Superimposed are posterior means obtained with the map and
 with MCMC, along with truth values of the weights.]
 {
 \includegraphics[width=4.8in]{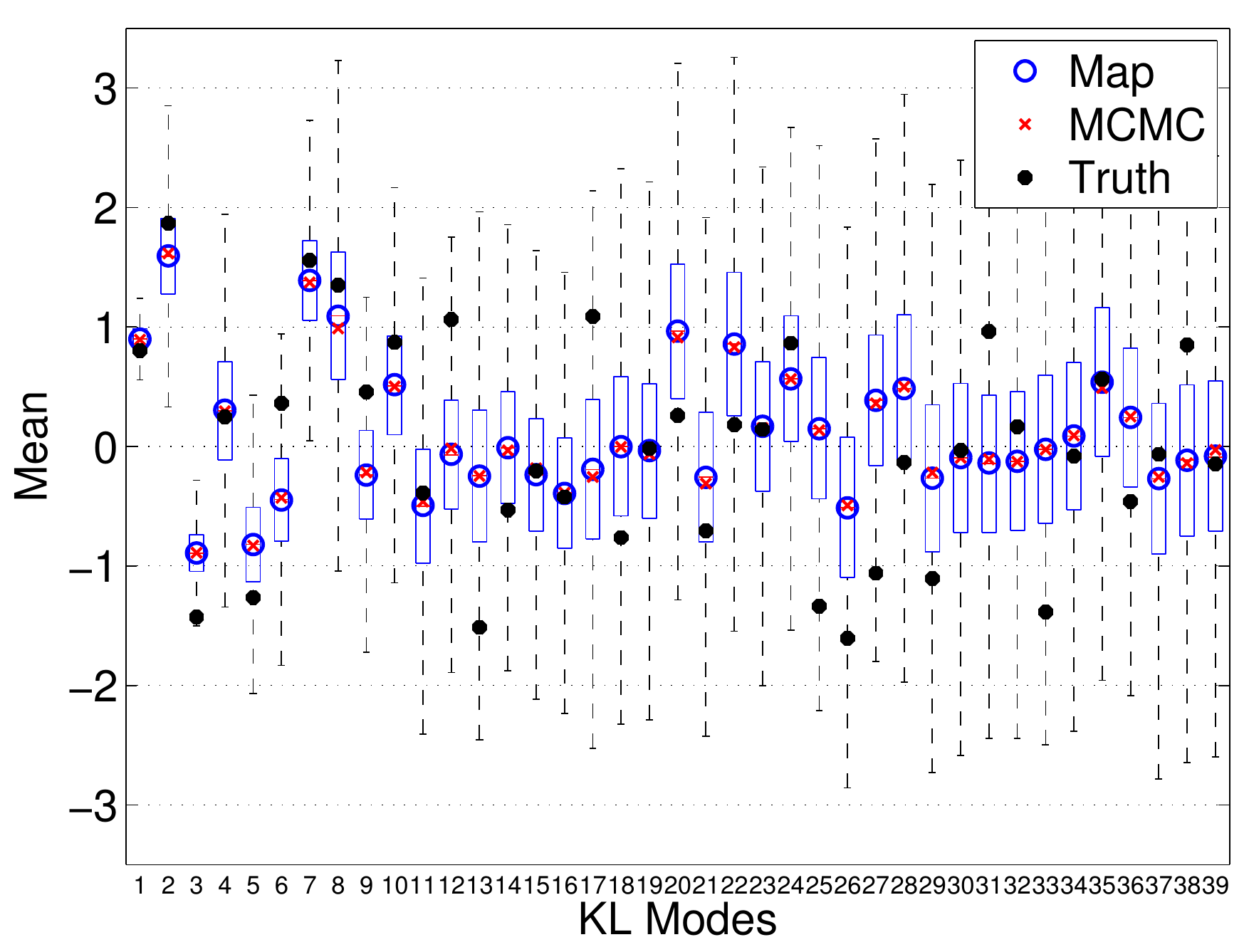}
\label{fig:twodpde_mdmnKLmean}
}
 \subfigure[Posterior standard deviation of the Karhunen-Lo\`{e}ve mode weights, map versus MCMC.]
 {
 \includegraphics[width=4.3in]{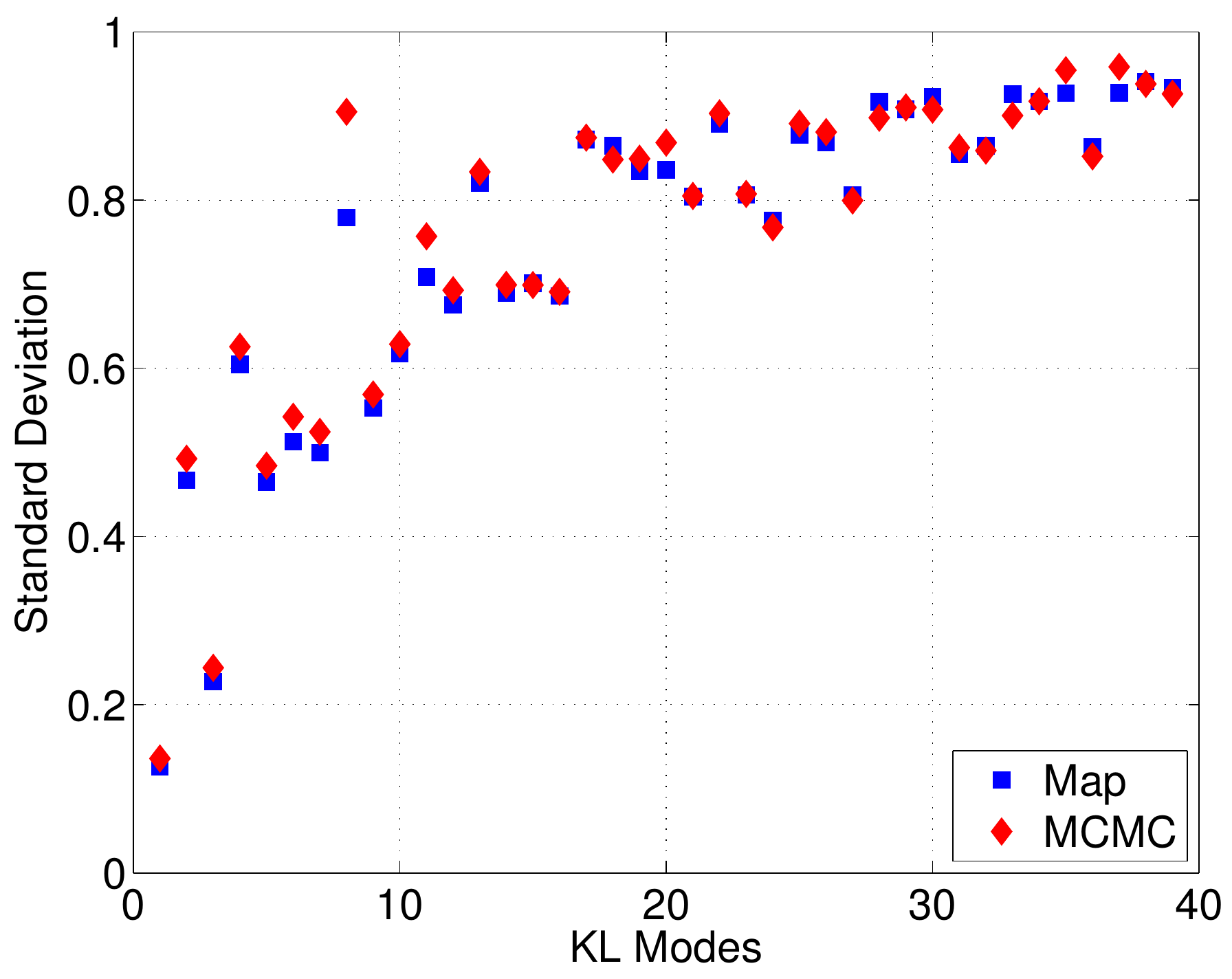}
\label{fig:twodpde_mdmnKLstd}
}
\caption{Two-dimensional elliptic PDE: 121 observations, $\sigma_n =
  0.08$. Posterior distribution of the Karhunen-Lo\`{e}ve modes of
  the log-permeability, as computed with the map and with MCMC.}
\end{figure}

We first focus our analysis on the posterior distribution of the
Karhunen-Lo\`{e}ve mode weights $\{z_i\}$, where $z = f(x)$ and $x_i
\sim N(0,1)$. Figure~\ref{fig:twodpde_mdmnKLmean} shows a boxplot of
the mode weights computed using the map.
The extent of each blue box marks the 25\% and 75\% quantiles of the
posterior marginal of each $z_i$, while the vertical lines or
``whiskers'' span the entire range of the posterior samples drawn via
the map.  We also plot the posterior mean of $z_i$ obtained with the
map (circle) and with MCMC (an $\times$). Finally we show the mode weights
corresponding to the ``true'' permeability field used to generate the
data, before the addition of observational noise. The map and MCMC
means agree reasonably well. Comparing the results of inference with
the true weights, it is observed that the first few modes are
accurately identified, whereas the higher-index modes are not. This is
because the higher-index modes are rougher; they correspond to
higher-frequency components of the permeability field, which are
smoothed by the elliptic operator and therefore difficult to identify
from observations of $u$. This trend is further demonstrated by
Figure~\ref{fig:twodpde_mdmnKLstd}, which shows the posterior standard
deviation of each mode, computed with the map and with MCMC. As the
mode indices increase, their posterior variances approach unity---which
is exactly the prior variance on $x_i$. Thus the data contained little
information to constrain the variability of these modes.

\begin{figure}[htb]
 \centering
 \includegraphics[width=4.8in]{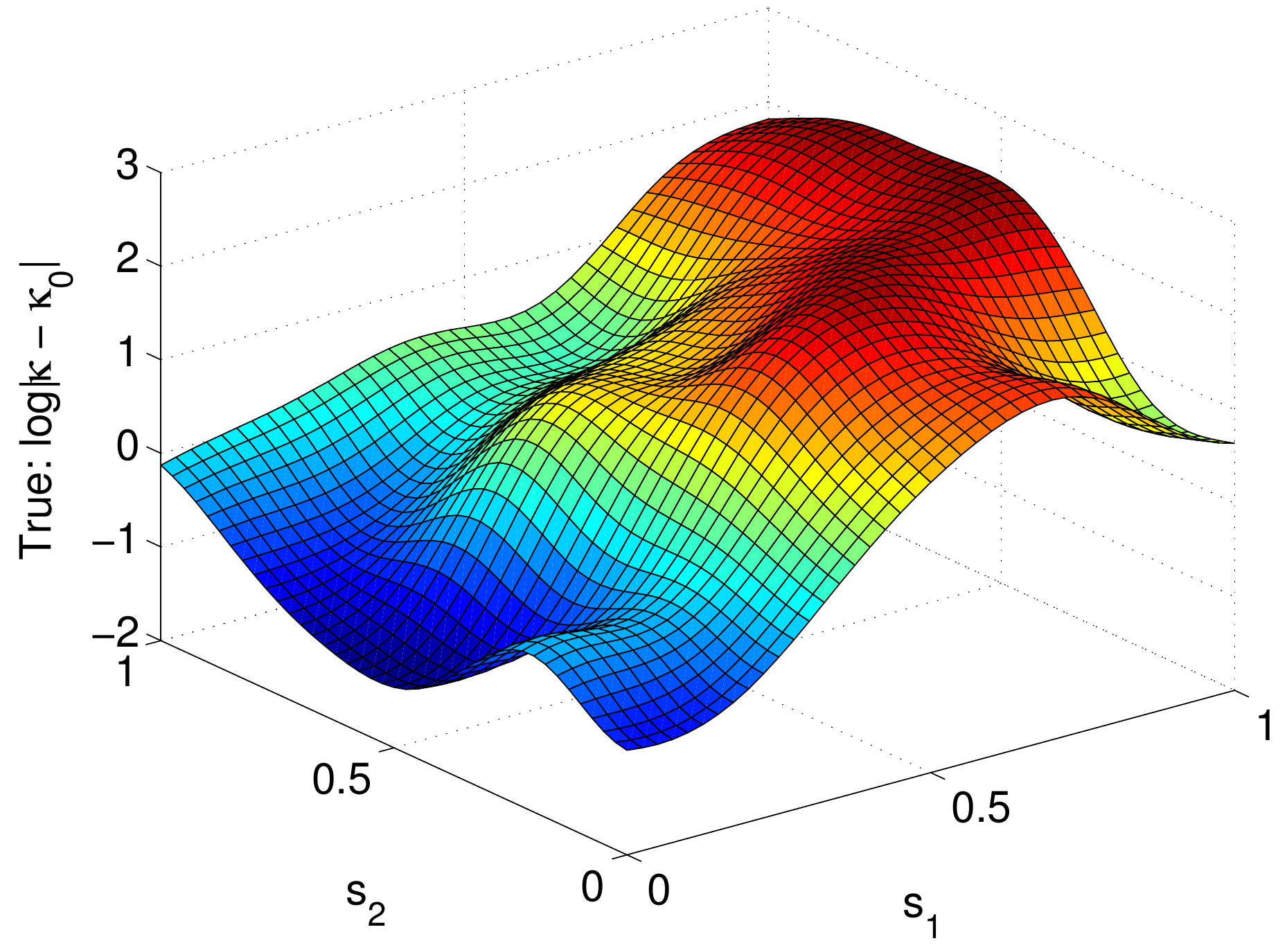}
 \caption{Two-dimensional elliptic PDE: truth $\log{(\kappa - \kappa_0)}$}
\label{fig:twodpde_mdmnTRUTHsurf}
\end{figure}

\begin{figure}[htb]
 \centering
  \subfigure[Posterior mean of $\log{(\kappa - \kappa_0)}$]
 {
 \includegraphics[width=4in]{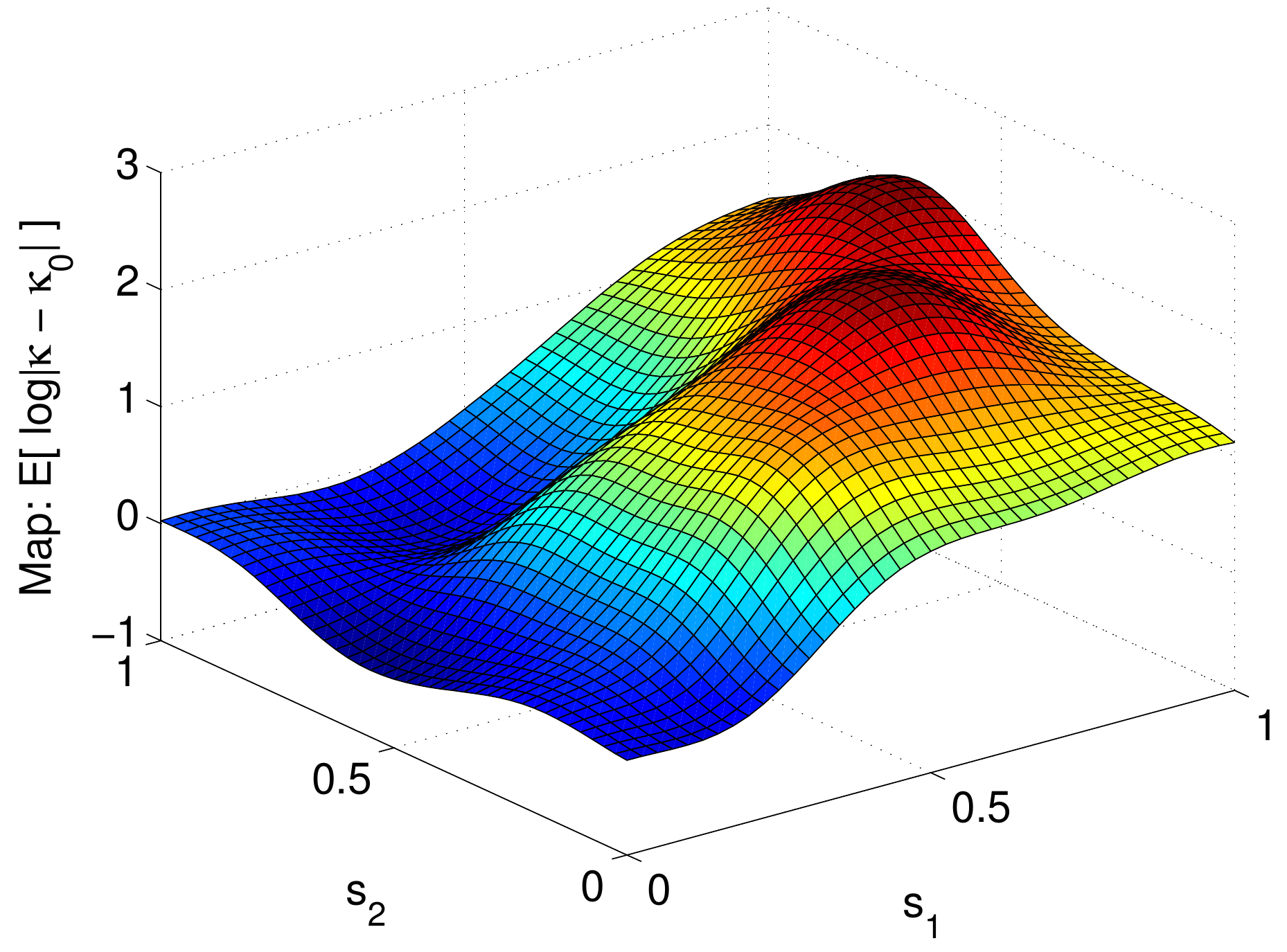}
\label{fig:twodpde_mdmnMAPsurf}
}
 \subfigure[Posterior standard deviation of $\log{(\kappa - \kappa_0)}$]
{
 \includegraphics[width=4in]{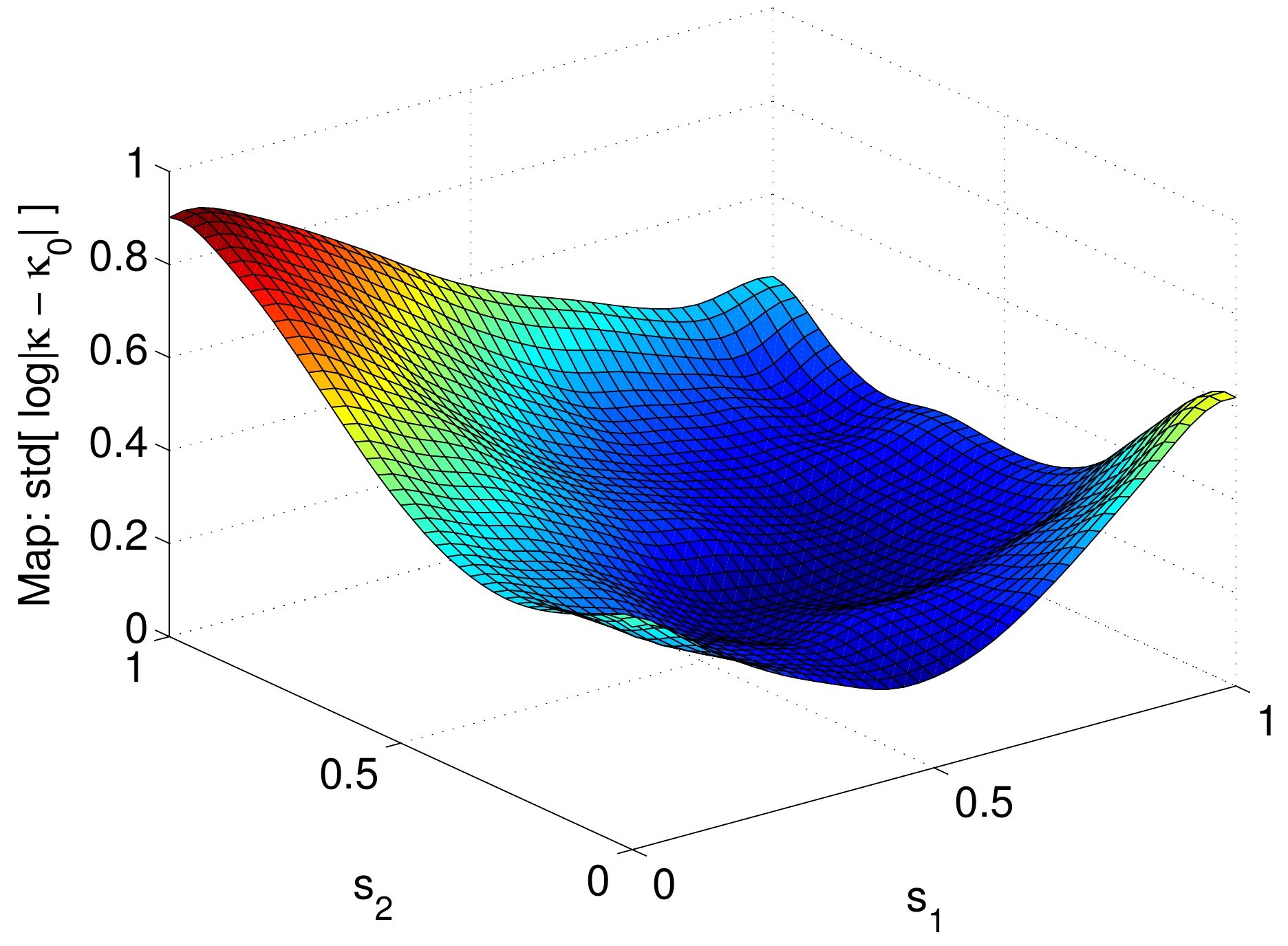}
\label{fig:twodpde_mdmnMAPsurfstd}
}
 \caption{Two-dimensional elliptic PDE: 121 observations, $\sigma_n =
  0.08$. Posterior mean and standard deviation of
   $\log{(\kappa - \kappa_0)}$ computed with the map.}
\end{figure}

\begin{figure}[htb]
 \centering
  \subfigure[Posterior mean of $\log{(\kappa - \kappa_0)}$]
 {
 \includegraphics[width=4in]{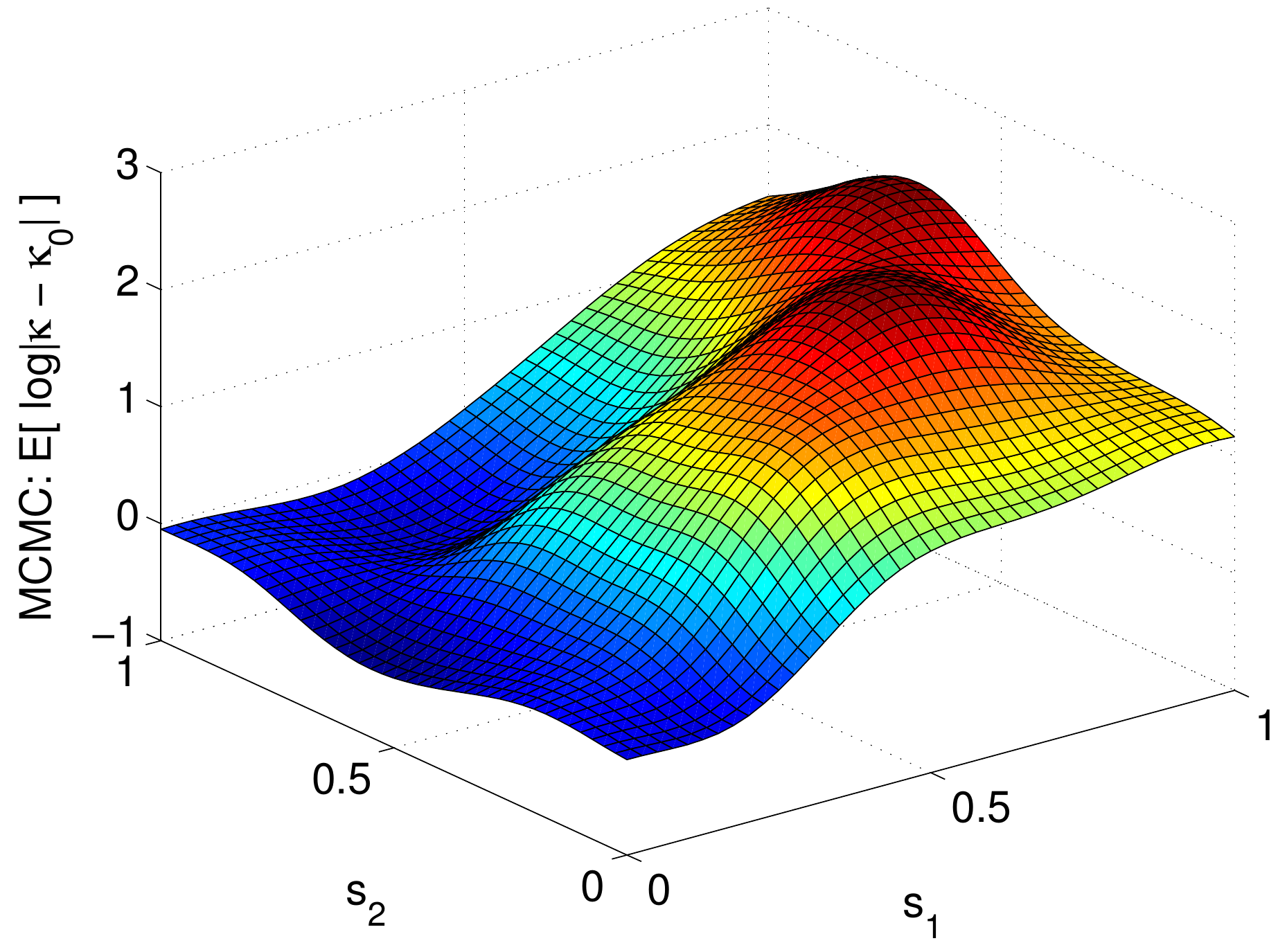}
\label{fig:twodpde_mdmnMCMCsurf}
}
 \subfigure[Posterior standard deviation of $\log{(\kappa - \kappa_0)}$]
{
 \includegraphics[width=4in]{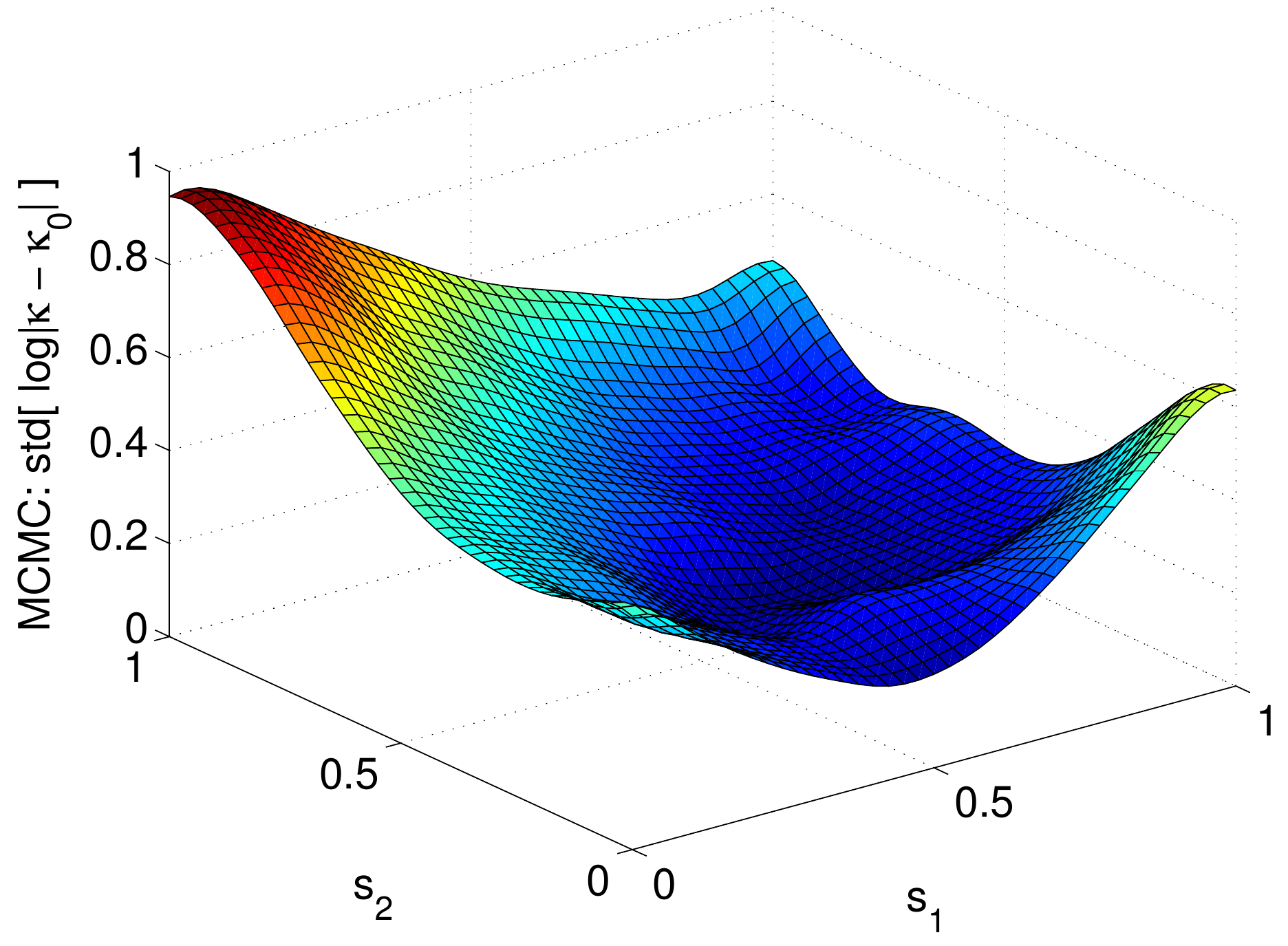}
\label{fig:twodpde_mdmnMCMCsurfstd}
}
\caption{Two-dimensional elliptic PDE: 121 observations, $\sigma_n =
  0.08$. Posterior mean and standard deviation of $\log{(\kappa -
    \kappa_0)}$ computed with MCMC.}
\end{figure}

Turning from analysis of the Karhunen-Lo\`{e}ve modes to analysis of
the permeability field itself, Figure~\ref{fig:twodpde_mdmnTRUTHsurf}
shows the truth log-permeability field used to generate the
data (which here reflects truncation to $n=39$ Karhunen-Lo\`{e}ve
modes). Figure~\ref{fig:twodpde_mdmnMAPsurf} shows the posterior mean
log-permeability obtained with the map in Case A. Given the smoothing character
of the forward model, the presence of noise on the data, and mismatch
between the full PDE model used to generate the noiseless data and the
PC approximation used for inversion, we should not expect the posterior
mean and the true permeability field to match exactly. Indeed, the
posterior mean matches the true field in its large-scale behavior, but
most of the localized or small-scale features are lost; the
corresponding mode weights necessarily revert to their prior mean of
zero. Figure~\ref{fig:twodpde_mdmnMAPsurfstd} shows the posterior
standard deviation of the log-permeability as computed with the map,
while Figures~\ref{fig:twodpde_mdmnMCMCsurf} and
\ref{fig:twodpde_mdmnMCMCsurfstd} show the posterior mean and standard
deviation fields computed with MCMC. Results obtained via the two
algorithms are very similar. The computational time required to solve
the optimization problem for the map, with tolerance $\delta =
0.5$, is equal to the time required for $5\times 10^5$ steps of MCMC.

\begin{figure}[htb]
\centering
 \subfigure[Boxplot of Karhunen-Lo\`{e}ve mode weights, obtained with
 the map. Superimposed are posterior means obtained with the map and
 with MCMC, along with truth values of the weights.]
 {
 \includegraphics[width=4.8in]{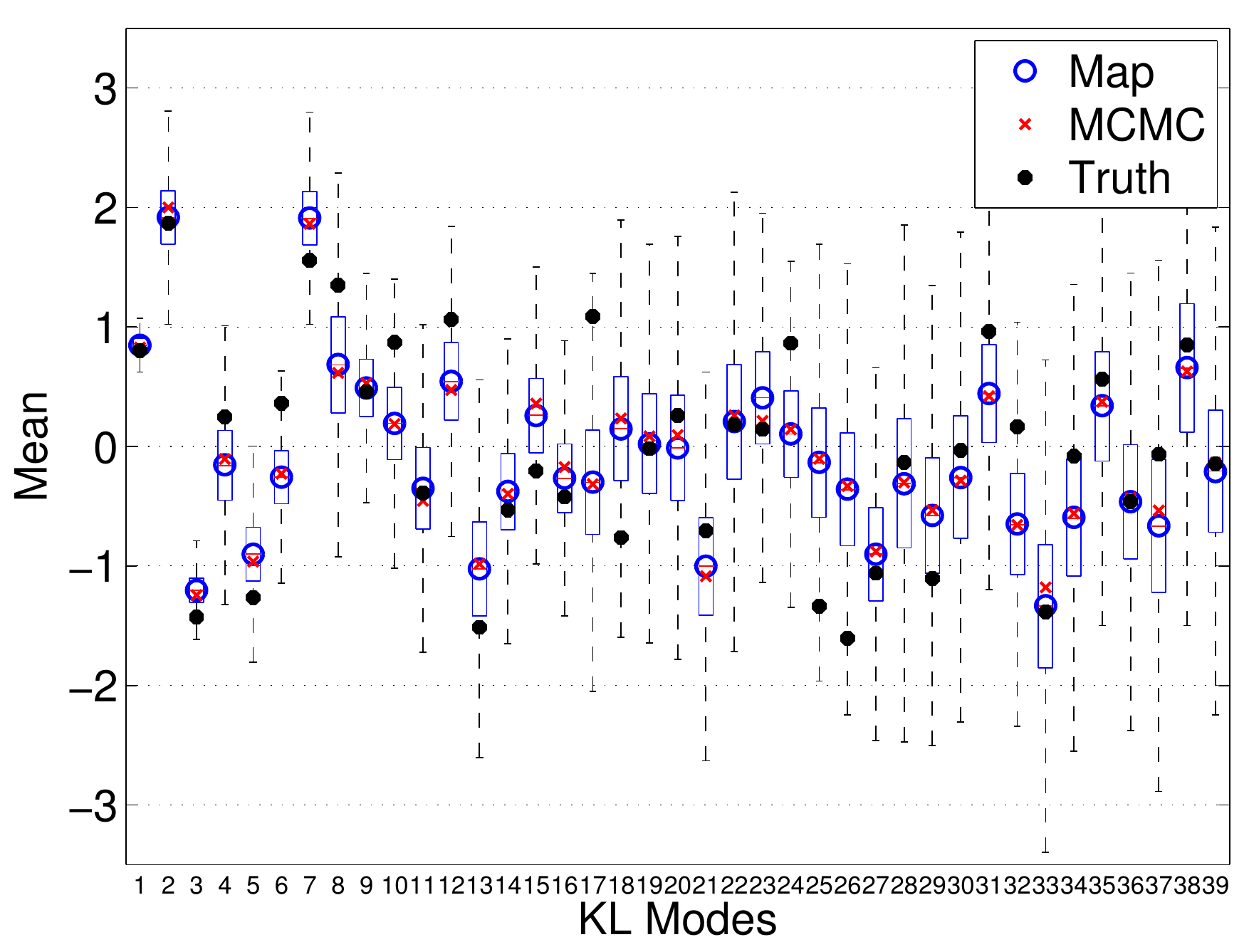}
\label{fig:twodpde_ldsnKLmean}
}
 \subfigure[Posterior standard deviation of the Karhunen-Lo\`{e}ve mode weights, map versus MCMC.]
 {
 \includegraphics[width=4.3in]{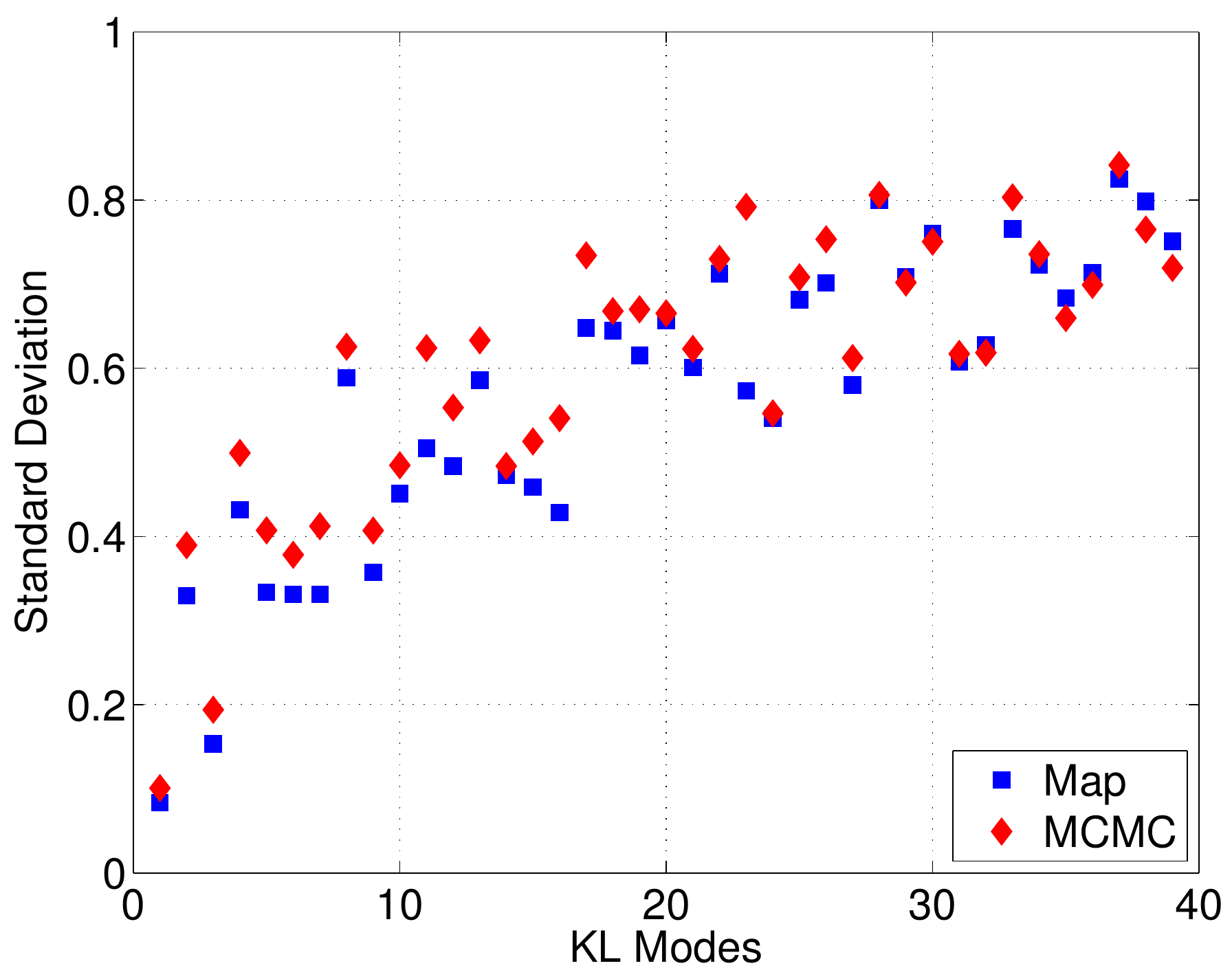}
\label{fig:twodpde_ldsnKLstd}
}
\caption{Two-dimensional elliptic PDE: 234 observations, $\sigma_n =
  0.04$. Posterior distribution of the Karhunen-Lo\`{e}ve modes of
  the log-permeability, as computed with the map and with MCMC.}
\label{fig:twodpde_ldsnKL}
\end{figure}
\begin{figure}[htb]
 \centering
  \subfigure[Posterior mean of $\log{(\kappa - \kappa_0)}$]
 {
 \includegraphics[width=4in]{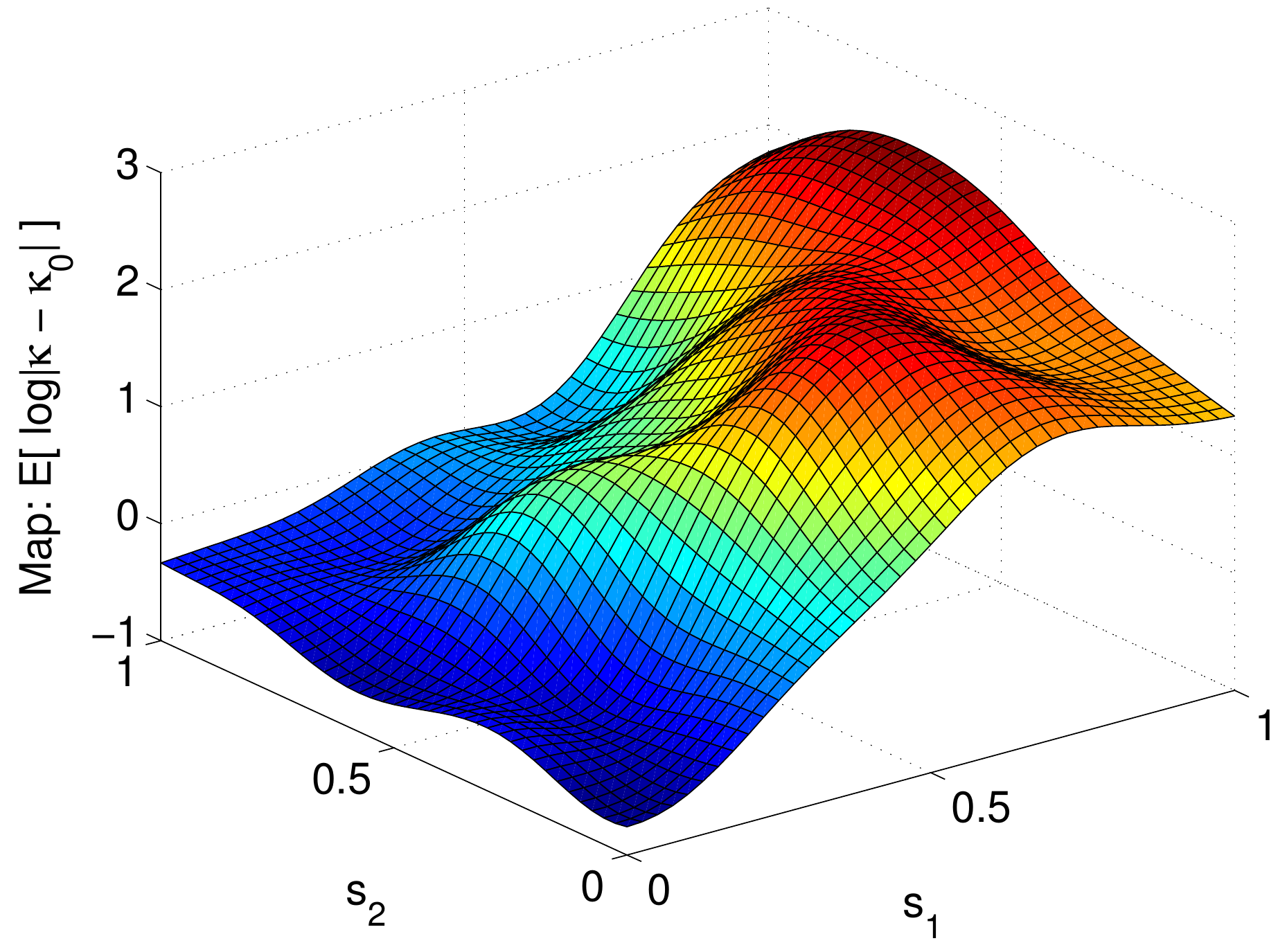}
\label{fig:twodpde_ldsnMAPsurf}
}
 \subfigure[Posterior standard deviation of $\log{(\kappa - \kappa_0)}$]
{
 \includegraphics[width=4in]{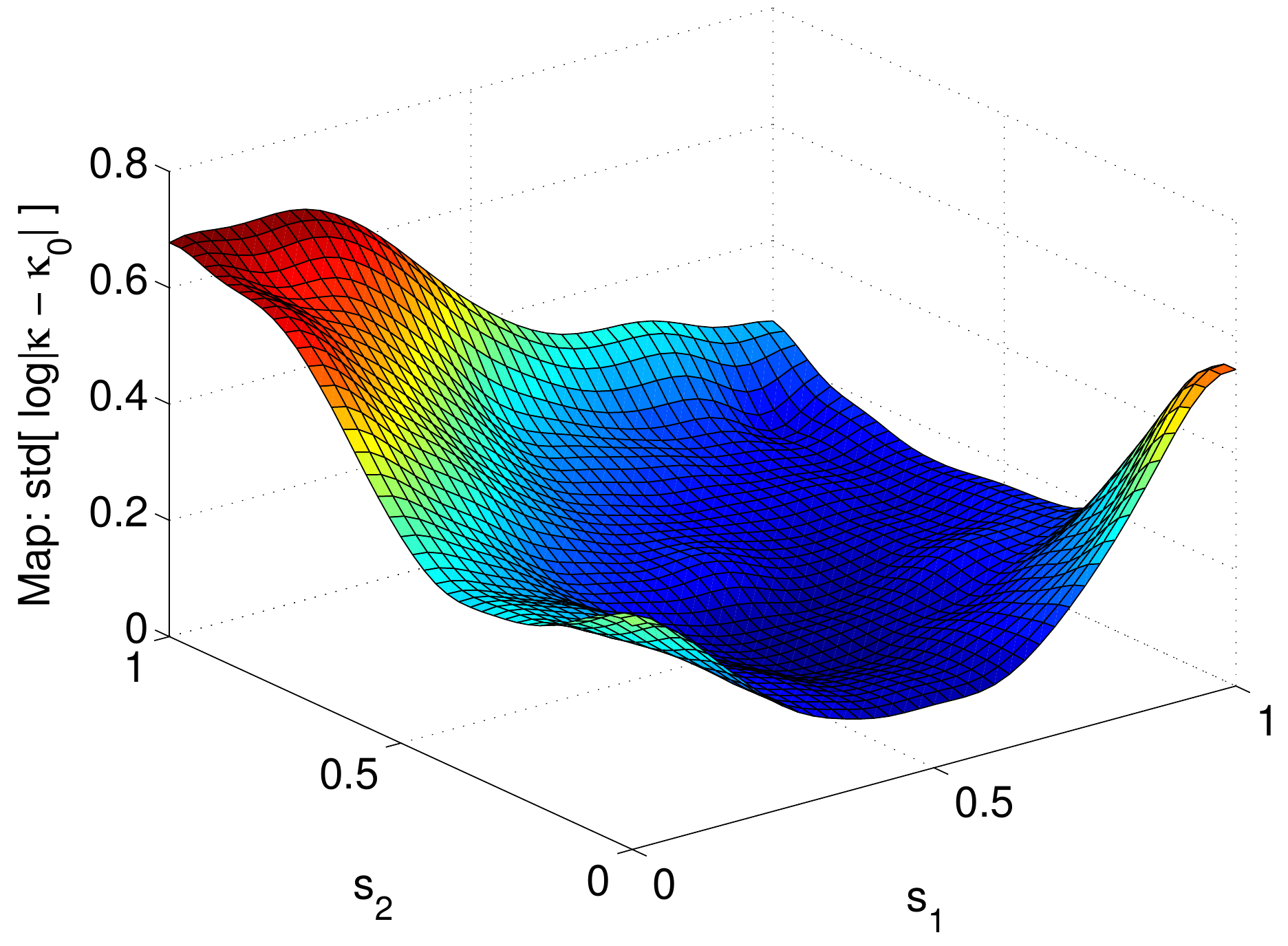}
\label{fig:twodpde_ldsnMAPsurfstd}
}
\caption{Two-dimensional elliptic PDE: 234 observations, $\sigma_n =
  0.04$. Posterior mean and standard deviation of $\log{(\kappa -
    \kappa_0)}$ computed with the map.}
\end{figure}
\begin{figure}[htb]
 \centering
  \subfigure[Posterior mean of $\log{(\kappa - \kappa_0)}$]
 {
 \includegraphics[width=4in]{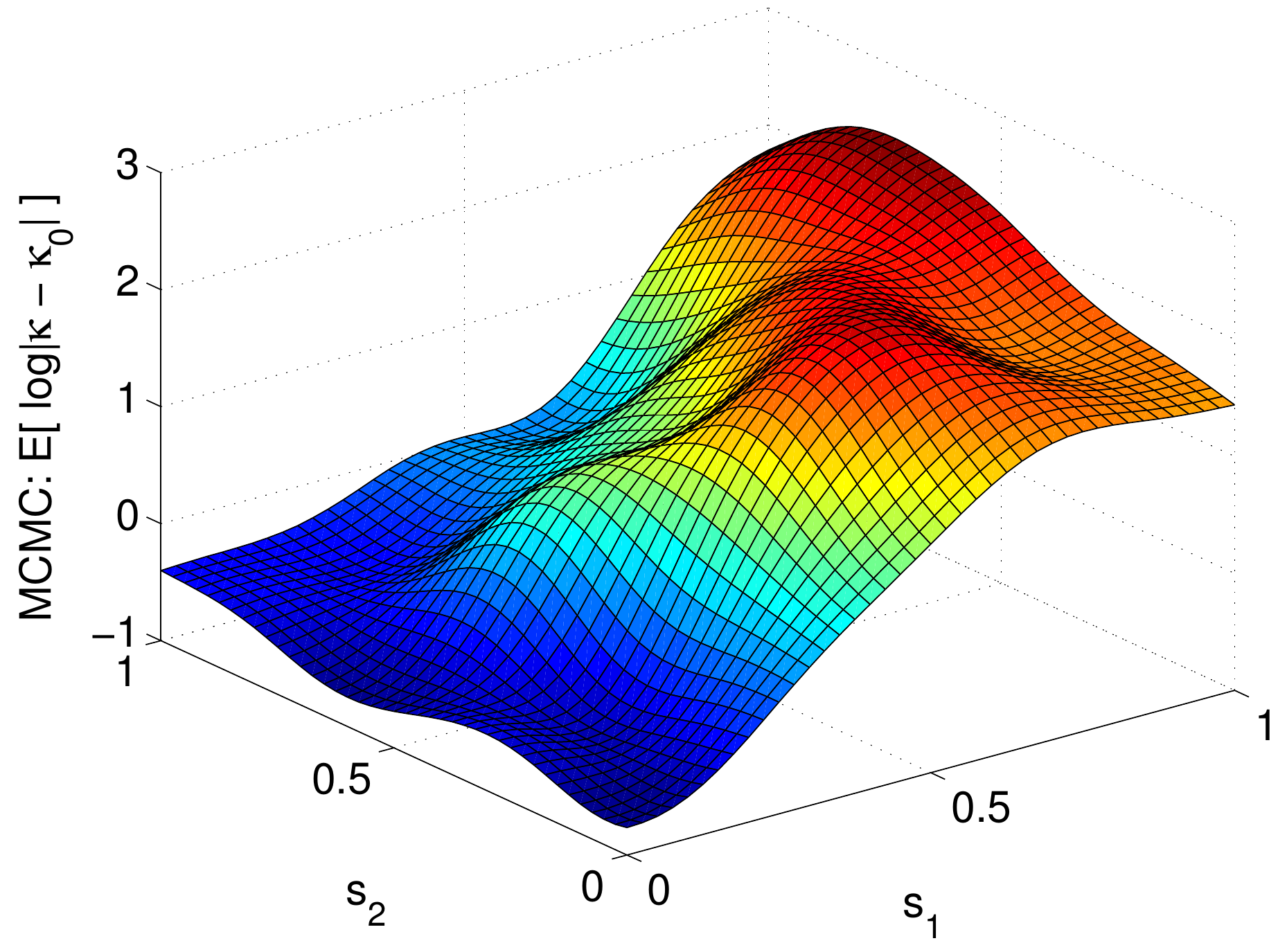}
\label{fig:twodpde_ldsnMCMCsurf}
}
 \subfigure[Posterior standard deviation of $\log{(\kappa - \kappa_0)}$]
{
 \includegraphics[width=4in]{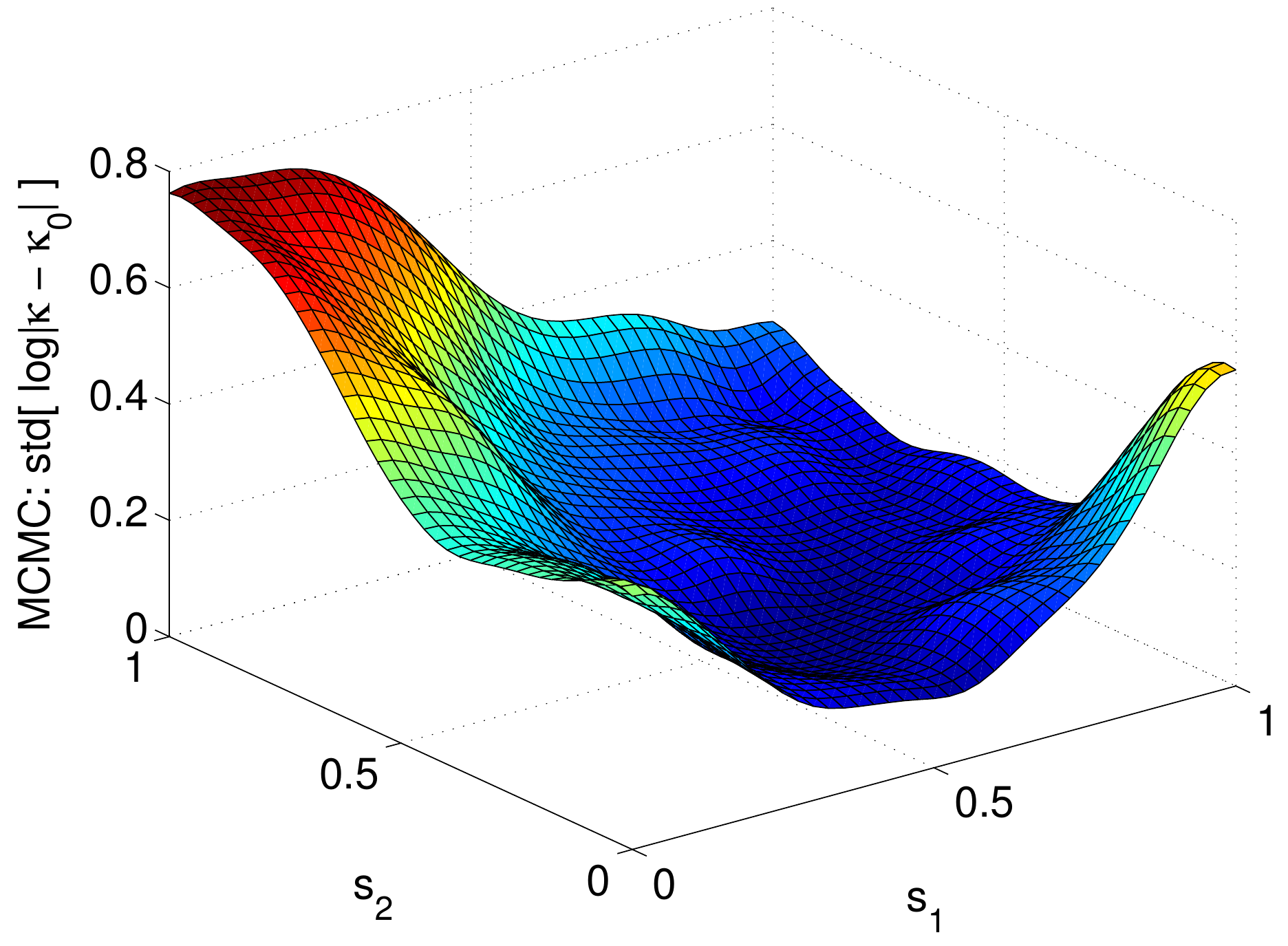}
\label{fig:twodpde_ldsnMCMCsurfstd}
}
 \caption{Two-dimensional elliptic PDE: 234 observations, $\sigma_n =
  0.04$. Posterior mean and standard deviation of $\log{(\kappa -
    \kappa_0)}$ computed with MCMC.}
\label{fig:twopde_ldsnMCMC}
\end{figure}

Figures~\ref{fig:twodpde_ldsnKL}--\ref{fig:twopde_ldsnMCMC} show
analogous results for Case B, with roughly twice as many observations
and half the noise standard deviation of Case A. The boxplot of
Karhunen-Lo\`{e}ve (KL) mode weights in
Figure~\ref{fig:twodpde_ldsnKLmean} shows that a larger number of
modes (at higher index) are accurately identified, compared to Case
A. In Figure~\ref{fig:twodpde_ldsnKLstd}, the posterior standard
deviation of the modes is less than in Case A, reflecting the
additional information carried by the
data. Figure~\ref{fig:twodpde_ldsnMAPsurf} shows the posterior mean
log-permeability obtained with the map for Case B; though this field
is still smooth, more of the small-scale features of the true
permeability field are captured. MCMC results shown
in Figure~\ref{fig:twopde_ldsnMCMC} are quite similar to those obtained with
the map.

\begin{figure}[htb]
 \centering
 \includegraphics[width=4.8in]{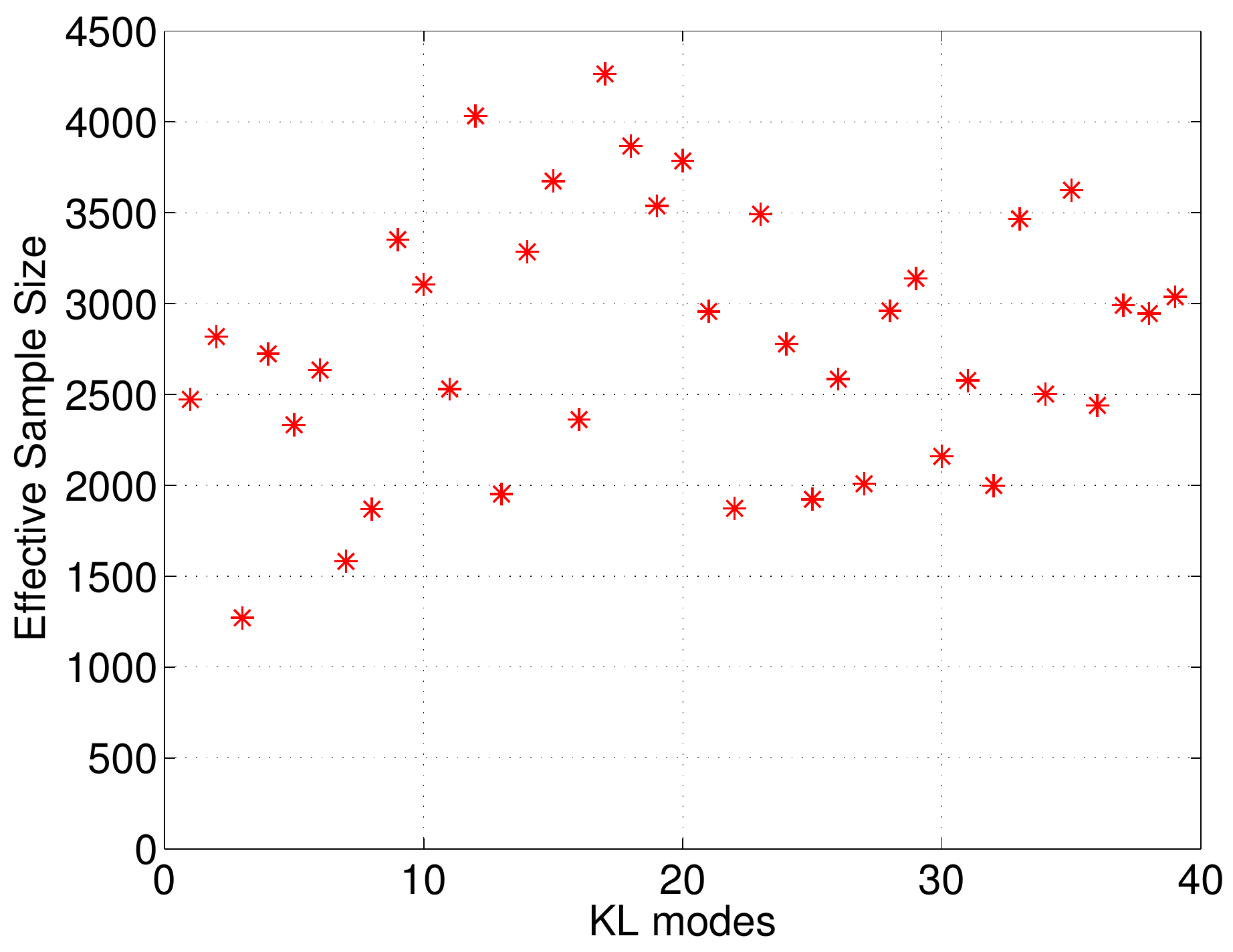}
 \caption{Two-dimensional elliptic PDE: 234 observations, $\sigma_n =
   0.04$. Effective sample size after $5 \times 10^5$ MCMC iterations.}
\label{fig:twodpde_ldsnESS}
\end{figure}

The MCMC results reported here were obtained using a DRAM chain of
$10^6$ samples, half of which were discarded as burn-in. We make no
claim that this is most efficient MCMC approach to this problem;
certainly a carefully hand-tuned algorithm could yield better
mixing. However, we do claim that it represents \textit{good} MCMC
performance. Because the inference problem has been transformed to the
Karhunen-Lo\`{e}ve modes, which have unit variance and are
uncorrelated in the prior, the adaptive Metropolis algorithm starts
from a favorable parameterization with known scaling. Even with a good
parameterization and an adaptive algorithm, however, MCMC mixing
remains a challenge in these ill-posed inverse
problems. Figure~\ref{fig:twodpde_ldsnESS} shows the effective sample
size (ESS) of each component of the chain corresponding to $N=5\times
10^5$ post burn-in samples. The effective sample size is computed by
integrating the chain autocorrelation $\rho_i$ at lag $i$:
\begin{equation}
\mathrm{ESS} = \frac{N}{ 1 + 2 \sum_{i=1}^\infty \rho_i} .
\end{equation}
The ESS for most of the chain components lies between 1500 and
4000. In order to obtain a reasonable number of posterior
samples---e.g., for an accurate estimate of a posterior moment, or to
propagate posterior uncertainty through a subsequent simulation---an
unreasonable number of MCMC steps is thus required. For instance, if
one desires $10^5$ effectively independent posterior samples, then at
least 30 million MCMC steps are needed (here, corresponding to about 2
days of simulation time). On the same computational platform and for
the same problem, using the map algorithm to generate $10^5$
independent samples requires about 45 minutes of wallclock time to
solve the optimization problem and construct $f$, followed by 5
minutes to pass $10^5$ prior samples through the map and generate the
desired posterior samples. This corresponds to a factor of 50
reduction in computational time.

We also note that when the noise standard deviation is reduced to
$\sigma_n = 0.01$, then the adaptive MCMC algorithm fails to converge,
producing a chain with near-zero acceptance rate regardless of how
many times we rerun it (starting from the prior mean). On the other
hand, the map algorithm has no trouble converging to the desired
accuracy $\delta$ in roughly 55 minutes of wallclock time (only 10 minutes
more than that required for Case B).


\vspace{1em}
 
As a final example we consider a more refined stochastic
discretization of the prior log-normal process. We now retain $n=139$
Karhunen-Lo\`{e}ve modes, such that 97.5\% of the input permeability
field's integrated variance is preserved. We make $m=227$ noisy
observations, with a noise standard deviation of $\sigma_n=0.04$. In
this example we focus on comparing the computational performance of
MCMC and map-based inference.

Two different MCMC chains are run, each of length $10^6$
samples. Both chains start from the prior mean and proceed with adaptive random-walk proposals.
It is observed that the burn-in period is
relatively long; we thus remove half the chain and use the remaining
$5 \times 10^5$ samples to compute all quantities below. Each chain
takes approximately 5 hours to produce. The map algorithm is run until
the variance of $T(X)$ is less than 1\% of the mean of $T(X)$; this
corresponds to a KL divergence $D_{\mathrm{KL}} \left ( p || \pt
\right )$ of 1 nat. The computational time required to solve the
optimization problem to this threshold is less than 3 hours.

\begin{figure}[htb]
 \centering
  \subfigure[Two different MCMC chains, each of length $10^6$.]
 {
 \includegraphics[width=4in]{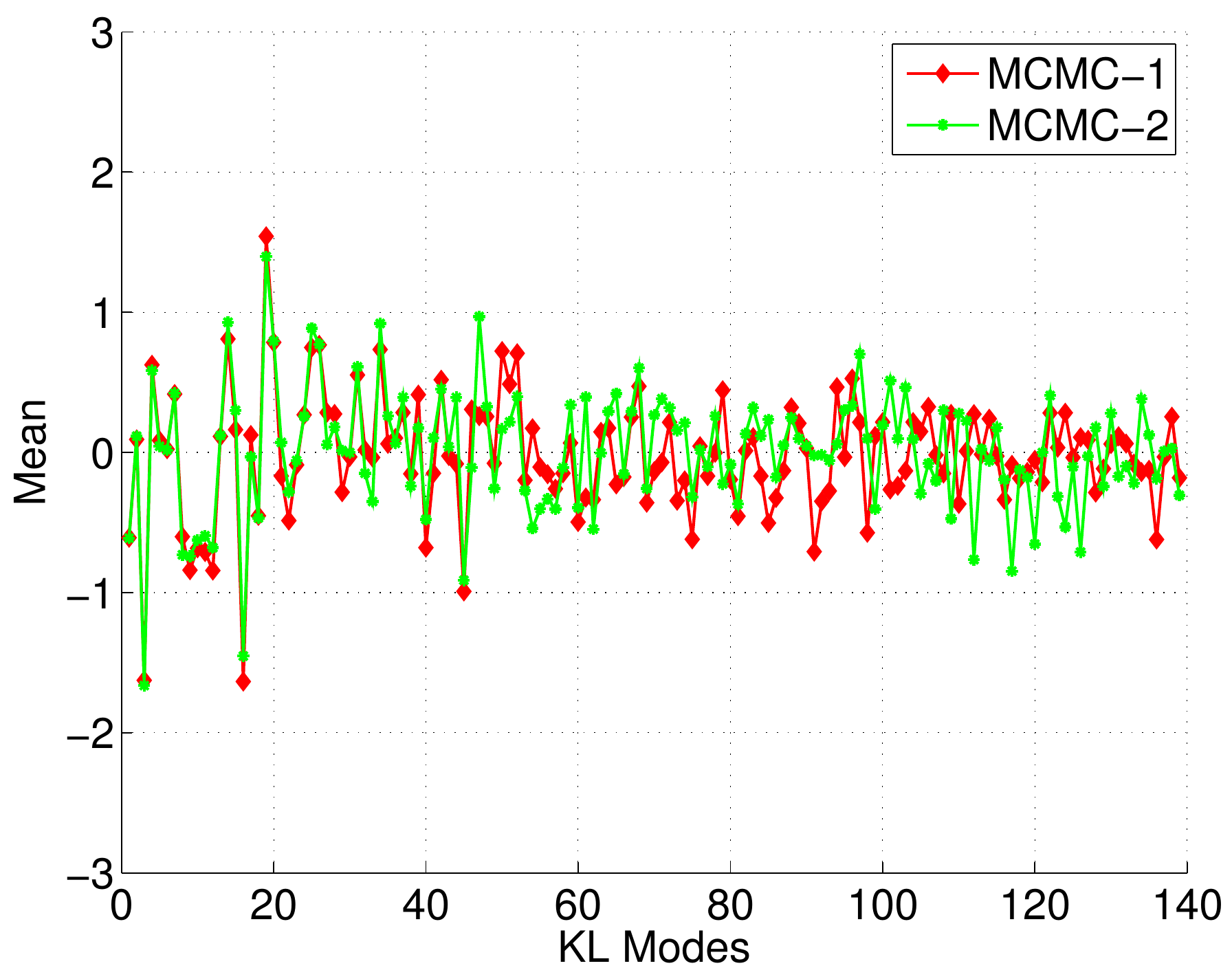}
\label{fig:twodpdelarge_KLMCMCmean}
}
 \subfigure[Map versus truth.]
{
 \includegraphics[width=4in]{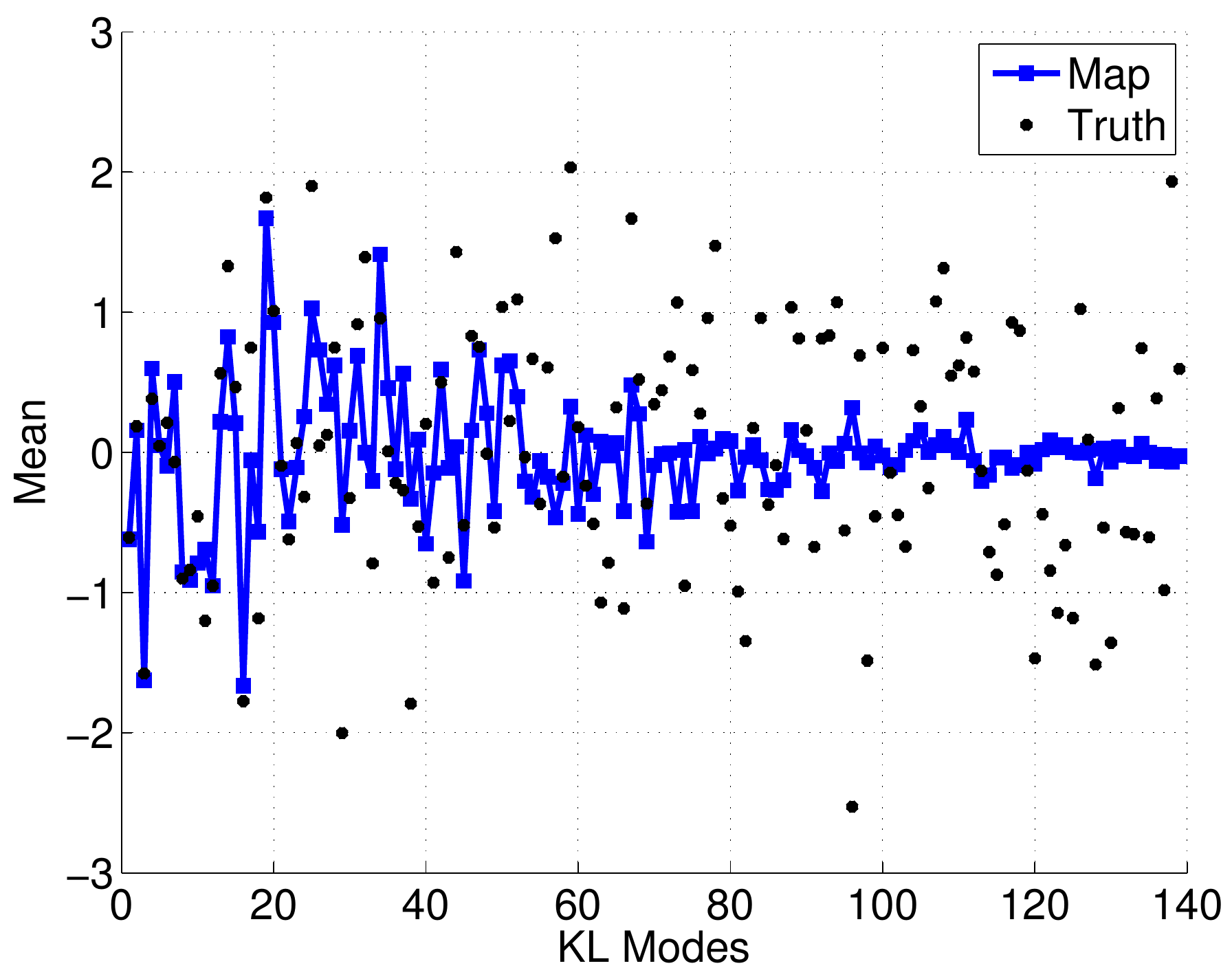}
\label{fig:twodpdelarge_KLMAPmean}
}
\caption{Two-dimensional elliptic PDE: 227 observations, $\sigma_n =
  0.04$, 139-dimensional posterior. Posterior mean of the
  Karhunen-Lo\`{e}ve modes as computed with MCMC and with the map.}
\label{fig:twodpdelarge_KLmean}
\end{figure}

\begin{figure}[htb]
 \centering
 \includegraphics[width=4.8in]{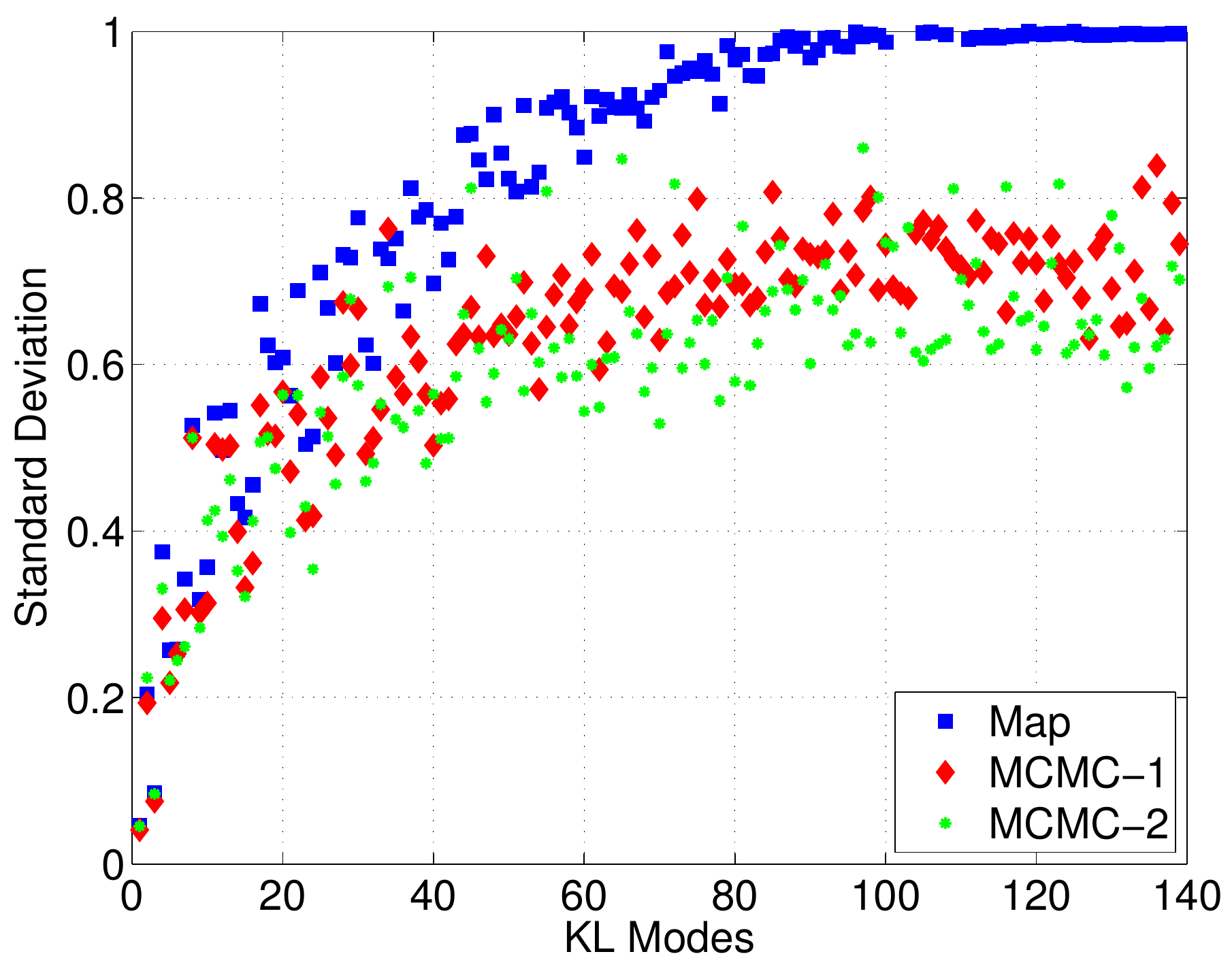}
 \caption{Two-dimensional elliptic PDE: 227 observations, $\sigma_n =
  0.04$, 139-dimensional posterior. Posterior standard deviation of
  the Karhunen-Lo\`{e}ve modes as computed with MCMC and with the map.}
\label{fig:twodpdelarge_KLstd}
\end{figure}

Figure~\ref{fig:twodpdelarge_KLmean} shows the posterior mean values
of the Karhunen-Lo\`{e}ve modes computed using MCMC and the map. The
two MCMC runs shown in Figure \ref{fig:twodpdelarge_KLMCMCmean} differ
significantly in their higher-index modes, indicating that these
components of the chain mix rather poorly. Comparing the map-based
posterior means with the ``truth'' shows, as usual, that the smoother
modes are well identified in the posterior while the rougher modes are
not, reverting instead to the prior. Poor mixing of the MCMC algorithm
is also evident in Figure~\ref{fig:twodpdelarge_KLstd}, which shows
the posterior standard deviation of each mode weight. For mode indices
larger than 10, the two MCMC runs yield very different standard
deviations. And the standard deviations of the higher-index modes
plateau below 0.8. The discrepancy between the chains suggests that
this value is not credible, and that the chains are in fact
\textit{not} exploring the full parameter space (despite using $10^6$
samples). One would instead expect the rough highest-index modes to
revert to the prior standard deviation of 1.0, exactly as observed in
the map-based results. Indeed, agreement of the prior and posterior distributions on the high-index KL modes is a consequence of absolute continuity of the posterior with respect to the prior~\cite{dashti:2012:uqa}. Limitations of standard MCMC algorithms in this context have been discussed in~\cite{cotter:2012:mmf}.

\begin{figure}[htb]
 \centering
 \includegraphics[width=4.8in]{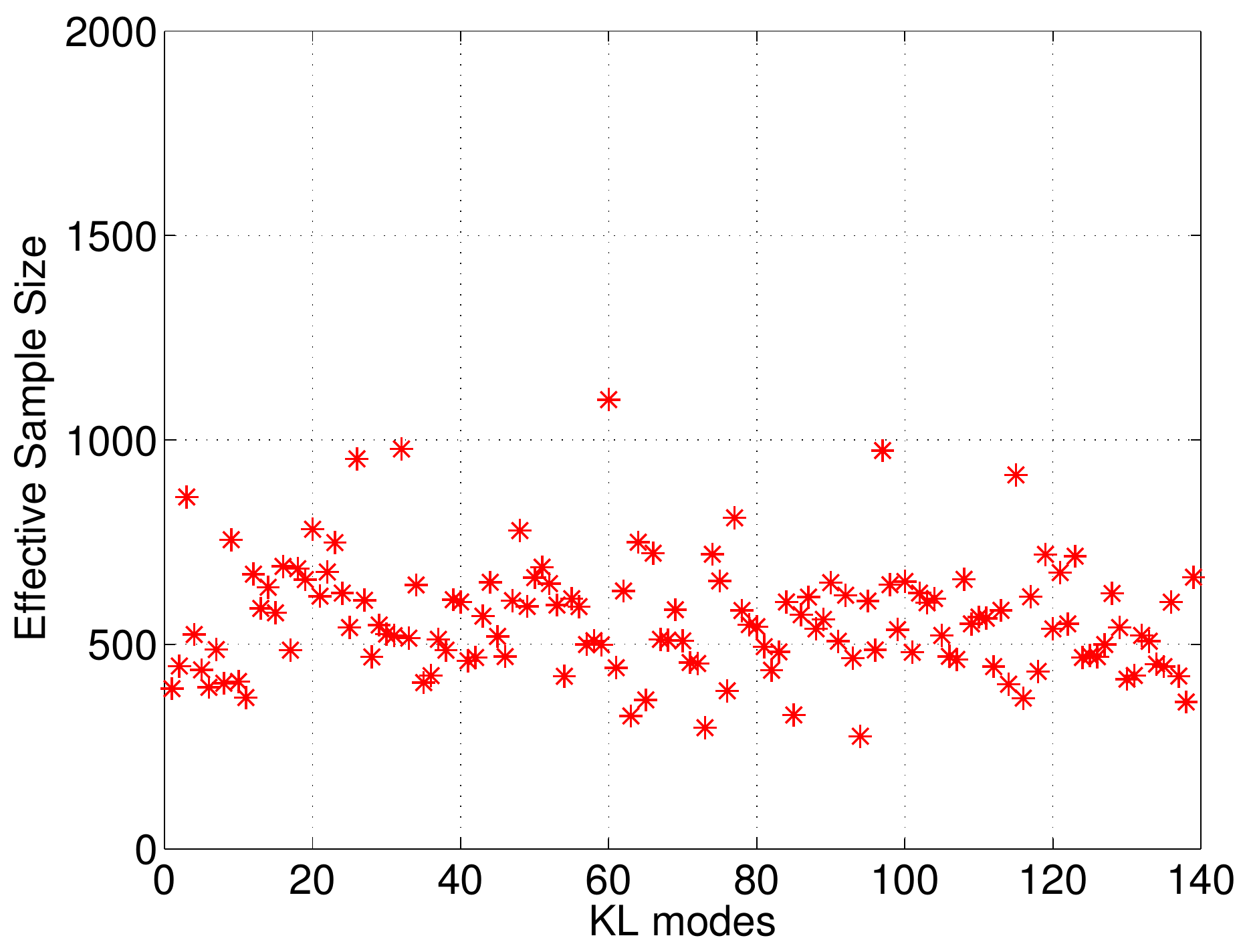}
 \caption{Two-dimensional elliptic PDE: 227 observations, $\sigma_n =
   0.04$, 139-dimensional posterior. Effective sample size after $5
   \times 10^5$ MCMC iterations.}
\label{fig:twodpdelarge_ESS}
\end{figure}

\begin{figure}[htb]
 \centering
 \includegraphics[width=4.8in]{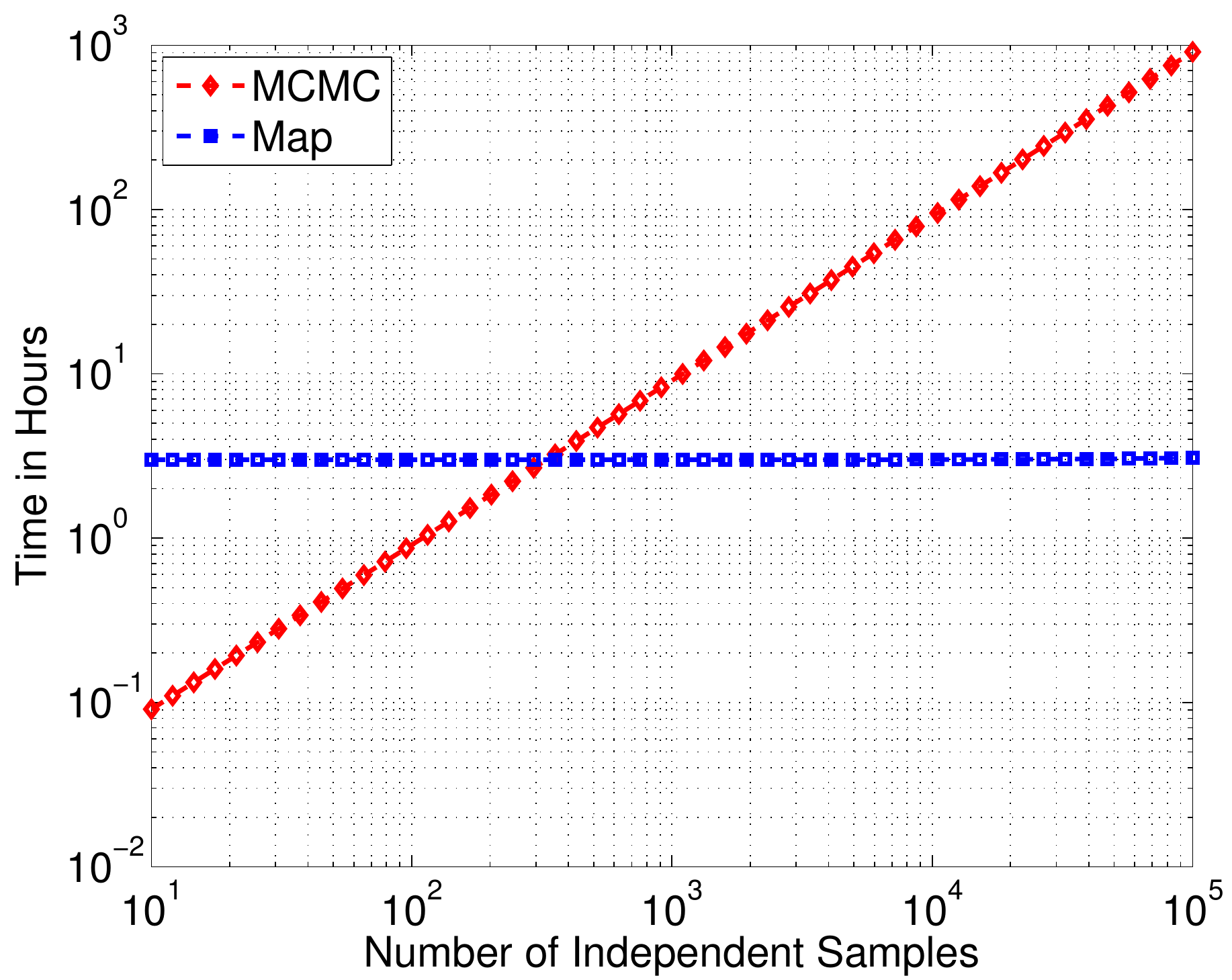}
 \caption{Two-dimensional elliptic PDE: 227 observations, $\sigma_n =
   0.04$, 139-dimensional posterior. Estimated wallclock time required
   to generate a particular number of independent posterior
   samples. The break-even point is at approximately 400 samples. MCMC
   burn-in time has not been included.}
\label{fig:twodpdelarge_samplesvstime}
\end{figure}

In Figure~\ref{fig:twodpdelarge_ESS} we compute effective sample sizes
for one of the MCMC chains from the previous figures. The minimum ESS
over all chain components is 275; the median, mean, and max ESS are
543, 561, 1100, respectively. Extrapolating these numbers lets us
determine the total computational (wallclock) time needed to compute
any number of effectively independent samples via MCMC.
Figure~\ref{fig:twodpdelarge_samplesvstime} thus compares the
computational effort of MCMC to that required by the map. We neglect
the computational time associated with any MCMC burn-in period,
effectively giving MCMC a significant boost in the performance
comparison. The time required by the MCMC algorithm grows linearly
with the number of independent samples. On the other hand, the map
requires a fixed amount of time at the outset, in order to solve the
optimization problem, while the computational cost for each new sample
point is almost negligible. (The latter still grows linearly with the
number of samples but with a very small slope.) Here, if one is
interested in generating fewer than 300 independent samples, then MCMC
is faster; otherwise finding the optimal map is more efficient.

\begin{figure}[htb]
 \centering
 \includegraphics[width=4.8in]{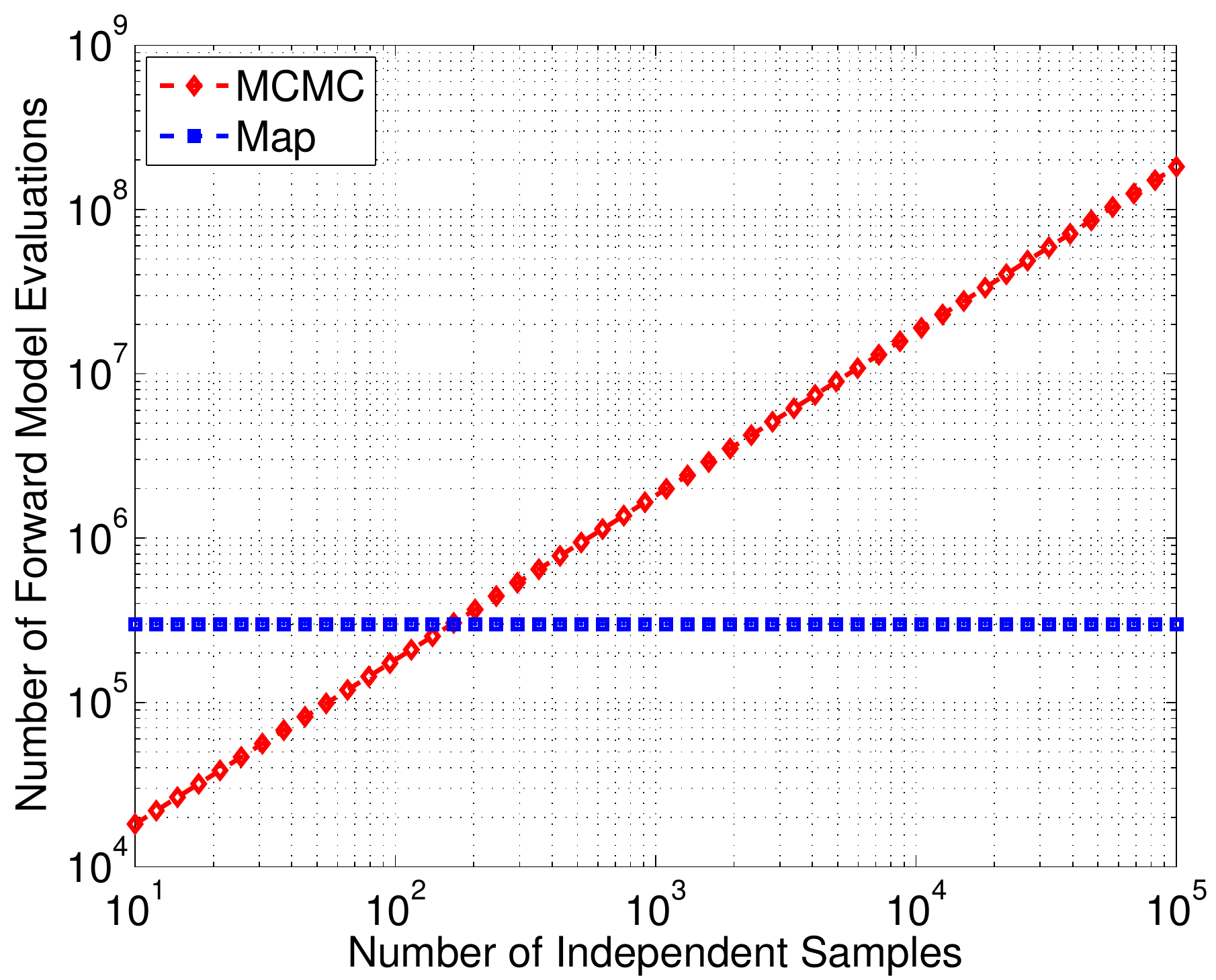}
 \caption{Two-dimensional elliptic PDE: 227 observations, $\sigma_n =
   0.04$, 139-dimensional posterior. Estimated number of forward model
   evaluations required to generate a particular number of independent
   posterior samples. The break-even point is at approximately 200
   samples. MCMC burn-in time has not been included.}
\label{fig:twodpdelarge_samplesvsFM}
\end{figure}

Figure~\ref{fig:twodpdelarge_samplesvsFM} shows a similar comparison,
but uses the number of forward model evaluations as a measure of
computational effort, rather than wallclock time. We assume that
derivatives of the forward model output with respect to the model
parameters can be computed with an adjoint method, which means that
the cost of computing the first derivatives is equivalent to
approximately two forward solves. Furthermore, per the current
implementation, we assume that second derivatives of the forward model
with respect to the parameters are not computed. In this comparison,
the break-even point is approximately 200 samples. If the desired
number of samples is smaller, then MCMC is more efficient; otherwise the
map algorithm can offer order-of-magnitude reductions in computational
effort.

Finally, we note that all of the computations above have used a serial
implementation of the map-based inference approach. It should be
emphasized, however, that the bulk of the computational effort
involved in solving the optimization problem for the map and
subsequently generating samples is embarrassingly parallel. 


\section{Conclusions}
\label{s:concs}

We have presented a new approach to Bayesian inference, based on the
explicit construction of a map that \textit{pushes forward} the prior
measure to the posterior measure. The approach is a significant
departure from Markov chain methods that characterize the posterior
distribution by generating correlated samples. Instead, the present
approach finds a deterministic map $f$ through the solution of an
\textit{optimization problem}. Existence and uniqueness of a monotone
measure-preserving map is established using results from optimal
transport theory. We adapt these results and propose two alternative
optimization formulations, one with an explicit regularization term in
the objective and one that regularizes the problem by constraining the
structure of $f$.
The new formulation offers several advantages over previous methods
for Bayesian computation:
\begin{itemize}
\item The optimization problem provides a clear convergence criterion,
  namely that $\Var[T(X)] \rightarrow 0$, with $T(X)$ defined in
  (\ref{eq:constantT}). Monitoring this criterion can be used to terminate
  iterations \textit{or} to adaptively enrich the function space used
  to describe the map, until a desired level of accuracy is reached.

\item The posterior normalizing constant, or evidence, is computed
  ``for free'' as an output of the optimization problem.

\item Because we describe the map using standard orthogonal
  polynomials, 
  posterior moments may be computed analytically from the polynomial
  coefficients.

\item Once a map is in hand, arbitrary numbers of \textit{independent}
  posterior samples may be generated with minimal computational cost,
  by applying the map to samples from the prior.

\item While the optimization objective involves prior expectations and
  is thus stochastic, efficient gradient-based methods (e.g.,
  Newton or quasi-Newton methods, with the full machinery of adjoints)
  can nonetheless be used to solve the optimization problem.

\item A sequence of low-order maps may be \textit{composed}
  to capture the transition from prior to posterior; this construction
  allows a complex change of measure to be captured more economically
  than with a single map.

\end{itemize}
We demonstrate the map-based approach on a range of examples. Fast
convergence to the exact solution is observed in a linear parameter
inference problem. We also infer parameters in a nonlinear ODE system,
using real experimental data, where the posterior is of complicated
form and differs in shape from the prior. Finally, we use the map to
tackle several high-dimensional, ill-posed, and nonlinear inverse
problems, involving inference of a spatially heterogeneous
diffusion coefficient in an elliptic PDE.

Overall, inference with maps proceeds with greater reliability and
efficiency than MCMC---particularly on high-dimensional inverse
problems. Speedup can be quantified in simple ways, such as counting
the computational effort required to produce a certain number of
effectively independent posterior samples. In the present problems,
the cost of computing the map with typical tolerances is equivalent to
obtaining roughly 200 independent samples with MCMC. But these
comparisons are necessarily insufficient, because inference with maps
provides \textit{more information} and more useful diagnostics than
MCMC, as detailed above.


Several promising avenues exist for future work. There is ample
opportunity to improve the efficiency of the optimization
algorithms. First, we note that each optimization step can be made
embarrassingly parallel, as it relies on prior sampling. Among the
most computationally intensive elements of the optimization procedure
is the evaluation of the forward model $h$, composed with the map, on
each prior sample.  Parallelizing these computations and their
corresponding adjoint solves would lead to immediate and substantial
computational gains. More efficient optimization approaches may also
employ importance sampling to compute the variance or mean of $T$, or
introduce stochastic expansions for $T$ itself.


The map itself is a \textit{polynomial chaos} expansion of the
posterior distribution, and thus it is readily suited to propagation
of posterior uncertainty through subsequent dynamics. With this
posterior propagation step comes an immediate extension to recursive
inference problems, i.e., filtering and prediction with sequential
data. Moreover, in the present work, we have focused on measure
transformations from prior to posterior, but one could also create
maps that push forward some third ``base'' measure to both the
posterior and prior. Such a construction could be useful when it is
not convenient or easy to generate independent prior samples, or if
the prior is improper.

There are several types of static inference problem for which the map
approach must be further developed. The examples presented in this
paper had only one `level'; extensions of map-based inference to
include hyperparameters, or more generally, to exploit the structure
of Bayesian hierarchical models, are currently ongoing.
%
%
Another challenge arises when the posterior has bounded support, with
significant probability mass accumulating near a boundary. The range
of the map must be appropriately bounded in such cases. Thus far we
have circumvented such problems by reparameterization, transforming
bounded domains into unbounded ones. This trick comes at a
computational price, e.g., greater nonlinearity. We ask, then, how
more directly to bound the range of the map, whether on a bounded or
unbounded input domain, via constraints in optimization or perhaps a
non-polynomial basis.
%

We would also like to better understand the limiting behavior of the
map in ill-posed and high-dimensional problems. When inferring
distributed parameters with some meaningful correlation structure---in
the elliptic inverse problem, for example---there is a natural
ordering of the random variables and adaptive enrichment of the map
works well. But in high-dimensional nonlinear problems with no natural
ordering (for instance, nonlinear ODE systems with hundreds of unknown
parameters), a high-order polynomial expansion in all the modes is
computationally prohibitive. Further exploration of adaptive methods,
perhaps coupled with dimensionality reduction, is needed. We ask, for
example, whether inferential maps have a low-rank tensorial
representation or a sparse representation in some basis.
%
%
%
Finally, we note that generating multi-modal posterior distributions
from unimodal priors will require more localized structure in the maps
than global polynomial basis functions can feasibly provide. Piecewise
polynomials, multi-element polynomial chaos~\cite{wan:2005:aam}, and
similar representations may be quite useful in such problems.





\section*{Acknowledgments}

The authors would like to acknowledge support from the US Department
of Energy, Office of Science, Office of Advanced Scientific Computing
Research (ASCR) under grant numbers DE-SC0002517 and DE-SC0003908.

\bibliographystyle{model1-num-names} 
\pdfbookmark{References}{s:references}
\bibliography{maprefs}

\begin{thebibliography}{60}
\expandafter\ifx\csname natexlab\endcsname\relax\def\natexlab#1{#1}\fi
\providecommand{\bibinfo}[2]{#2}
\ifx\xfnm\relax \def\xfnm[#1]{\unskip,\space#1}\fi
\bibitem[{Stuart(2010)}]{stuart:2010:ipa}
\bibinfo{author}{A.~M. Stuart},
\newblock \bibinfo{title}{Inverse problems: a {B}ayesian perspective},
\newblock \bibinfo{journal}{Acta Numerica} \bibinfo{volume}{19}
  (\bibinfo{year}{2010}) \bibinfo{pages}{451--559}.
\bibitem[{Gelman et~al.(2003)Gelman, Carlin, Stern, and
  Rubin}]{gelman:2003:bda}
\bibinfo{author}{A.~Gelman}, \bibinfo{author}{J.~B. Carlin},
  \bibinfo{author}{H.~S. Stern}, \bibinfo{author}{D.~B. Rubin},
  \bibinfo{title}{{Bayesian Data Analysis}}, \bibinfo{publisher}{Chapman and
  Hall/CRC}, \bibinfo{edition}{2nd} edition, \bibinfo{year}{2003}.
\bibitem[{Kaipio and Somersalo(2004)}]{kaipio:2004:bip}
\bibinfo{author}{J.~Kaipio}, \bibinfo{author}{E.~Somersalo},
  \bibinfo{title}{{Statistical and Computational Inverse Problems}},
  \bibinfo{publisher}{Springer}, \bibinfo{year}{2004}.
\bibitem[{Sivia and Skilling(2006)}]{sivia:2006:daa}
\bibinfo{author}{D.~S. Sivia}, \bibinfo{author}{J.~Skilling},
  \bibinfo{title}{{Data Analysis: A Bayesian Tutorial}},
  \bibinfo{publisher}{Oxford Science Publications}, \bibinfo{edition}{2nd}
  edition, \bibinfo{year}{2006}.
\bibitem[{Bernardo and Smith(1994)}]{bernardo:1994:bt}
\bibinfo{author}{J.~M. Bernardo}, \bibinfo{author}{A.~F.~M. Smith},
  \bibinfo{title}{Bayesian Theory}, \bibinfo{publisher}{Wiley},
  \bibinfo{year}{1994}.
\bibitem[{Metropolis et~al.(1953)Metropolis, Rosenbluth, Rosenbluth, Teller,
  and Teller}]{metropolis:1953:eos}
\bibinfo{author}{N.~Metropolis}, \bibinfo{author}{A.~W. Rosenbluth},
  \bibinfo{author}{M.~N. Rosenbluth}, \bibinfo{author}{A.~H. Teller},
  \bibinfo{author}{E.~Teller},
\newblock \bibinfo{title}{Equation of state calculations by fast computing
  machines},
\newblock \bibinfo{journal}{The Journal of Chemical Physics}
  \bibinfo{volume}{21} (\bibinfo{year}{1953}) \bibinfo{pages}{1087--1092}.
\bibitem[{Hastings(1970)}]{hastings:1970:mcs}
\bibinfo{author}{W.~K. Hastings},
\newblock \bibinfo{title}{{M}onte {C}arlo sampling methods using {M}arkov
  chains and their applications},
\newblock \bibinfo{journal}{Biometrika} \bibinfo{volume}{57}
  (\bibinfo{year}{1970}) \bibinfo{pages}{97--109}.
\bibitem[{Gilks et~al.(1996)Gilks, Gilks, Richardson, and
  Spiegelhalter}]{gilks:1996:mcm}
\bibinfo{author}{W.~Gilks}, \bibinfo{author}{W.~Gilks},
  \bibinfo{author}{S.~Richardson}, \bibinfo{author}{D.~Spiegelhalter},
  \bibinfo{title}{{M}arkov Chain {M}onte {C}arlo in Practice},
  \bibinfo{publisher}{Chapman \& Hall}, \bibinfo{year}{1996}.
\bibitem[{Robert and Casella(2004)}]{robert:1999:mcs}
\bibinfo{author}{C.~P. Robert}, \bibinfo{author}{G.~Casella},
  \bibinfo{title}{{Monte Carlo Statistical Methods}},
  \bibinfo{publisher}{Springer-Verlag}, \bibinfo{edition}{2nd} edition,
  \bibinfo{year}{2004}.
\bibitem[{Liu(2008)}]{junliu:2008:mcs}
\bibinfo{author}{J.~S. Liu}, \bibinfo{title}{{Monte Carlo Strategies in
  Scientific Computing}}, \bibinfo{publisher}{Springer}, \bibinfo{year}{2008}.
\bibitem[{Haario et~al.(2006)Haario, Laine, Mira, and
  Saksman}]{haario:2006:dra}
\bibinfo{author}{H.~Haario}, \bibinfo{author}{M.~Laine},
  \bibinfo{author}{A.~Mira}, \bibinfo{author}{E.~Saksman},
\newblock \bibinfo{title}{{DRAM}: Efficient adaptive {MCMC}},
\newblock \bibinfo{journal}{Statistics and Computing} \bibinfo{volume}{16}
  (\bibinfo{year}{2006}) \bibinfo{pages}{339--354}.
\bibitem[{Roberts and Rosenthal(2009)}]{roberts:2009:eof}
\bibinfo{author}{G.~O. Roberts}, \bibinfo{author}{J.~S. Rosenthal},
\newblock \bibinfo{title}{Examples of adaptive {MCMC}},
\newblock \bibinfo{journal}{Journal of Computational and Graphical Statistics}
  \bibinfo{volume}{18} (\bibinfo{year}{2009}) \bibinfo{pages}{349--367}.
\bibitem[{Stramer and Tweedie(1999)}]{stramer:1999:ltm}
\bibinfo{author}{O.~Stramer}, \bibinfo{author}{R.~Tweedie},
\newblock \bibinfo{title}{{L}angevin-type models {II}: Self-targeting
  candidates for {MCMC} algorithms},
\newblock \bibinfo{journal}{Methodology and Computing in Applied Probability}
  \bibinfo{volume}{1} (\bibinfo{year}{1999}) \bibinfo{pages}{307--328}.
\bibitem[{Apte et~al.(2007)Apte, Hairer, Stuart, and Voss}]{apte:2007:stp}
\bibinfo{author}{A.~Apte}, \bibinfo{author}{M.~Hairer},
  \bibinfo{author}{A.~Stuart}, \bibinfo{author}{J.~Voss},
\newblock \bibinfo{title}{Sampling the posterior: an approach to non-{G}aussian
  data assimilation},
\newblock \bibinfo{journal}{Physica D: Nonlinear Phenomena}
  \bibinfo{volume}{230} (\bibinfo{year}{2007}) \bibinfo{pages}{50--64}.
\bibitem[{Girolami and Calderhead(2011)}]{girolami:2011:rml}
\bibinfo{author}{M.~Girolami}, \bibinfo{author}{B.~Calderhead},
\newblock \bibinfo{title}{Riemann manifold {L}angevin and {H}amiltonian {M}onte
  {C}arlo methods},
\newblock \bibinfo{journal}{Journal of the Royal Statistical Society: Series B
  (Statistical Methodology)} \bibinfo{volume}{73} (\bibinfo{year}{2011})
  \bibinfo{pages}{123--214}.
\bibitem[{Martin et~al.(2012)Martin, Wilcox, Burstedde, and
  Ghattas}]{martin:2011:asn}
\bibinfo{author}{J.~Martin}, \bibinfo{author}{L.~Wilcox},
  \bibinfo{author}{C.~Burstedde}, \bibinfo{author}{O.~Ghattas},
\newblock \bibinfo{title}{A stochastic {N}ewton {MCMC} method for large-scale
  statistical inverse problems},
\newblock \bibinfo{journal}{{SIAM} Journal on Scientific Computing}
  (\bibinfo{year}{2012}). \bibinfo{note}{To appear}.
\bibitem[{Neal(2011)}]{neal:2011:mcm}
\bibinfo{author}{R.~M. Neal},
\newblock \bibinfo{title}{{MCMC} using {H}amiltonian dynamics},
\newblock in: \bibinfo{editor}{S.~Brooks}, \bibinfo{editor}{A.~Gelman},
  \bibinfo{editor}{G.~Jones}, \bibinfo{editor}{X.~Meng} (Eds.),
  \bibinfo{booktitle}{Handbook of Markov Chain Monte Carlo},
  \bibinfo{publisher}{Chapman and Hall/CRC Press}, \bibinfo{year}{2011}, pp.
  \bibinfo{pages}{113--162}.
\bibitem[{Higdon et~al.(2002)Higdon, Lee, and Bi}]{higdon:2002:aba}
\bibinfo{author}{D.~Higdon}, \bibinfo{author}{H.~Lee}, \bibinfo{author}{Z.~Bi},
\newblock \bibinfo{title}{A {B}ayesian approach to characterizing uncertainty
  in inverse problems using coarse and fine-scale information},
\newblock \bibinfo{journal}{{IEEE} Transactions on Signal Processing}
  \bibinfo{volume}{50} (\bibinfo{year}{2002}) \bibinfo{pages}{389--399}.
\bibitem[{Vrugt et~al.(2009)Vrugt, ter Braak, Diks, Robinson, Hyman, and
  Higdon}]{vrugt:2009:amc}
\bibinfo{author}{J.~A. Vrugt}, \bibinfo{author}{C.~J.~F. ter Braak},
  \bibinfo{author}{C.~G.~H. Diks}, \bibinfo{author}{B.~A. Robinson},
  \bibinfo{author}{J.~M. Hyman}, \bibinfo{author}{D.~Higdon},
\newblock \bibinfo{title}{Accelerating {M}arkov chain {M}onte {C}arlo
  simulation by differential evolution with self-adaptive randomized subspace
  sampling},
\newblock \bibinfo{journal}{International Journal of Nonlinear Sciences and
  Numerical Simulation} \bibinfo{volume}{10} (\bibinfo{year}{2009})
  \bibinfo{pages}{273--290}.
\bibitem[{Craiu et~al.(2009)Craiu, Rosenthal, and Yang}]{craiu:2009:lft}
\bibinfo{author}{R.~V. Craiu}, \bibinfo{author}{J.~Rosenthal},
  \bibinfo{author}{C.~Yang},
\newblock \bibinfo{title}{Learn from thy neighbor: Parallel-chain and regional
  adaptive {MCMC}},
\newblock \bibinfo{journal}{Journal of the American Statistical Association}
  \bibinfo{volume}{104} (\bibinfo{year}{2009}) \bibinfo{pages}{1454--1466}.
\bibitem[{Christen and Fox(2005)}]{christen:2005:mcm}
\bibinfo{author}{J.~A. Christen}, \bibinfo{author}{C.~Fox},
\newblock \bibinfo{title}{{M}arkov chain {M}onte {C}arlo using an
  approximation},
\newblock \bibinfo{journal}{Journal of Computational and Graphical Statistics}
  \bibinfo{volume}{14} (\bibinfo{year}{2005}) \bibinfo{pages}{795--810}.
\bibitem[{Efendiev et~al.(2006)Efendiev, Hou, and Luo}]{efendiev:2006:pmc}
\bibinfo{author}{Y.~Efendiev}, \bibinfo{author}{T.~Y. Hou},
  \bibinfo{author}{W.~Luo},
\newblock \bibinfo{title}{Preconditioning {M}arkov chain {M}onte {C}arlo
  simulations using coarse-scale models},
\newblock \bibinfo{journal}{{SIAM} Journal on Scientific Computing}
  \bibinfo{volume}{28} (\bibinfo{year}{2006}) \bibinfo{pages}{776--803}.
\bibitem[{Marzouk et~al.(2007)Marzouk, Najm, and Rahn}]{marzouk:2007:ssm}
\bibinfo{author}{Y.~M. Marzouk}, \bibinfo{author}{H.~N. Najm},
  \bibinfo{author}{L.~A. Rahn},
\newblock \bibinfo{title}{Stochastic spectral methods for efficient {B}ayesian
  solution of inverse problems},
\newblock \bibinfo{journal}{Journal of Computational Physics}
  \bibinfo{volume}{224} (\bibinfo{year}{2007}) \bibinfo{pages}{560--586}.
\bibitem[{Marzouk and Najm(2009)}]{marzouk:2009:dra}
\bibinfo{author}{Y.~M. Marzouk}, \bibinfo{author}{H.~N. Najm},
\newblock \bibinfo{title}{Dimensionality reduction and polynomial chaos
  acceleration of {B}ayesian inference in inverse problems},
\newblock \bibinfo{journal}{Journal of Computational Physics}
  \bibinfo{volume}{228} (\bibinfo{year}{2009}) \bibinfo{pages}{1862--1902}.
\bibitem[{Lieberman et~al.(2010)Lieberman, Willcox, and
  Ghattas}]{lieberman:2010:psm}
\bibinfo{author}{C.~Lieberman}, \bibinfo{author}{K.~Willcox},
  \bibinfo{author}{O.~Ghattas},
\newblock \bibinfo{title}{Parameter and state model reduction for large-scale
  statistical inverse problems},
\newblock \bibinfo{journal}{{SIAM} Journal on Scientific Computing}
  \bibinfo{volume}{32} (\bibinfo{year}{2010}) \bibinfo{pages}{2485--2496}.
\bibitem[{Frangos et~al.(2010)Frangos, Marzouk, Willcox, and van
  Bloemen~Waanders}]{frangos:2011:sar}
\bibinfo{author}{M.~Frangos}, \bibinfo{author}{Y.~Marzouk},
  \bibinfo{author}{K.~Willcox}, \bibinfo{author}{B.~van Bloemen~Waanders},
\newblock \bibinfo{title}{Surrogate and reduced-order modeling: a comparison of
  approaches for large-scale statistical inverse problems},
\newblock in: \bibinfo{editor}{L.~Biegler}, \bibinfo{editor}{G.~Biros},
  \bibinfo{editor}{O.~Ghattas}, \bibinfo{editor}{M.~Heinkenschloss},
  \bibinfo{editor}{D.~Keyes}, \bibinfo{editor}{B.~Mallick},
  \bibinfo{editor}{Y.~Marzouk}, \bibinfo{editor}{L.~Tenorio},
  \bibinfo{editor}{B.~van Bloemen~Waanders}, , \bibinfo{editor}{K.~Willcox}
  (Eds.), \bibinfo{booktitle}{Computational Methods for Large-Scale Inverse
  Problems and Quantification of Uncertainty}, \bibinfo{publisher}{Wiley},
  \bibinfo{year}{2010}, pp. \bibinfo{pages}{123--149}.
\bibitem[{Wang and Zabaras(2005)}]{wang:2005:ubs}
\bibinfo{author}{J.~Wang}, \bibinfo{author}{N.~Zabaras},
\newblock \bibinfo{title}{Using {B}ayesian statistics in the estimation of heat
  source in radiation},
\newblock \bibinfo{journal}{International Journal of Heat and Mass Transfer}
  \bibinfo{volume}{48} (\bibinfo{year}{2005}) \bibinfo{pages}{15--29}.
\bibitem[{Gelman and Shirley(2011)}]{gelman:2011:ifs}
\bibinfo{author}{A.~Gelman}, \bibinfo{author}{K.~Shirley},
\newblock \bibinfo{title}{Inference from simulations and monitoring
  convergence},
\newblock in: \bibinfo{editor}{S.~Brooks}, \bibinfo{editor}{A.~Gelman},
  \bibinfo{editor}{G.~Jones}, \bibinfo{editor}{X.~Meng} (Eds.),
  \bibinfo{booktitle}{Handbook of Markov Chain Monte Carlo},
  \bibinfo{publisher}{Chapman and Hall/CRC Press}, \bibinfo{year}{2011}, pp.
  \bibinfo{pages}{163--174}.
\bibitem[{Ghanem and Spanos(2003)}]{ghanem:2003:sfe}
\bibinfo{author}{R.~Ghanem}, \bibinfo{author}{P.~Spanos},
  \bibinfo{title}{{Stochastic Finite Elements: a Spectral Approach}},
  \bibinfo{publisher}{Dover Publications}, \bibinfo{year}{2003}.
\bibitem[{LeMa{\^\i}tre and Knio(2010)}]{maître:2010:smf}
\bibinfo{author}{O.~P. LeMa{\^\i}tre}, \bibinfo{author}{O.~M. Knio},
  \bibinfo{title}{Spectral Methods for Uncertainty Quantification: with
  Applications to Computational Fluid Dynamics}, \bibinfo{publisher}{Springer},
  \bibinfo{year}{2010}.
\bibitem[{Xiu(2010)}]{xiu:2010:nmf}
\bibinfo{author}{D.~Xiu}, \bibinfo{title}{Numerical Methods for Stochastic
  Computations: A Spectral Method Approach}, \bibinfo{publisher}{Princeton
  University Press}, \bibinfo{year}{2010}.
\bibitem[{Doucet et~al.(2001)Doucet, Freitas, and Gordon}]{doucet:2001:smc}
\bibinfo{author}{A.~Doucet}, \bibinfo{author}{N.~Freitas},
  \bibinfo{author}{N.~Gordon}, \bibinfo{title}{{Sequential {M}onte {C}arlo
  Methods in Practice}}, \bibinfo{publisher}{Springer}, \bibinfo{year}{2001}.
\bibitem[{McCann(1995)}]{mccann:1995:eau}
\bibinfo{author}{R.~J. McCann},
\newblock \bibinfo{title}{Existence and uniqueness of monotone
  measure-preserving maps},
\newblock \bibinfo{journal}{Duke Mathematical Journal} \bibinfo{volume}{80}
  (\bibinfo{year}{1995}) \bibinfo{pages}{309--323}.
\bibitem[{Xiu and Karniadakis(2002)}]{xiu:2002:twa}
\bibinfo{author}{D.~Xiu}, \bibinfo{author}{G.~Karniadakis},
\newblock \bibinfo{title}{The {W}iener-{A}skey polynomial chaos for stochastic
  differential equations},
\newblock \bibinfo{journal}{SIAM Journal on Scientific Computing}
  \bibinfo{volume}{24} (\bibinfo{year}{2002}) \bibinfo{pages}{619--644}.
\bibitem[{Kass and Raftery(1995)}]{kass:1995:bf}
\bibinfo{author}{R.~E. Kass}, \bibinfo{author}{A.~E. Raftery},
\newblock \bibinfo{title}{Bayes factors},
\newblock \bibinfo{journal}{Journal of the American Statistical Association}
  \bibinfo{volume}{90} (\bibinfo{year}{1995}) \bibinfo{pages}{773--795}.
\bibitem[{Jaakkola and Jordan(2000)}]{jaakkola:2000:bpe}
\bibinfo{author}{T.~Jaakkola}, \bibinfo{author}{M.~Jordan},
\newblock \bibinfo{title}{Bayesian parameter estimation via variational
  methods},
\newblock \bibinfo{journal}{Statistics and Computing} \bibinfo{volume}{10}
  (\bibinfo{year}{2000}) \bibinfo{pages}{25--37}.
\bibitem[{Chorin and Tu(2009)}]{chorin:2009:isf}
\bibinfo{author}{A.~J. Chorin}, \bibinfo{author}{X.~Tu},
\newblock \bibinfo{title}{Implicit sampling for particle filters},
\newblock \bibinfo{journal}{Proceedings of the National Academy of Sciences
  {USA}} \bibinfo{volume}{106} (\bibinfo{year}{2009})
  \bibinfo{pages}{17249--17254}.
\bibitem[{Chorin et~al.(2010)Chorin, Morzfeld, and Tu}]{chorin:2010:ipf}
\bibinfo{author}{A.~Chorin}, \bibinfo{author}{M.~Morzfeld},
  \bibinfo{author}{X.~Tu},
\newblock \bibinfo{title}{Implicit particle filters for data assimilation},
\newblock \bibinfo{journal}{Communications in Applied Mathematics and
  Computational Science} \bibinfo{volume}{5} (\bibinfo{year}{2010})
  \bibinfo{pages}{221--240}.
\bibitem[{Reich(2011)}]{reich:2011:ads}
\bibinfo{author}{S.~Reich},
\newblock \bibinfo{title}{A dynamical systems framework for intermittent data
  assimilation},
\newblock \bibinfo{journal}{BIT Numerical Mathematics} \bibinfo{volume}{51}
  (\bibinfo{year}{2011}) \bibinfo{pages}{235--249}.
\bibitem[{Tarantola(2005)}]{tarantola:2005:ipt}
\bibinfo{author}{A.~Tarantola}, \bibinfo{title}{Inverse problem theory and
  methods for model parameter estimation}, \bibinfo{publisher}{Society for
  Industrial and Applied Mathematics}, \bibinfo{year}{2005}.
\bibitem[{McShane(1937)}]{mcshane:1937:ji}
\bibinfo{author}{E.~J. McShane},
\newblock \bibinfo{title}{Jensen's inequality},
\newblock \bibinfo{journal}{Bulletin of the American Mathematical Society}
  \bibinfo{volume}{43} (\bibinfo{year}{1937}) \bibinfo{pages}{521--527}.
\bibitem[{Brenier(1991)}]{brenier:1991:pfa}
\bibinfo{author}{Y.~Brenier},
\newblock \bibinfo{title}{Polar factorization and monotone rearrangement of
  vector-valued functions},
\newblock \bibinfo{journal}{Communications on Pure and Applied Mathematics}
  \bibinfo{volume}{44} (\bibinfo{year}{1991}) \bibinfo{pages}{375--417}.
\bibitem[{Caffarelli(1992)}]{caffarelli:1992:tro}
\bibinfo{author}{L.~A. Caffarelli},
\newblock \bibinfo{title}{The regularity of mappings with a convex potential},
\newblock \bibinfo{journal}{Journal of the American Mathematical Society}
  \bibinfo{volume}{5} (\bibinfo{year}{1992}) \bibinfo{pages}{99--104}.
\bibitem[{Gangbo and McCann(1996)}]{gangbo:1996:tgo}
\bibinfo{author}{W.~Gangbo}, \bibinfo{author}{R.~J. McCann},
\newblock \bibinfo{title}{The geometry of optimal transportation},
\newblock \bibinfo{journal}{Acta Mathematica} \bibinfo{volume}{177}
  (\bibinfo{year}{1996}) \bibinfo{pages}{113--161}.
\bibitem[{Villani(2009)}]{villani:2009:oto}
\bibinfo{author}{C.~Villani}, \bibinfo{title}{{Optimal Transport: Old and
  New}}, \bibinfo{publisher}{Springer}, \bibinfo{year}{2009}.
\bibitem[{Rosenblatt(1952)}]{rosenblatt:1952:roa}
\bibinfo{author}{M.~Rosenblatt},
\newblock \bibinfo{title}{Remarks on a multivariate transformation},
\newblock \bibinfo{journal}{The Annals of Mathematical Statistics}
  \bibinfo{volume}{23} (\bibinfo{year}{1952}) \bibinfo{pages}{470--472}.
\bibitem[{Knothe(1957)}]{knothe:1957:ctt}
\bibinfo{author}{H.~Knothe},
\newblock \bibinfo{title}{Contributions to the theory of convex bodies},
\newblock \bibinfo{journal}{Michigan Mathematical Journal} \bibinfo{volume}{4}
  (\bibinfo{year}{1957}) \bibinfo{pages}{39--52}.
\bibitem[{Carlier et~al.(2010)Carlier, Galichon, and
  Santambrogio}]{carlier:2010:fkt}
\bibinfo{author}{G.~Carlier}, \bibinfo{author}{A.~Galichon},
  \bibinfo{author}{F.~Santambrogio},
\newblock \bibinfo{title}{From {K}nothe's transport to {B}renier's map and a
  continuation method for optimal transport},
\newblock \bibinfo{journal}{SIAM Journal on Mathematical Analysis}
  \bibinfo{volume}{41} (\bibinfo{year}{2010}).
\bibitem[{Nemirovski et~al.(2009)Nemirovski, Juditsky, Lan, and
  Shapiro}]{nemirovski:2009:rsa}
\bibinfo{author}{A.~Nemirovski}, \bibinfo{author}{A.~Juditsky},
  \bibinfo{author}{G.~Lan}, \bibinfo{author}{A.~Shapiro},
\newblock \bibinfo{title}{Robust stochastic approximation approach to
  stochastic programming},
\newblock \bibinfo{journal}{SIAM Journal on Optimization} \bibinfo{volume}{19}
  (\bibinfo{year}{2009}) \bibinfo{pages}{1574--1609}.
\bibitem[{Wiener(1938)}]{wiener:1938:thc}
\bibinfo{author}{N.~Wiener},
\newblock \bibinfo{title}{The homogeneous chaos},
\newblock \bibinfo{journal}{American Journal of Mathematics}
  \bibinfo{volume}{60} (\bibinfo{year}{1938}) \bibinfo{pages}{897--936}.
\bibitem[{Soize and Ghanem(2004)}]{soize:2004:psw}
\bibinfo{author}{C.~Soize}, \bibinfo{author}{R.~Ghanem},
\newblock \bibinfo{title}{Physical systems with random uncertainties: Chaos
  representations with arbitrary probability measure},
\newblock \bibinfo{journal}{{SIAM} Journal on Scientific Computing}
  \bibinfo{volume}{26} (\bibinfo{year}{2004}) \bibinfo{pages}{395--410}.
\bibitem[{Rupert and Miller(2007)}]{rupert:2007:aao}
\bibinfo{author}{C.~Rupert}, \bibinfo{author}{C.~Miller},
\newblock \bibinfo{title}{An analysis of polynomial chaos approximations for
  modeling single-fluid-phase flow in porous medium systems},
\newblock \bibinfo{journal}{Journal of Computational Physics}
  \bibinfo{volume}{226} (\bibinfo{year}{2007}) \bibinfo{pages}{2175--2205}.
\bibitem[{Gardner et~al.(2000)Gardner, Cantor, and Collins}]{gardner:2000:coa}
\bibinfo{author}{T.~S. Gardner}, \bibinfo{author}{C.~R. Cantor},
  \bibinfo{author}{J.~J. Collins},
\newblock \bibinfo{title}{Construction of a genetic toggle switch in
  escherichia coli.},
\newblock \bibinfo{journal}{Nature} \bibinfo{volume}{403}
  (\bibinfo{year}{2000}) \bibinfo{pages}{339--342}.
\bibitem[{Marzouk and Xiu(2009)}]{marzouk:2009:asc}
\bibinfo{author}{Y.~M. Marzouk}, \bibinfo{author}{D.~Xiu},
\newblock \bibinfo{title}{A stochastic collocation approach to {B}ayesian
  inference in inverse problems},
\newblock \bibinfo{journal}{Communications in Computational Physics}
  \bibinfo{volume}{6} (\bibinfo{year}{2009}) \bibinfo{pages}{826--847}.
\bibitem[{Dashti and Stuart(2012)}]{dashti:2012:uqa}
\bibinfo{author}{M.~Dashti}, \bibinfo{author}{A.~Stuart},
\newblock \bibinfo{title}{Uncertainty quantification and weak approximation of
  elliptic inverse},
\newblock \bibinfo{journal}{{SIAM} Journal on Numerical Analysis}
  (\bibinfo{year}{2012}). \bibinfo{note}{To appear}.
\bibitem[{Rasmussen and Williams(2006)}]{rasmussen:2006:gpf}
\bibinfo{author}{C.~E. Rasmussen}, \bibinfo{author}{C.~K.~I. Williams},
  \bibinfo{title}{Gaussian processes for machine learning},
  \bibinfo{publisher}{MIT Press}, \bibinfo{year}{2006}.
\bibitem[{Papaspiliopoulos et~al.(2012)Papaspiliopoulos, Pokern, Roberts, and
  Stuart}]{Papaspiliopoulos:2012:neo}
\bibinfo{author}{O.~Papaspiliopoulos}, \bibinfo{author}{Y.~Pokern},
  \bibinfo{author}{G.~O. Roberts}, \bibinfo{author}{A.~Stuart},
\newblock \bibinfo{title}{Nonparametric estimation of diffusions: a
  differential equations approach},
\newblock \bibinfo{journal}{Biometrika}  (\bibinfo{year}{2012}).
  \bibinfo{note}{To appear}.
\bibitem[{Moselhy and Marzouk(2011)}]{moselhy:2011:aai}
\bibinfo{author}{T.~A. Moselhy}, \bibinfo{author}{Y.~M. Marzouk},
  \bibinfo{title}{An adaptive iterative method for high-dimensional stochastic
  {PDE}s}, \bibinfo{year}{2011}. \bibinfo{note}{Preprint}.
\bibitem[{Cotter et~al.(2012)Cotter, Roberts, Stuart, and
  White}]{cotter:2012:mmf}
\bibinfo{author}{S.~Cotter}, \bibinfo{author}{G.~Roberts},
  \bibinfo{author}{A.~Stuart}, \bibinfo{author}{D.~White},
\newblock \bibinfo{title}{{MCMC} methods for functions: Modifying old
  algorithms to make them faster}  (\bibinfo{year}{2012}).
  \bibinfo{note}{ArXiv:1202.0709}.
\bibitem[{Wan and Karniadakis(2005)}]{wan:2005:aam}
\bibinfo{author}{X.~Wan}, \bibinfo{author}{G.~E. Karniadakis},
\newblock \bibinfo{title}{An adaptive multi-element generalized polynomial
  chaos method for stochastic differential equations},
\newblock \bibinfo{journal}{Journal of Computational Physics}
  \bibinfo{volume}{209} (\bibinfo{year}{2005}) \bibinfo{pages}{617--642}.

\end{thebibliography}

\cleardoublepage

%
%

\cleardoublepage

%

\cleardoublepage

\cleardoublepage

\cleardoublepage

\cleardoublepage

\cleardoublepage

\cleardoublepage

\pdfbookmark{Appendices}{s:appendices}
\cleardoublepage
\appendix

\section{Relation between variance and Kullback-Leibler divergence near convergence}
\label{app:KLversusVar}

When the variance of $T(X)$ is small, the Kullback Leibler-divergence
from $p$ to $\tilde{p}$ is small as well. In this case, the
relationship between these two quantities is asymptotically linear.
To demonstrate this, we start from the KL divergence:
\begin{equation}
D_{\mathrm{KL}}\left (p || \tilde{p} \right) =  \log \EE \left[ \exp
     \( T \) \right] -  \EE \left[ T \right].
\end{equation}
For small perturbations of $T$ around $T_0 =   {\mathbb E}\left[ T \right] $, we obtain
\begin{eqnarray}
D_{\mathrm{KL}}\left(p || \tilde{p} \right ) 
& \approx & \log \EE \left [ \exp{ (T_0) } \left(1 + (T-T_0) + \frac{1}{2} (T-T_0)^2  \right) \right] -
\EE\left[ T_0 \right] \nonumber \\
& = &  \log{\EE \left[ \left(1 + (T-T_0) + \frac{1}{2} (T-T_0)^2  \right) \right]} \nonumber\\
& = &  \log \left(1 + \frac{1}{2} \Var \left[ T \right]  \right )  \nonumber\\
& \approx & \Var \left[ T \right ] / \, 2  
\end{eqnarray}

\section{Linear-Gaussian model}
\label{app:lineargaussian}
The special case of a linear forward model $h(x) = A x$ with additive
Gaussian noise $\varepsilon \sim N(0, \Sigma_n)$ and a Gaussian prior
$x \sim N(0, \Sigma_P)$ admits a closed form solution, where the
posterior is also Gaussian.
Without loss of generality, we assume a prior mean of zero and write
the posterior density $q$ as:
\begin{eqnarray}
q(z) &=& \frac{1}{\beta}\exp{\left( -\frac{1}{2} \left( \|  A z -d \|^2_{\Sigma_n} + \|z \|^2_{\Sigma_P }\right) \right )} \nonumber\\
 &=& \frac{1}{\beta}\exp{\left( -\frac{1}{2} \left( \| d  \|^2_{\Sigma_n}  +
       z^T \left( A^T {\Sigma_n^{-1}}A + {\Sigma_P^{-1}}  \right) z +
       2 d^T  {\Sigma_n^{-1}} A  z \right) \right)} \nonumber\\
 &=& \frac{1}{\beta}\exp{\left( -\frac{1}{2} \left( \| d \|^2_{\Sigma_n}  + z^T\Sigma^{-1} z -2 \mu^T \Sigma^{-1} z \right) \right)}
\end{eqnarray}
where $\mu$, $\Sigma$, and ${\beta}$ are the posterior mean,
covariance, and evidence, respectively. Equating 
the second and third lines above, we obtain the following relations:
\begin{eqnarray}
\Sigma &=& \left( A^T {\Sigma_n^{-1}} A + {\Sigma_P^{-1}}   \right)^{-1} \nonumber\\
 \mu  &=&  \Sigma A^T {\Sigma_n^{-1}} d    \nonumber \\
 \beta &=& \exp{\left( -\frac{1}{2} \left( \| d \|^2_{{\Sigma_n}}  -
       \mu^T \Sigma^{-1} \mu^T \right) \right) } \sqrt{\left |
     \det{\Sigma} \right |}.
\end{eqnarray}

The map from the prior to the posterior is then given by
\begin{equation}
f(x) = z_0 + Z_1 x 
\end{equation}
where $z_0 = \mu$ and the only constraint on $Z_1$ is that $\Sigma =
Z_1 \Sigma_P Z_1^T$.

This result is informative because it indicates that the matrix $Z_1$
is not uniquely determined. Indeed, if the prior covariance is the
identity, then any orthonormal matrix $Q$ can result in a new map with
$Z_1 Q$ replacing $Z_1$.

\section{Nonlinear model with Gaussian prior and Gaussian additive noise}
\label{app:additivegaussian}

In this appendix we provide detailed expressions for derivatives with
respect to optimization parameters in the case of a \textit{nonlinear}
forward model $h$, additive Gaussian observational error $\varepsilon
\sim N(0, \Sigma_n)$, and a Gaussian prior on the parameters
$x$.\footnote{These assumptions are typical for PDE-constrained
  inverse problems, and include the case of Gaussian priors derived
  from differential operators.} For simplicity we let the prior have
zero mean and identity covariance, $x \sim N(0, I)$.
In particular, we obtain derivative expressions needed to solve the
optimization problem when the map $f$ is represented with multivariate
polynomials.

Let $p$ be the prior density, $q$ the posterior density, and $d$ the
data. The problem setup is summarized as follows:
\begin{eqnarray*}
p(x) & \propto &  \exp{\left ( - \frac{1}{2} x^T x \right )} \\
d &=& h(z) + \varepsilon \\
q(z) &=&  \frac{1}{\beta} 
\exp{ \left( -\frac{1}{2} \left( \left \| h(z) - d \right \|_{\Sigma_n}^2 
      + z^T z \right) \right)}
\end{eqnarray*}
Using the map:
\begin{equation*}
z = f(x) 
\end{equation*}
we obtain the following $\tilde{p}$
\begin{equation*}
\tilde{p}(x) = \frac{1}{\beta} \exp{\left( -\frac{1}{2} 
\left(\|h \left ( f(x) \right )-d\|_{\Sigma_n}^2  + f(x)^Tf(x) -2 \log
  {\left|\det D_x f \right|} \right) 
\right ) } 
\end{equation*}
where 
\begin{equation*}
\|h \left ( f(x) \right )-d\|_{\Sigma_n}^2 \equiv 
\left(h \left ( f(x) \right ) - d\right)^T
 \Sigma_n^{-1} 
\left(h \left ( f(x) \right ) - d \right)
\end{equation*}
and $\det D_x f$ is the determinant of the Jacobian of the map.
Following (\ref{eq:constantT}), $T$ is given by
\begin{equation}
- 2 T(x) =  \| h ( f(x) ) - d \|^2_{\Sigma_n} + 
\| f(x) \|^2 - 2 \log \left| \det D_x f \right| - \| x \|^2 
\label{e:Tnonlinpoly}
\end{equation}

Assume that the map is given by the polynomial chaos expansion
(\ref{e:pcmap})
\begin{equation*}
f(x) = F^T \Psi(x).
\end{equation*}
We need to compute derivatives of $T(x; F)$ with respect to
elements of the matrix $F$.  Recall that $n$ is the dimension of the
model parameter vector. Hence
\begin{eqnarray}
z  =  f(x) & = & \( \Psi(x)^T  F I_n \) ^T = \left(I_n \otimes
  \Psi(x)^T \right) F(:) \nonumber\\[9pt]
\frac{\partial z}{\partial F} = \frac{\partial f(x)}{\partial F}  &=& I_n \otimes \Psi^T \label{eq:dzdF}\\[9 pt]
 \frac{\partial h \( f(x) \) }{\partial F} = \frac{\partial h \( F^T
   \Psi \)}{\partial F}  &=& 
\frac{\pd h(z) }{\pd z} 
\frac{\partial z}{\partial F} = 
\frac{\pd h}{\pd z} 
\( I_n \otimes \Psi^T \) \label{eq:dhdF}
\end{eqnarray}
and
\begin{eqnarray}
D_x f \equiv \frac{\partial f}{\partial x}  &=& F^T \frac{ \pd \Psi }{\pd x} \nonumber\\[9pt]
\frac{\partial \( \log\left| \det D_x f \right| \) }{\partial F} &=& \trace \left( \left[ D_x f \right]^{-1} \frac{ \partial  } {\partial F_{ij}  } \left[ D_x f \right] \right) \nonumber \\
&=& \left[ \frac{\pd \Psi}{\pd x} \left(F^T  \frac{\pd \Psi}{\pd x}  \right)^{-1} \right](:)
\end{eqnarray}
where the notation $\left[ A \right] (:)$ signifies a rearrangement of the
elements of the rectangular matrix $A$ into a single column vector.
Returning to (\ref{e:Tnonlinpoly}) above, we obtain the following expression for
the first derivatives of $T$ evaluated at $x$:
\begin{equation*}
 \frac{\partial T(x; F)}{\partial F} =   
- \( \frac {\partial h \( f(x) \)}{\partial F} \)^T \Sigma_n^{-1} 
 \( h \( f\(x\) \) - d \) 
-  \( \frac{\partial f(x)}{\partial F} \)^T f(x) +  \left[ \frac{\pd
    \Psi}{\pd x} \left(F^T  \frac{\pd \Psi}{\pd x}
  \right)^{-1}\right] (:)  %
\end{equation*}
Using~(\ref{eq:dzdF}) and (\ref{eq:dhdF})
\begin{eqnarray}
 \frac{\partial T(x; F)}{\partial F} & = &   
-  \(  \frac{\pd h(z)}{\pd z}^T \Sigma_n^{-1} 
 \( h \( f\(x\) \) - d \)  +f(x)  \) \otimes \Psi \nonumber \\
  & &  +  \left[ \frac{\pd
    \Psi}{\pd x} \left(F^T  \frac{\pd \Psi}{\pd x}
  \right)^{-1}\right] (:) , \label{eq:dTdF}
\end{eqnarray}
where derivatives with respect to $z$ are evaluated at $z= f(x)$.

To evaluate the second derivatives needed for Newton's method, one
needs to compute the second derivative of the logarithm of the
Jacobian determinant:
\begin{eqnarray*}
\frac{\partial \log | \det D_x f | }{\partial F_{ij}} &=& \left( \frac{\pd \Psi}{\pd x}(i, :) \left( D_x f \right)^{-1} \right) [j] \\
\frac{\partial \log{|  \det D_x f  |}}{\partial F_{ij}} &=& \frac{\pd \Psi}{\pd x}(i, :) \left( D_x f \right)^{-1} e_j \\
\frac{\partial^2 \log{|  \det D_x f  |}}{\partial F_{ij} \partial F_{mn}} &=& - \frac{d\Psi}{d x}(i, :)  \left(  D_x f \right)^{-1}  \left(e_n \frac{\pd \Psi}{\pd x} (m, :) \right)  \left( D_x f \right)^{-1}  e_j \\
\frac{\partial^2 \log{|  \det D_x f |}}{\partial F_{ij} \partial F_{mn}} &=& - \left( \frac{\pd \Psi}{\pd x}(i, :)  \left(   D_x f  \right)^{-1} \right)  e_n \left( \frac{\pd \Psi}{\pd x} (m, :) \left(  D_x f  \right)^{-1}  \right)  e_j 
\end{eqnarray*}
where $e_j$ is a vector of all zeros except for a 1 in the $j$th
location. It is clear that the only expensive quantity to be computed
is the matrix
\begin{equation*}
\mathcal{M} =  \frac{\pd \Psi}{\pd x} \left(  D_x f  \right)^{-1}   
\end{equation*}
\begin{equation*}
\frac{\partial^2 \log| \det D_x f |}{\partial F_{ij}\partial F_{mn}} = - \mathcal{M}(i,n) \mathcal{M}(m,j) 
\end{equation*}

\begin{eqnarray}
 \frac{\partial^2 T}{\partial F_{ij}\pd F_{mn}} &  =  &   -
  \( \frac{\pd h( f(x))}{\pd F_{ij}} \)^T\Sigma_n^{-1} \frac{\pd h( f(x)) }{\pd F_{mn}} 
-  \left( \frac{\pd^2 h( f(x))}{\pd F_{ij}  \pd F_{mn}} \right)^T\Sigma_n^{-1} (h( f(x))-d) \nonumber\\
&&  - \left( \frac{\pd f(x)}{\pd F_{ij}} \right)^T \frac{\pd f(x)}{\pd F_{mn}} - \mathcal{M}(i,n) \mathcal{M}(m,j) .
\end{eqnarray}
Notice that storage of the object ${\pd^2 h( f(x))}/{\pd F \pd F}$
requires special care in implementation, since it is the second
derivative of a vector-valued function with respect to a vector. We
find that it is best stored as a third-order tensor of dimension $m
\times \ell \times \ell$, where $m$ is the number of observations and
$\ell$ is the number of optimization variables. Alternatively, one
could instead retain only the second derivative of the term $\| h(z) -
d \|_{\Sigma_n}^2$ with respect to $F$, which can be stored in
standard matrix format. For example, consider the special case in
which the forward model $h$ is approximated using the polynomial chaos
expansion
\begin{equation*}
h(z) = C^T \Psi_h(z) 
\end{equation*}
where $C$ is the output matrix and has $m$ columns.
The second derivative of $\| h(z) - d \|_{\Sigma_n}^2$ with
respect to $F$ is computed as follows:
\begin{eqnarray*}
\( \frac{\pd h( f(x))}{\pd F} \)^T\Sigma_n^{-1} \( \frac{\pd h( f(x)) }{\pd F} \)
& + & \sum\limits_{j=1}^{m} \sum\limits_{i=1}^{m} \( \frac{\pd^2 h_i( f(x))}{\pd F \pd F}\)^T\left[\Sigma_n^{-1}\right](i,j) (h_j( f(x))-d_j)  \\
& = &  \left( C^T \frac{\pd \Psi_h(z)}{\pd z} \frac{\pd f(x)}{\pd F}
\right)^T \Sigma_n^{-1} \left(  C^T \frac{\pd \Psi_h(z)}{\pd z}
  \frac{\pd f(x)}{\pd F} \right)   \\
&& + \sum\limits_{j=1}^{n} \left( C^T \frac{\pd^2 \Psi_h(z)}{\pd z \pd
     z_j}   \frac{\pd f(x)}{\pd F} \right)^T \Sigma_n^{-1} (h( f(x))-d)
 \left( \frac{\pd f_j(x)}{\pd F} \right)   .
\end{eqnarray*}
Putting everything together we arrive at the following expression for ${\pd^2 T(x)}/{\pd F \pd F}$:
\begin{eqnarray}
\frac{\partial^2 T}{\partial F \pd F} & = & - \( \( \frac{\pd \Psi_h(f(x))}{\pd f(x)} \)^T C \Sigma_n^{-1} C^T \frac{\pd \Psi_h(f(x))}{\pd f(x)}  \) \otimes \Psi \Psi^T \nonumber\\
&&  - \( \sum\limits_{j=1}^{n} \left(  \frac{\pd^2 \Psi_h(f(x))}{\pd f(x) \pd
     f_j(x)}   \right)^T C \Sigma_n^{-1} ( C^T \Psi_h(f(x)) -d) e_j^T+ I_n \) \otimes \Psi \Psi^T   \nonumber\\  && - {\mathcal M}^T \otimes {\mathcal M}   \label{d2TdF2}.
\end{eqnarray}

\end{document}